\documentclass[aps,prd,reprint,preprintnumbers,showpacs,floatfix,nofootinbib,superscript address]{revtex4-2}
\usepackage[utf8]{inputenc}
\usepackage{amssymb}
\usepackage{stix}
\usepackage{amsmath}
\usepackage{mathtools}
\usepackage[dvipsnames]{xcolor}
\usepackage{xspace}
\usepackage{siunitx}
\usepackage{multirow}
\usepackage{graphicx}
\usepackage{xstring}
\usepackage{etoolbox}
\usepackage{notoccite}
\usepackage{natbib}
\usepackage{mathrsfs}
\usepackage{lineno}
\usepackage{tensor}
\usepackage{accents}
\allowdisplaybreaks
\DeclareSIUnit\parsec{pc}

\usepackage{xspace}
\usepackage{xstring}
\usepackage{titlesec}
\usepackage{stix}
\usepackage{soul}
\allowdisplaybreaks

\newrobustcmd{\pea}[1]{%
	\emph{#1}\textbf{\ \ \ ---}
}
\titleformat{\paragraph}[runin]{\normalfont\normalsize\bfseries}{\emph\theparagraph}{1em}{\pea}

\newcommand{\myDel}[1]{{\color{red}\ifmmode\cancel{#1}\else\st{#1}\fi}}

\def\tfrac#1#2{{\textstyle\frac{#1}{#2}}}

\usepackage{etoolbox}
\usepackage{tensor}

\newrobustcmd{\PSALTer}{\textit{PSALTer}\xspace}
\newrobustcmd{\MPl}{%
	{m_{\mathrm{P}}} 
}
\newrobustcmd{\LPl}{%
	{\ell_{\mathrm{P}}} 
}
\newrobustcmd{\LPhy}{%
{\ell_{\mathrm{Phy}}} 
}
\newrobustcmd{\LPhyprime}{%
{\ell^\prime_{\mathrm{Phy}}} 
}
\newrobustcmd{\FieldOmega}[1]{%
  \tensor{\omega}{#1}
}
\newrobustcmd{\FieldTau}[1]{%
  \tensor{\tau}{#1}
}
\newrobustcmd{\FieldSpin}[1]{%
  \tensor{\sigma}{#1}
}
\newrobustcmd{\FieldPert}[1]{%
  \tensor{f}{#1}
}
\newrobustcmd{\FieldMomentum}[1]{%
  \tensor{k}{#1}
}
\newrobustcmd{\FieldVScalar}[1]{%
  \tensor{\delta B_\perp}{#1}
}
\newrobustcmd{\FieldVScalarDagger}[1]{%
  \tensor{\delta B_\perp^*}{#1}
}
\newrobustcmd{\FieldVVector}[1]{%
  \tensor{\delta B_\parallel}{#1}
}
\newrobustcmd{\FieldVVectorDagger}[1]{%
  \tensor{\delta B_\parallel^*}{#1}
}
\newrobustcmd{\FieldF}[1]{%
  \tensor{F}{#1}
}
\newrobustcmd{\SCovD}[1]{%
	\tensor{\mathscr{D}}{#1}
}
\newrobustcmd{\FieldH}[1]{%
  \tensor{h}{#1}
}
\newrobustcmd{\FieldHHat}[1]{%
  \tensor{\hat h}{#1}
}
\newrobustcmd{\FieldB}[1]{%
  \tensor{b}{#1}
}
\newrobustcmd{\FieldBHat}[1]{%
  \tensor{\hat b}{#1}
}
\newrobustcmd{\FieldV}[1]{%
  \tensor{B}{#1}
}
\newrobustcmd{\FieldVHat}[1]{%
  \tensor{\hat B}{#1}
}
\newrobustcmd{\FieldVV}[1]{%
  \tensor{V}{#1}
}
\newrobustcmd{\FieldSigma}[1]{%
  \tensor{\Sigma}{#1}
}
\newrobustcmd{\FieldATilde}[1]{%
  \tensor{\tilde A}{#1}
}
\newrobustcmd{\Curtright}[1]{%
  \tensor{C}{#1}
}
\newrobustcmd{\FieldAHat}[1]{%
  \tensor{\hat A}{#1}
}
\newrobustcmd{\Alp}[1]{%
  \tensor{\alpha}{_#1}
}
\newrobustcmd{\Bet}[1]{%
  \tensor{\beta}{_#1}
}
\newrobustcmd{\Gam}[1]{%
  \tensor{\gamma}{_#1}
}
\newrobustcmd{\MassSquared}[1]{%
	\left(\tensor{m}{_{#1}}\right)^2
}
\newrobustcmd{\FieldA}[1]{%
  \tensor{A}{#1}
}
\newrobustcmd{\FieldDelta}[1]{%
	\tensor*{\delta}{#1}
}
\newrobustcmd{\FieldEta}[1]{%
	\tensor{\eta}{#1}
}
\newrobustcmd{\FieldX}[1]{%
	\tensor{x}{#1}
}
\newrobustcmd{\FieldY}[1]{%
	\tensor{y}{#1}
}
\newrobustcmd{\FieldG}[1]{%
	\tensor{g}{#1}
}
\newrobustcmd{\FieldGHat}[1]{%
	\tensor{\hat g}{#1}
}
\newrobustcmd{\FieldLambda}[1]{%
	\tensor{\Lambda}{#1}
}
\newrobustcmd{\PD}[1]{%
	\tensor{\partial}{#1}%
}
\newrobustcmd{\Danger}[1]{%
	{\color{red}{#1}}	
}
\newrobustcmd{\Coupling}[1]{%
	{a_{#1}}	
}
\newrobustcmd{\CouplingB}[1]{%
	{b_{#1}}	
}
\newrobustcmd{\MAGg}[1]{%
	\tensor{g}{#1}
}
\newrobustcmd{\MAGl}[1]{%
	\tensor{\xi}{#1}
}
\newrobustcmd{\MAGP}[1]{%
  \tensor{P}{#1}
}
\newrobustcmd{\MAGC}[1]{%
  \tensor{C}{#1}
}
\newrobustcmd{\ShiftA}[1]{%
  \tensor{A}{#1}
}
\newrobustcmd{\ShiftB}[1]{%
  \tensor{B}{#1}
}
\newrobustcmd{\ShiftC}[1]{%
  \tensor{C}{#1}
}
\newrobustcmd{\MAGA}[1]{%
  \tensor{\Gamma}{#1}
}
\newrobustcmd{\Ja}[1]{%
	\tensor{\smash{\overset{\scalebox{0.6}{$\mathrm{(M)}$}}{\mathcal{J}}}}{#1}
}
\newrobustcmd{\Jv}[1]{%
	\tensor{\smash{\overset{\scalebox{0.6}{$\mathrm{(N)}$}}{\mathcal{J}}}}{#1}
}
\newrobustcmd{\MAGH}[1]{%
	\tensor{H}{#1}
}
\newrobustcmd{\MAGHHat}[1]{%
	\tensor{\hat H}{#1}
}
\newrobustcmd{\MAGF}[1]{%
	\tensor{R}{#1}
}
\newrobustcmd{\MAGFt}[1]{%
	\tensor{\tilde{R}}{#1}
}
\newrobustcmd{\MAGFh}[1]{%
	\tensor{\hat{R}}{#1}
}
\newrobustcmd{\MAGFa}[1]{%
	\tensor{\check{R}}{#1}
}
\newrobustcmd{\MAGFb}[1]{%
	\tensor{R}{#1}
}
\newrobustcmd{\MAGT}[1]{%
	\tensor{T}{#1}
}
\newrobustcmd{\MAGTTilde}[1]{%
	\tensor{\tilde T}{#1}
}
\newrobustcmd{\MAGTt}[1]{%
	\tensor{\hat{T}}{#1}
}
\newrobustcmd{\MAGt}[1]{%
	\tensor{t}{#1}
}
\newrobustcmd{\MAGQ}[1]{%
	\tensor{Q}{#1}
}
\newrobustcmd{\MAGQt}[1]{%
	\tensor{\hat{Q}}{#1}
}
\newrobustcmd{\MAGq}[1]{%
	\tensor{q}{#1}
}
\newrobustcmd{\MAGV}[1]{%
	\tensor{Z}{#1}
}
\newrobustcmd{\MAGVt}[1]{%
	\tensor{\tilde{Z}}{#1}
}

\newrobustcmd{\xMAGA}[1]{%
  \tensor{A}{#1}
}
\newrobustcmd{\xMAGF}[1]{%
	\tensor{\mathcal{F}}{#1}
}
\newrobustcmd{\xMAGFt}[1]{%
	\tensor{\tilde{\mathcal{F}}}{#1}
}
\newrobustcmd{\xMAGFa}[1]{%
	\tensor{\mathcal{F}}{^{(14)}#1}
}
\newrobustcmd{\xMAGFb}[1]{%
	\tensor{\mathcal{F}}{^{(13)}#1}
}
\newrobustcmd{\xMAGT}[1]{%
	\tensor{\mathcal{T}}{#1}
}
\newrobustcmd{\xMAGTt}[1]{%
	\tensor{\tilde{\mathcal{T}}}{#1}
}
\newrobustcmd{\xMAGQ}[1]{%
	\tensor{\mathcal{Q}}{#1}
}
\newrobustcmd{\xMAGQt}[1]{%
	\tensor{\tilde{\mathcal{Q}}}{#1}
}
\newrobustcmd{\xMAGTh}[1]{%
	\tensor{\hat{\mathcal{T}}}{#1}
}
\newrobustcmd{\xMAGV}[1]{%
	\tensor{\mathcal{V}}{#1}
}
\newrobustcmd{\g}[1]{%
	\tensor{g}{#1}%
}
\newrobustcmd{\rcCon}[1]{%
	\tensor*{\Gamma}{#1}%
}
\newrobustcmd{\rCon}[1]{%
	\tensor{\mathring{\Gamma}}{#1}%
}
\newrobustcmd{\Con}[1]{%
	\tensor{\mathring{\Gamma}}{#1}%
}
\newrobustcmd{\B}[1]{%
	\tensor{B}{#1}%
}
\newrobustcmd{\rD}[1]{%
	\tensor{\mathring{\nabla}}{#1}%
}
\newrobustcmd{\rcD}[1]{%
	\tensor{\nabla}{#1}%
}
\newrobustcmd{\rR}[1]{%
	\tensor{\mathring{R}}{#1}%
}
\newrobustcmd{\rcR}[1]{%
	\tensor{R}{#1}%
}

\usepackage{hyperref}
\hypersetup{%
     colorlinks = true,%
     linkcolor = Blue,%
     citecolor = Blue,%
     filecolor = Blue,%
     urlcolor = Blue%
     }%

\usepackage[capitalize]{cleveref}

\usepackage{layouts}

\begin{document}

\title{Every Poincar\'e gauge theory is conformal: a compelling case for dynamical vector torsion}

\author{Will Barker}
\email{wb263@cam.ac.uk}
\affiliation{Astrophysics Group, Cavendish Laboratory, JJ Thomson Avenue, Cambridge CB3 0HE, UK}
\affiliation{Kavli Institute for Cosmology, Madingley Road, Cambridge CB3 0HA, UK}

\author{Michael Hobson}
\email{mph13@cam.ac.uk}
\affiliation{Astrophysics Group, Cavendish Laboratory, JJ Thomson Avenue, Cambridge CB3 0HE, UK}

\author{Anthony Lasenby}
\email{a.n.lasenby@mrao.cam.ac.uk}
\affiliation{Astrophysics Group, Cavendish Laboratory, JJ Thomson Avenue, Cambridge CB3 0HE, UK}
\affiliation{Kavli Institute for Cosmology, Madingley Road, Cambridge CB3 0HA, UK}

\author{Yun-Cherng Lin}
\email{yunclin54@gmail.com}
\affiliation{Astrophysics Group, Cavendish Laboratory, JJ Thomson Avenue, Cambridge CB3 0HE, UK}
\affiliation{Kavli Institute for Cosmology, Madingley Road, Cambridge CB3 0HA, UK}

\author{Zhiyuan Wei}
\email{zhiyuan.wei@uni-ulm.de}
\affiliation{Astrophysics Group, Cavendish Laboratory, JJ Thomson Avenue, Cambridge CB3 0HE, UK}
\affiliation{Institute for Complex Quantum Systems, Ulm University, Albert-Einstein-Allee, 89069 Ulm, DE}

\begin{abstract}
	The Poincar\'e gauge theory (PGT) of gravity provides a viable formulation of general relativity (Einstein--Cartan theory), and a popular model-building framework for modified gravity with torsion. Notoriously, however, the PGT terms which propagate vector torsion lead to strongly-coupled ghosts: the modern view is that only scalar torsion can propagate. To fix this, we revisit the concept of embedding explicit mass scales in scale-invariant theories, showing how the Klein--Gordon theory naturally leads to a slowly-rolling inflaton. We then show that the unique scale-invariant embedding of PGT leads to two new terms, one of which is the Maxwell term for vector torsion. We provide the full spectrum of quantum particles in the resulting theory. Our result means that every PGT is conformal and -- after a two-decade hiatus -- vector torsion is back on the menu.
\end{abstract}

\maketitle

\paragraph*{Introduction to torsion} General relativity (GR) offers a remarkably successful description of gravity as spacetime curvature~\cite{Einstein:1915}.\footnote{The term `GR' refers strictly to the textbook, metric-based theory with an Einstein--Hilbert action in Riemann spacetime.} An open problem is the degree -- if any -- to which spacetime \emph{torsion} also participates in the gravitational interaction. There are excellent theoretical reasons to consider torsion: it was shown by Utiyama~\cite{Utiyama:1956sy}, Kibble~\cite{Kibble:1961ba} and Sciama~\cite{Sciama:1962} that when fermions are coupled to gravity, the natural result is a local gauge theory not only of spacetime translations (i.e. the diffeomorphisms of GR), but also of spatial rotations and Lorentz boosts. In this \emph{Poincar\'e gauge theory} (PGT) of gravity~\cite{Blagojevic:2002}, the fundamental fields can be\footnote{By contrast with GR, the term `PGT' encompasses \emph{any} action formulated with the 40 d.o.f of the tetrad and spin connection in Riemann--Cartan spacetime. In the absence of fermionic matter, the dynamics of PGT are also fully captured by the alternative \emph{post-Riemannian} field parameterisation of 34 d.o.f in~$\big\{\FieldG{_{\mu\nu}},\MAGT{^\lambda_{\mu\nu}}\big\}$, though the Poincar\'e symmetry is hidden.} the metric tensor~$\FieldG{_{\mu\nu}}$ and an a priori independent affine connection~$\MAGA{^\alpha_{\mu}_{\nu}}$ which gives rise (as we will show later) to torsion and curvature as the field strength tensors of spacetime translations and rotations, respectively:
\begin{subequations}
\begin{align}
	\MAGT{^\alpha_\mu_\nu}&\equiv 2\MAGA{^\alpha_{[\mu}_{\nu]}},\label{ECTorsion}\\
	\MAGF{^\rho_\sigma_{\mu\nu}}&\equiv 2\left(\PD{_{[\mu|}}\MAGA{^\rho_{|\nu]}_\sigma}+\MAGA{^\rho_{[\mu|}_\alpha}\MAGA{^\alpha_{|\nu]}_\sigma}\right).\label{ECCurvature}
\end{align}
\end{subequations}
To connect this to textbook GR, if~$\MAGT{^\alpha_\mu_\nu}\to 0$ in~\cref{ECTorsion} then~$\MAGA{^\mu_\nu_\rho}$ loses its independence from the metric\footnote{We assume the metricity condition.} and is fixed by the Christoffel formula~$\MAGA{^\mu_\nu_\rho}\to\rCon{^\mu_{\nu\rho}}\equiv\FieldG{^{\mu\lambda}}\big(\PD{_{(\nu}}\g{_{\rho)\lambda}}-\frac{1}{2}\PD{_{\lambda}}\g{_{\nu\rho}}\big)$. In this torsion-free limit,~\cref{ECCurvature} defines the \emph{Riemannian} curvature~$\MAGF{^\rho_\sigma_{\mu\nu}}\to\rR{^\rho_\sigma_{\mu\nu}}$. Despite concrete theoretical provenance from PGT, there is uncertainty over what a signal for torsion might look like in the phenomena. This is in contrast to curvature, whose presence can be inferred e.g. through the geodesic motion of test particles. To understand this uncertainty, notice that we can always perform the field reparameterisation\footnote{Note~$\MAGA{^\mu_\nu_\rho}-\rCon{^\mu_\nu_\rho}$ is sometimes called \emph{contortion}, and is linear in torsion.}~$\MAGA{}\mapsto\rCon{}+\MAGT{}$ (leading to~$\MAGF{}\mapsto\rR{}+\PD{}\MAGT{}+\MAGT{}^2$ with indices suppressed). This \emph{post-Riemannian expansion} from variables~$\big\{\FieldG{_{\mu\nu}},\MAGA{^\alpha_{\mu}_{\nu}}\big\}$ to~$\big\{\FieldG{_{\mu\nu}},\MAGT{^\lambda_{\mu\nu}}\big\}$ reveals all PGT models to be nothing more than metric-based gravity coupled -- minimally or otherwise -- to the field~$\MAGT{^\lambda_{\mu\nu}}\equiv\MAGT{^\lambda_{[\mu\nu]}}$ as if it were an exotic kind of matter. Notice how there are infinitely many such models: if there are models for all experimental outcomes, then the theoretical framework ceases to be useful. The classic example of a useful PGT which establishes a predictive baseline is the minimal Einstein--Cartan theory (ECT)~\cite{Cartan:1922,Cartan:1923,Cartan:1924,Cartan:1925,Einstein:1925, Einstein:1928,Einstein:19282}. ECT is defined as the specific case of PGT which shares the Einstein--Hilbert action of GR
\begin{equation}\label{ECT}
	S_{\text{ECT}}\equiv\int\mathrm{d}^4x\sqrt{-g}\Big[-\frac{\MPl{}^2}{2}\MAGF{}
	-\MPl{}^2\Lambda+\text{matter}\Big],
\end{equation}
where~$\MAGF{}\equiv\MAGF{^\mu_\mu}\equiv\MAGF{^{\mu\nu}_{\mu\nu}}$ and~$\MPl{}\equiv 1/\sqrt{\kappa}$ is the Planck mass, for~$\kappa$ the Einstein constant. With~$\Lambda\equiv 0$ and in the absence of matter,~\cref{ECT} implies the vacuum Einstein equations~$\MAGF{_{\mu\nu}}= 0$ and the auxiliary equation~$\MAGT{^\lambda_{\mu\nu}}= 0$. Inclusion of matter leads to a contact spin-torsion coupling; the torsion integrates out to leave effective four-Fermi interactions~\cite{Kibble:1961ba, Rodichev:1961,Freidel:2005sn,Alexandrov:2008iy,Shaposhnikov:2020frq,Karananas:2021zkl,Rigouzzo:2023sbb}. Such interactions have many pheonomenological applications~\cite{Freidel:2005sn,Bauer:2008zj,Poplawski:2011xf,Diakonov:2011fs,Khriplovich:2012xg,Magueijo:2012ug,Khriplovich:2013tqa,Markkanen:2017tun,Carrilho:2018ffi,Enckell:2018hmo,Rasanen:2018fom,BeltranJimenez:2019hrm,Rubio:2019ypq,Shaposhnikov:2020geh,Karananas:2020qkp,Langvik:2020nrs,Shaposhnikov:2020gts,Mikura:2020qhc,Shaposhnikov:2020aen,Kubota:2020ehu,Enckell:2020lvn,Iosifidis:2021iuw,Bombacigno:2021bpk,Racioppi:2021ynx,Cheong:2021kyc,Dioguardi:2021fmr,Piani:2022gon,Dux:2022kuk,Rigouzzo:2022yan,Pradisi:2022nmh,Salvio:2022suk,Rasanen:2022ijc,Gialamas:2022xtt,Gialamas:2023emn,Gialamas:2023flv,Piani:2023aof,Poisson:2023tja,Rigouzzo:2023sbb,Barker:2023fem,Karananas:2023zgg,Martini:2023apm,He:2024wqv}, such as to fermionic dark matter production. Thus, non-propagating torsion might reasonably be constrained (albeit weakly) by dark matter abundances. Propagating torsion, on the other hand, can come from any number of next-to-minimal~$\big\{\FieldG{_{\mu\nu}},\MAGT{^\lambda_{\mu\nu}}\big\}$ models: it could be sufficiently heavy to evade current bounds, or conveniently account for any new scalar or vector bosons that might be observed (note that~$\MAGT{^\lambda_{\mu\nu}}$ contains various bosons up to spin-two). In other words, the phenomological richness of the full~$\big\{\FieldG{_{\mu\nu}},\MAGT{^\lambda_{\mu\nu}}\big\}$ theory-space, which has infinitely many parameters, renders it non-predictive.

\paragraph*{No-go theorem for vector torsion} Happily, the literature has a convincing way to deal with this problem. Just as PGT motivates torsion, so the usually assumed structure of its action also endows torsionful model-building with predictivity. The logic is that PGT should be restricted (see e.g.~\cite{Hayashi:1967se,Hayashi:1980qp,Yo:1999ex,Yo:2001sy,Puetzfeld:2004yg}) to a \emph{Yang--Mills-type} action
\begin{align}\label{PGTAction}
	&S_{\text{PGT}}\equiv\int\mathrm{d}^4x\sqrt{-g}\Big[
	-\frac{\MPl^2}{2}\MAGF{}
	-\MPl{}^2\Lambda
	+\Alp{1}\MAGF{}^2
	+\MAGF{_{\mu\nu}}\big(
	\Alp{2}\MAGF{^{\mu\nu}}
	\nonumber\\ &
	+\Alp{3}\MAGF{^{\nu\mu}}\big)
	+\MAGF{_{\mu\nu\sigma\lambda}}\big(
	\Alp{4}\MAGF{^{\mu\nu\sigma\lambda}}
	+\Alp{5}\MAGF{^{\mu\sigma\nu\lambda}}
	+\Alp{6}\MAGF{^{\sigma\lambda\mu\nu}}\big)
	\nonumber\\ &
	+\MAGT{_{\mu\nu\sigma}}\big(\Bet{1}\MAGT{^{\mu\nu\sigma}}
	+\Bet{2}\MAGT{^{\nu\mu\sigma}}\big)
	+\Bet{3}\MAGT{_{\mu}}\MAGT{^{\mu}}
	+\text{matter}
	\Big],
\end{align}
where~$\MAGT{_\mu}\equiv\MAGT{^\nu_{\nu\mu}}$, and~\cref{PGTAction} extends~\cref{ECT} by only admitting quadratic field strength invariants.\footnote{In this letter we omit the parity-odd invariants only out of simplicity; there are no convincing theoretical grounds for excluding them~\cite{Karananas:2014pxa,Blagojevic:2018dpz}.} Evidently,~\cref{PGTAction} occupies just a small corner of the~$\big\{\FieldG{_{\mu\nu}},\MAGT{^\lambda_{\mu\nu}}\big\}$ theory-space: the Yang--Mills structure is similar to the gauge boson sector of the standard model (SM), and it imposes surprising limits on the phenomenology. Firstly, the parameters in~\cref{PGTAction} must be carefully tuned to eliminate ghosts and tachyons from the linearised spectrum~\cite{Sezgin:1981xs,Blagojevic:1983zz,Blagojevic:1986dm,Kuhfuss:1986rb,Yo:1999ex,Yo:2001sy,Blagojevic:2002,Puetzfeld:2004yg,Yo:2006qs,Shie:2008ms,Nair:2008yh,Nikiforova:2009qr,Chen:2009at,Ni:2009fg,Baekler:2010fr,Ho:2011qn,Ho:2011xf,Ong:2013qja,Puetzfeld:2014sja,Karananas:2014pxa,Ni:2015poa,Ho:2015ulu,Karananas:2016ltn,Obukhov:2017pxa,Blagojevic:2017ssv,Blagojevic:2018dpz,Tseng:2018feo,Lin:2018awc,BeltranJimenez:2019acz,Zhang:2019mhd,Aoki:2019rvi,Zhang:2019xek,Jimenez:2019qjc,Lin:2019ugq,Percacci:2019hxn,Barker:2020gcp,BeltranJimenez:2020sqf,MaldonadoTorralba:2020mbh,Barker:2021oez,Marzo:2021esg,Marzo:2021iok,delaCruzDombriz:2021nrg,Baldazzi:2021kaf,Annala:2022gtl,Mikura:2023ruz,Mikura:2024mji,Barker:2024ydb,Karananas:2024xja}. Even then, inconsistencies usually reappear at non-linear order due to strongly coupled particles~\cite{Moller:1961,Pellegrini:1963,Hayashi:1967se,Cho:1975dh, Hayashi:1979qx,Hayashi:1979qx,Dimakis:1989az,Dimakis:1989ba,Lemke:1990su,Hecht:1990wn,Hecht:1991jh,Yo:2001sy,Afshordi:2006ad,Magueijo:2008sx,Charmousis:2008ce,Charmousis:2009tc,Papazoglou:2009fj,Baumann:2011dt,Baumann:2011dt,DAmico:2011eto,Gumrukcuoglu:2012aa,Wang:2017brl,Mazuet:2017rgq,BeltranJimenez:2020lee,JimenezCano:2021rlu,Barker:2022kdk,Delhom:2022vae,Annala:2022gtl,Barker:2022kdk,Barker:2023fem,Karananas:2024hoh} (for general strong coupling see~\cite{Vainshtein:1972sx,Deffayet:2001uk,Deffayet:2005ys,Charmousis:2008ce,Charmousis:2009tc,Papazoglou:2009fj,deRham:2014zqa,Deser:2014hga,Wang:2017brl}) and the breaking of so-called `accidental' symmetries which only survive linearly~\cite{Velo:1969txo,Aragone:1971kh,Cheng:1988zg,Hecht:1996np,Chen:1998ad,Yo:1999ex,Yo:2001sy,Blixt:2018znp,Blixt:2019ene,Blixt:2020ekl,Krasnov:2021zen,Bahamonde:2021gfp,Delhom:2022vae} (see also~\cite{Hayashi:1980qp,Blagojevic:1983zz,Blagojevic:1986dm,Yo:2001sy,Blagojevic:2002,Ong:2013qja,Blagojevic:2013dea,Blagojevic:2013taa,Blagojevic:2018dpz,BeltranJimenez:2019hrm,Aoki:2020rae,Barker:2021oez}). Weakly coupled, consistent PGTs are very rare: the current consensus is that only spin-zero scalar torsion is allowed to propagate, and then only with very specific non-linear interactions~\cite{Hecht:1996np,Chen:1998ad,Yo:1999ex,Yo:2001sy} (see e.g.~\cite{Yo:2006qs,Shie:2008ms,Chen:2009at,Baekler:2010fr,Ho:2011qn,Ho:2011xf,Ho:2015ulu,Tseng:2018feo,Zhang:2019mhd,Zhang:2019xek,MaldonadoTorralba:2020mbh,delaCruzDombriz:2021nrg} for applications,~\cite{Puetzfeld:2004yg,Ni:2009fg,Puetzfeld:2014sja,Ni:2015poa,Barker:2020gcp} for reviews, and~\cite{BeltranJimenez:2019acz,BeltranJimenez:2019esp,Percacci:2020ddy,BeltranJimenez:2020sqf,Marzo:2021esg,Piva:2021nyj,Marzo:2021iok,Iosifidis:2021xdx,Jimenez-Cano:2022sds,Iosifidis:2023pvz} for similar analyses of spacetime non-metricity). Spin-one vector torsion, on the other hand, is ruled out~\cite{Yo:1999ex,Yo:2001sy}. Purely vector torsion can only be linearly propagated by the~$\MAGF{_{[\mu\nu]}}\MAGF{^{[\mu\nu]}}$ term, reached by setting~$\Alp{2}+\Alp{3}=0$ in~\cref{PGTAction}. However, that term carries in strongly-coupled ghosts, so the resulting theory is inconsistent~\cite{Yo:1999ex,Yo:2001sy}. In a recent attempt to evade this no-go theorem, we introduced multiplier fields which pacify this strong coupling~\cite{Barker:2023fem}. A fair criticism of our approach, however, was that it effectively reduced~$\MAGF{_{[\mu\nu]}}\MAGF{^{[\mu\nu]}}$ to the post-Riemannian Maxwell term~$\PD{_{[\mu}}\MAGT{_{\nu]}}\PD{^{[\mu}}\MAGT{^{\nu]}}$ which -- though non-linearly consistent -- is absent from~\cref{PGTAction}. This Maxwell term can still be added by hand, as was done for example in~\cite{Mikura:2023ruz}. The problem with adding terms manually is that, since almost any desired dynamics can then be fabricated, the resulting models are less predictive and compelling. To summarise: every PGT in~\cref{PGTAction} has a post-Riemannian expansion, but not every covariant model of~$\big\{\FieldG{_{\mu\nu}},\MAGT{^\lambda_{\mu\nu}}\big\}$ is the post-Riemannian expansion of a PGT in~\cref{PGTAction}.

\paragraph*{In this letter} We show that the~$\PD{_{[\mu}}\MAGT{_{\nu]}}\PD{^{[\mu}}\MAGT{^{\nu]}}$ term is a Yang--Mills-type term in disguise, without which~\cref{PGTAction} is actually \emph{incomplete}. Our reasoning is that PGT is not the most fundamental gauge theory of gravity. Recently we developed an \emph{extended}\footnote{Given the considerations to be presented in this letter, perhaps it would be more appropriate to replace `extended' with `economical'. Nonetheless, we retain the original nomenclature for consistency with the existing literature.} version of \emph{Weyl gauge theory} (eWGT), which gauges the full conformal group~\cite{Lasenby:2015dba,Hobson:2020bms,Hobson:2020doi,Hobson:2021iwy,Hobson:2023rdw}. We will show that the covariant derivative of eWGT reduces precisely to that of PGT when it is expressed in terms of scale-invariant variables (equivalent to fixing the scale gauge, and so breaking the conformal symmetry to Poincar\'e symmetry). The Yang--Mills-type action of eWGT reduces to~\cref{PGTAction} plus new terms (to be shown in~\cref{eWGTAction}). This provides a concrete motivation for vector torsion, which was hitherto lacking. Before obtaining these results, we introduce scale-invariant embeddings with a simple toy model, which is nonetheless interesting.

 \begin{figure}
 	\includegraphics[width=1\linewidth]{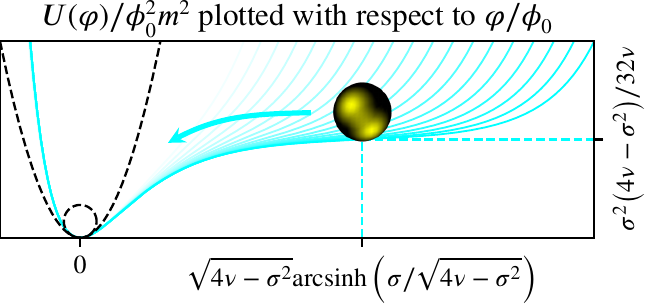}
	 \caption{Massive scalar perturbations in~\cref{EinsteinKleinGordon} have a unique, non-linear, scale-invariant embedding given in~\cref{ToyModel}. This embedding can lead to only one qualitative alteration in~\cref{FullTheory,Potential}: the quadratic potential (dashed, with a unit-circle illustrating the mass scale~$m$) develops a non-linear plateau (cyan) whose depth and height depend on model parameters~$\nu$ and~$\sigma$. The plateau is consistent with current CMB constraints on a slowly-rolling inflaton (arrow).}
 	\label{PotentialFig}
 \end{figure}

\paragraph*{Scale-invariant embedding} For our toy model, we work just with the matter sector. Let us suppose we know that some scalar perturbations~$\delta\varphi$ have a Klein--Gordon mass~$m$ 
\begin{equation}\label{EinsteinKleinGordon}
	S=\int\mathrm{d}^4x\sqrt{-g}\left[\tfrac{1}{2}\PD{_\mu}\delta\varphi\ \PD{^\mu}\delta\varphi-\tfrac{1}{2}m^2\delta\varphi^2+\text{gravity}\right],
\end{equation}
where we allow the theory to be minimally coupled to a metric-based gravity theory (e.g. GR). This situation already arises in cosmology: we typically infer that the canonical inflaton potential is approximated by a mass term~$U(\varphi)=m^2\varphi^2/2+\dots$ in order to facilitate reheating into the SM plasma after the end of inflation\footnote{Strictly, this is only true for `textbook' slow-roll inflation~\cite{Hobson:2006se,Vazquez:2018qdg,Achucarro:2022qrl}. We particularly note that Higgs inflation does not require a reheating mass~\cite{Bezrukov:2007ep}.}~\cite{Allahverdi:2010xz}. This inference does not, however, place constraints on~$U(\varphi)$ beyond the~$\delta\varphi$ linear regime. For a host of reasons the most favoured non-linear completion envisages~$\varphi$ rolling off a plateau in~$U(\varphi)$ sufficiently slowly to drive~$N\sim 55$ e-folds of inflation~\cite{Planck:2013jfk,Planck:2015sxf,Planck:2018jri}. Quite how this plateau should be motivated through fundamental theory is not yet settled. What we want to show here is that, starting purely from the linear model in~\cref{EinsteinKleinGordon}, there is a way to build up the non-linear plateau by asking that~$m$ be a dynamically acquired mass scale in a locally scale-invariant `embedding' theory (see~\cref{PotentialFig}). There is precedent here, since~\cref{EinsteinKleinGordon} also describes pions~\cite{Georgi:1984zwz,Donoghue:1992dd,Manohar:2000dt,Scherer:2012xha}. Pion masses, however, emerge long after reheating, through chiral and electroweak symmetry breaking~\cite{Higgs:1964pj,Guralnik:1964eu}. Moreover, the absence of explicit mass scales only makes the SM \emph{globally} scale-invariant. Under \emph{local} rescalings, the metric deforms as~$\FieldG{_{\mu\nu}}\mapsto e^{2\rho}\FieldG{_{\mu\nu}}$ and the scalar as~$\varphi\mapsto e^{-\rho}\varphi$ for local~$\rho\equiv\rho(x)$. Considering the first term in~\cref{EinsteinKleinGordon}, we must therefore introduce a \emph{Weyl} vector~$\FieldV{_\mu}\mapsto\FieldV{_\mu}-\PD{_\mu}\rho$ to define a new covariant derivative on scalars~$\SCovD{_\mu}\varphi\equiv\big(\PD{_\mu}-\FieldV{_\mu}\big)\varphi$. The kinematic d.o.f are thus increased from~$\big\{\FieldG{_{\mu\nu}},\varphi\big\}$ to~$\big\{\FieldG{_{\mu\nu}},\varphi,\FieldV{_\mu}\big\}$. Next, considering the second term in~\cref{EinsteinKleinGordon}, we must introduce a \emph{compensator} scalar~$\phi\mapsto e^{-\rho}\phi$ to make up the mass dimension of~$m$, which now becomes a dimensionless parameter. In the effective field theory approach, the non-linear completion of~\cref{EinsteinKleinGordon} includes all scale-invariant operators of mass-dimension four in the new variables~$\big\{\FieldG{_{\mu\nu}},\varphi,\FieldV{_\mu},\phi\big\}$, namely
\begin{align}\label{ToyModel}
		S=\int\mathrm{d}^4x\sqrt{-g}\Big[&\frac{1}{2}\SCovD{_\mu}\varphi\SCovD{^\mu}\varphi
		-\frac{\sigma}{2}\SCovD{_\mu}\varphi\SCovD{^\mu}\phi
		+\frac{\nu}{2}\SCovD{_\mu}\phi\SCovD{^\mu}\phi
		\nonumber\\&
		\hspace{-25pt}
		-\frac{\mu^2}{2}\phi^2\varphi^2
		-\frac{\xi}{16}\MAGH{_{\mu\nu}}\MAGH{^{\mu\nu}}+\text{gravity}\Big],
\end{align}
where~$\mu$,~$\sigma$,~$\nu$ and~$\xi$ are dimensionless model parameters, and~$\MAGH{_{\mu\nu}}\equiv 2\PD{_{[\mu}}\FieldV{_{\nu]}}$ is the natural Weyl field-strength tensor, i.e.~$\SCovD{_{[\mu}}\SCovD{_{\nu]}}\varphi\equiv-\frac{1}{2}\MAGH{_{\mu\nu}}\varphi$. For brevity we omit the~$\varphi^4$ and~$\phi^4$ self-interaction terms, though we still expect inflation to be a viable outcome with these terms present.\footnote{We will find that the~$\varphi^4$ operator deforms the potential without (visually) compromising the desired slow-roll plateau, though a detailed comparison with the CMB is lacking; the~$\phi^4$ operator provides a cosmological constant, which can be neglected when considering the early-Universe physics.}

\paragraph*{Scale-invariant variables} One can compare~$\FieldV{_\mu}$ to the electromagnetic four-vector potential. But unlike in scalar electrodynamics, where the matter fields carry a non-physical phase as well as a physical magnitude, the fields~$\varphi$ and~$\phi$ each have only one real d.o.f which merges entirely with the choice of scale gauge. This means that the field equations of~\cref{ToyModel} cannot simultaneously propagate~$\varphi$ and~$\phi$ from the initial data:\footnote{The scale-invariance of this model is local, i.e. it constitutes a gauge symmetry. To propagate the fields in the presence of a gauge symmetry, not only must the initial data be specified at some time-slice, but also the choice of gauge (in our case, the value of~$\phi$) at all future times must be agreed upon~\cite{Henneaux:1992ig}. The same scenario arises in classical electromagnetism and in GR: it does not detract from the well-posedness of the model.} we must first declare a global solution for one of these fields. Remembering that we want to recover~\cref{EinsteinKleinGordon}, we may take the opportunity to eliminate~$\phi$ by assuming constant~$\phi=\phi_0$. This is the \emph{Einstein} choice of scale gauge. Rather than imposing the Einstein gauge, it is instead possible to eliminate~$\phi$ completely from~\cref{ToyModel} through the field reparameterisations~$\FieldG{_{\mu\nu}}\mapsto(\phi_0/\phi)^2\FieldGHat{_{\mu\nu}}$, $\varphi\mapsto\phi\hat\varphi/\phi_0$ and~$\FieldV{_\mu}\mapsto\FieldVHat{_\mu}+\PD{_\mu}\ln\phi$, resulting in
\begin{align}\label{ScaleInvariantVariables}
	S=\int\mathrm{d}^4x\sqrt{-\hat g}\Big[&\frac{1}{2}\left(\PD{_\mu}\hat\varphi-\left(\hat\varphi-\sigma\phi_0\right)\FieldVHat{_\mu}\right)\left(\PD{^\mu}\hat\varphi-\hat\varphi\FieldVHat{^\mu}\right)
		\nonumber\\&
		\hspace{-60pt}
		-\frac{\mu^2\phi_0^2}{2}\hat\varphi^2
		-\frac{\xi}{16}\MAGHHat{_{\mu\nu}}\MAGHHat{^{\mu\nu}}
		+\frac{\nu\phi_0^2}{2}\FieldVHat{_\mu}\FieldVHat{^\mu}
		+\text{gravity}\Big],
\end{align}
where we note~$\MAGHHat{_{\mu\nu}} \equiv 2\PD{_{[\mu}}\FieldVHat{_{\nu]}}\equiv 2\PD{_{[\mu}}\FieldV{_{\nu]}}\equiv \MAGH{_{\mu\nu}}$. By construction, the new fields~$\big\{\FieldGHat{_{\mu\nu}},\hat\varphi,\FieldVHat{_\mu}\big\}$ are \emph{scale-invariant variables}. The gauge symmetry in~\cref{ScaleInvariantVariables} seems to be destroyed, but we now understand that the rescaling transformations are still happening internally within the new fields. Scale-invariant variables and the Einstein gauge choice both result in~\cref{ScaleInvariantVariables}, a key feature of which is that the dimensionless couplings are now accompanied by powers of~$\phi_0$. We will return to the natural values of these couplings later.

\paragraph*{Bootstrapping inflation} After dropping the hats, shifting the vector by the scalar gradient to remove the cross-term and rescaling both fields\footnote{Note during the rescaling~$\FieldV{_\mu}\mapsto\FieldV{_\mu}/\sqrt{\xi}$ that unitarity requires~$\xi\geq 0$.}, and for perturbations~$\delta\varphi$ and~$\delta\FieldV{_\mu}$ around the vacuum~$\varphi=\FieldV{_\mu}= 0$, we find that~\cref{ScaleInvariantVariables} reduces to
\begin{align}\label{PerturbativeTheory}
S&=\int\mathrm{d}^4x\sqrt{-g}\Big[\frac{1}{2}\PD{_\mu}\delta\varphi\PD{^\mu}\delta\varphi-\frac{m^2}{2}\delta\varphi^2\nonumber\\
	&-\frac{1}{4}\PD{_{[\mu}}\delta\FieldV{_{\nu]}}\PD{^{[\mu}}\delta\FieldV{^{\nu]}}+\frac{M^2}{2} \delta\FieldV{_\mu}\delta\FieldV{^\mu}+\text{gravity}\Big],
\end{align}
where~$m^2\equiv4\nu\mu^2\phi_0^2/(4\nu-\sigma^2)$ and~$M^2\equiv\nu\phi_0^2/\xi$. This result is very interesting. Not only do we recover~\cref{EinsteinKleinGordon} in the first line of~\cref{PerturbativeTheory}, but we see how local scale invariance additionally points to the presence of a new Proca field. To compare, the Higgs mechanism pointed to the presence of a new massive scalar, which was later observed experimentally~\cite{CMS:2012qbp}. Arguably, there is no need for a new vector in particle physics or cosmology. But whilst~$\nu\to 0$ predicts two radiative d.o.f which are not observed, taking instead~$\nu\to \infty$ nonetheless gives a heavy-enough Proca to evade all observational bounds (even better, it may be a dark matter candidate~\cite{Lasenby:2015dba}). Beyond the perturbative regime, the Proca and Klein--Gordon fields acquire non-linear interactions. For sufficiently large~$\nu$ the Proca can be eliminated as~$\FieldV{_\mu}=\frac{1}{2}\PD{_\mu}\ln\left(\varphi^2-\sigma\phi_0\varphi+\nu\phi_0^2\right)$ on-shell,\footnote{A specific regime is~$\nu\gg\sigma^2$ where the gauge boson is weakly coupled by~$\xi\gtrsim 1$. We work in Planck units, assuming Weyl symmetry breaking at~$\phi_0\sim\MPl{}$. An effective scalar theory emerges at~$\mu^2\ll\mathcal{E}^2\ll\nu/\xi$ from which we retain only the zeroth-order correction in the~$\mathcal{E}^2/M^2$ expansion to reach~\cref{FullTheory,Potential}. A visual comparison between~\cref{PotentialFig} and the potential in~\cite{Salvio:2022suk} suggests agreement with the CMB when the field value at slow roll is~$\sigma\sim 10$ and the scale of inflation is~$\sigma^2\mu^2\sim 10^{-10}$. This yields an inflaton of mass~$\mu\sim 10^{-6}$ or~$m\sim 10^{13}\SI{}{\giga\electronvolt}$. Meanwhile, the vector mass~$M$ approaches the Planck scale for stronger couplings at smaller~$\xi$.} and~\cref{ScaleInvariantVariables} becomes
\begin{subequations}
\begin{align}
	&S=\int\mathrm{d}^4x\sqrt{-g}\left[\tfrac{1}{2}\PD{_\mu}\varphi\PD{^\mu}\varphi-U\left(\varphi\right)+\text{gravity}\right],\label{FullTheory}\\
	&U\left(\varphi\right)\equiv\frac{\mu^2\phi_0^4}{2}\left[\frac{\sigma}{2}+\sqrt{\nu-\frac{\sigma^2}{4}}\text{sinh}\left(\frac{(\varphi-c)/\phi_0}{\sqrt{\nu-\frac{\sigma^2}{4}}}\right)\right]^{\ 2},\label{Potential}
\end{align}
\end{subequations}
where~$c$ is an integration constant.\footnote{Note that the scalar in~\cref{FullTheory,Potential} has been recanonicalised, so that all the physical implications may be read off from the potential~$U$. If a~$\varphi^4$ operator had been included in~\cref{ToyModel}, then~\cref{FullTheory} would feature an additional~$U^2$ operator. In this case, model parameters may still be found which reproduce the plateau associated with slow-roll inflation; we are not, however, aware of any CMB constraints on this extended model.} The non-linear potential (see~\cref{PotentialFig}) is identical to that found in~\cite{Barker:2024dhb,Salvio:2022suk}, though the motivations are unrelated. The predictions from the inflationary plateau approach the universal values for the spectral index~$n_s\approx 1-2/N$ and also the tensor-to-scalar ratio~$r\approx12/N^2$ for appropriate values of the model parameters~\cite{Salvio:2022suk,Karananas:2025xcv}.\footnote{It remains to be investigated, however, whether the universal regime is consistent with a weakly-coupled gauge boson.} This agrees with current cosmic microwave background (CMB) measurements~\cite{Planck:2018jri,BICEP:2021xfz}, pending future experiments~\cite{arXiv:1110.3193,Weltman:2018zrl,Abazajian:2019eic,LiteBIRD:2022cnt}. The inverse problem of constraining (through Bayesian inference) the model parameters from the CMB observations is well within the scope of precision cosmology~\cite{Handley:2019fll}. We will develop this toy model in future work, but first need to make the `$+\ \text{gravity}$' term more concrete.

\paragraph*{Recap and extension} Collecting our results so far:
\begin{enumerate}
	\item Gravitational coupling to fermions naturally leads to PGT as a leading formulation of gravity. PGT does not have local scale invariance.
	\item The~$\PD{_{[\mu}}\MAGT{_{\nu]}}\PD{^{[\mu}}\MAGT{^{\nu]}}$ term for vector torsion~$\MAGT{_\mu}$ is notoriously absent from the PGT action in~\cref{PGTAction}.
	\item Theories with mass scales have locally-scale-invariant embeddings which can lead to excellent phenomenology (inflation) besides a new massive vector~$\FieldV{_\mu}$.
\end{enumerate}
We will now connect these three observations by showing that eWGT is the unique scale-invariant embedding of PGT. This will identify~$\MAGT{_\mu}/3$ with the vector~$\FieldV{_\mu}$ when expressed in scale-invariant variables, and thereby reveal~$\PD{_{[\mu}}\MAGT{_{\nu]}}\PD{^{[\mu}}\MAGT{^{\nu]}}$ to be a Yang--Mills-type term.

\paragraph*{Poincar\'e gauge theory (PGT)} Gauge theories of gravity can be formulated on completely flat Minkowski spacetime, just like the SM~\cite{Lasenby:2015dba}. In this formulation, we use holonomic Greek indices and Lorentz Roman indices. Translational gauge fields~$\FieldB{^i_\mu}$ and inverses~$\FieldH{_i^\mu}$ obey~$\FieldB{^i_\mu}\FieldH{_i^\nu}\equiv\FieldDelta{_\mu^\nu}$ and~$\FieldB{^i_\mu}\FieldH{_j^\mu}\equiv\FieldDelta{_j^i}$, and give rise to the curved-space metric~$\FieldG{_{\mu\nu}}$ through~$\FieldB{^i_\mu}\FieldB{^j_\nu}\FieldEta{_{ij}}\equiv\FieldG{_{\mu\nu}}$. We assume contraction with these fields eliminates all coordinate indices of general (i.e. not necessarily scalar) matter fields~$\varphi$, which have only suppressed Lorentz (or spinor) indices. A general~$\varphi$ with Weyl weight~$w$ and belonging to the~$\mathrm{SL}(2,\mathbb{C})$ representation with Lorentz generators~$\FieldSigma{_{ij}}\equiv\FieldSigma{_{[ij]}}$ transforms as~$\varphi\mapsto e^{w\rho+\FieldOmega{^{ij}}\FieldSigma{_{ij}}/2}\varphi$, where~$\FieldOmega{^{ij}}\equiv\FieldOmega{^{[ij]}}(x)$ are finite angles. The rotational gauge field is~$\FieldA{^{ij}_\mu}\equiv\FieldA{^{[ij]}_\mu}$. The transformations~$\FieldB{^i_\mu}\mapsto e^{\rho}\FieldLambda{^i_j}\FieldB{^j_\mu}$ and~$\FieldA{^{ij}_\mu}\mapsto\FieldLambda{^i_k}\FieldLambda{^j_l}\FieldA{^{kl}_\mu}-\FieldLambda{^{jk}}\PD{_\mu}\FieldLambda{^i_k}$ (where~$\FieldLambda{}\equiv e^{\FieldOmega{}}$ is the Lorentz matrix) ensure that the derivative~$\SCovD{_i}\equiv\FieldH{_i^\mu}\SCovD{_\mu}$ with
\begin{equation}\label{PGTDer}
\SCovD{_\mu}\varphi\equiv\left(\PD{_\mu}+\tfrac{1}{2}\FieldA{^{ij}_\mu}\FieldSigma{_{ij}}\right)\varphi,
\end{equation}
is Lorentz-covariant. The field strengths from the commutator~$\FieldB{^{[i|}_\mu}\FieldB{^{|j]}_\nu}\SCovD{_{[i}}\SCovD{_{j]}}\varphi\equiv\frac{1}{4}\FieldB{^{i}_\sigma}\FieldB{^j_\lambda}\MAGF{^{\sigma\lambda}_{\mu\nu}}\FieldSigma{_{ij}}\varphi-\frac{1}{2}\MAGT{^\lambda_{\mu\nu}}\SCovD{_\lambda}\varphi$ are
\begin{subequations}
\begin{align}	\MAGT{^\lambda_{\mu\nu}}&\equiv2\FieldH{_i^\lambda}(\PD{_{[\mu|}}\FieldB{^i_{|\nu]}}+\FieldA{^i_j_{[\mu|}}\FieldB{^j_{|\nu]}}),\label{Torsion}\\
\MAGF{^{\rho\sigma}_{\mu\nu}}&\equiv2\FieldH{_i^\rho}\FieldH{_j^\sigma}(\PD{_{[\mu|}}\FieldA{^{ij}_{|\nu]}}+\FieldA{^i_k_{[\mu|}}\FieldA{^{kj}_{|\nu]}}),\label{Curvature}
\end{align}
\end{subequations}
where~\cref{Torsion,Curvature} are equivalent to~\cref{ECTorsion,ECCurvature}; the reparameterisation from gauge-theoretic~$\big\{\FieldB{^i_\mu},\FieldA{^{ij}_\mu}\big\}$ to geometric~$\big\{\FieldG{_{\mu\nu}},\MAGA{^\alpha_{\mu}_{\nu}}\big\}$ is described in~\cite{Blagojevic:2002,Lasenby:2015dba}.

\paragraph*{Weyl gauge theory (WGT)} From our toy model, the Weyl-covariant extension of~\cref{PGTDer} is 
\begin{equation}\label{WGTDer}
\SCovD{_\mu}\varphi\equiv\left(\PD{_\mu}+\tfrac{1}{2}\FieldA{^{ij}_\mu}\FieldSigma{_{ij}}+w\FieldV{_\mu}\right)\varphi,
\end{equation}
and this construction defines the well-known \emph{Weyl gauge theory} (WGT) of gravity~\cite{Weyl:1918ib,Weyl:1931}. Whilst~$\MAGF{^{\rho\sigma}_{\mu\nu}}$ in~\cref{Curvature} is already Weyl-covariant, we have under local dilations
\begin{equation}\label{TorsionTransform}
	\MAGT{^\lambda_{\mu\nu}}\mapsto \MAGT{^\lambda_{\mu\nu}}-2\FieldDelta{^\lambda_{[\mu}}\PD{_{\nu]}}\rho\implies\MAGT{_\mu}\mapsto\MAGT{_\mu}-3\PD{_\mu}\rho,
\end{equation}
so from~\cref{TorsionTransform} only the combination
\begin{equation}\label{TracelessTorsion}
\MAGTTilde{^\lambda_{\mu\nu}}\equiv\MAGT{^\lambda_{\mu\nu}}-2\FieldDelta{^\lambda_{[\mu}}\FieldV{_{\nu]}},
\end{equation}
is covariant, indeed~$\FieldB{^{[i|}_\mu}\FieldB{^{|j]}_\nu}\SCovD{_{[i}}\SCovD{_{j]}}\varphi\equiv\frac{1}{4}\FieldB{^{i}_\sigma}\FieldB{^j_\lambda}\MAGF{^{\sigma\lambda}_{\mu\nu}}\FieldSigma{_{ij}}\varphi-\frac{1}{2}\MAGTTilde{^\lambda_{\mu\nu}}\SCovD{_\lambda}\varphi+\frac{1}{2}w\MAGH{_{\mu\nu}}\varphi$. Following~\cref{ToyModel}, the parity-even Yang--Mills-type embedding of~\cref{PGTAction} must then be strictly
\begin{align}\label{WGTAction}
	&S_{\text{WGT}}\equiv\int\mathrm{d}^4x\sqrt{-g}\Big[
	-\frac{\zeta\phi^2}{2}\MAGF{}
	-\lambda\phi^4
	+\Alp{1}\MAGF{}^2
	+\MAGF{_{\mu\nu}}\big(
	\Alp{2}\MAGF{^{\mu\nu}}
	\nonumber\\ &
	+\Alp{3}\MAGF{^{\nu\mu}}\big)
	+\MAGF{_{\mu\nu\sigma\lambda}}\big(
	\Alp{4}\MAGF{^{\mu\nu\sigma\lambda}}
	+\Alp{5}\MAGF{^{\mu\sigma\nu\lambda}}
	+\Alp{6}\MAGF{^{\sigma\lambda\mu\nu}}\big)
	\nonumber\\ &
	+\phi^2\MAGTTilde{_{\mu\nu\sigma}}\big(\Gam{1}\MAGTTilde{^{\mu\nu\sigma}}
	+\Gam{2}\MAGTTilde{^{\nu\mu\sigma}}\big)
	+\phi^2\Gam{3}\MAGTTilde{_{\mu}}\MAGTTilde{^{\mu}}
	+\frac{\nu}{2}\SCovD{_\mu}\phi\SCovD{^\mu}\phi
	\nonumber\\ &
	-\frac{\xi}{16}\MAGH{_{\mu\nu}}\MAGH{^{\mu\nu}}
	+\frac{\chi}{2}\MAGF{_{\mu\nu}}\MAGH{^{\mu\nu}}
	+\text{matter}
	\Big],
\end{align}
with dimensionless~$\zeta$,~$\lambda$,~$\Gam{1}$,~$\Gam{2}$,~$\Gam{3}$,~$\nu$,~$\xi$,~$\chi$ and~$\alpha_1$ through~$\alpha_6$. Crucially, everything is quadratic in field strengths.

\paragraph*{Extended Weyl gauge theory (eWGT)} In moving from PGT to WGT, we have enlarged the kinematic space from~$\big\{\FieldB{^i_\mu},\FieldA{^{ij}_\mu}\big\}$ to~$\big\{\FieldB{^i_\mu},\FieldA{^{ij}_\mu},\FieldV{_\mu},\phi\big\}$, just like in our toy model. Unlike for our toy model, however, the enlargement proves unnecessary: it is more economical to recycle~$\FieldV{_\mu}$ from within the rotational gauge field. Let the traceless part of the latter field be~$\FieldATilde{^{ij}_{\mu}}\equiv\FieldATilde{^{[ij]}_{\mu}}$ with only 20 d.o.f, such that~$\FieldH{_j^\nu}\FieldATilde{^{ij}_\nu}\equiv 0$. Then, using~$\big\{\FieldB{^i_\mu},\FieldATilde{^{ij}_\mu},\FieldV{_\mu},\phi\big\}$ there is one (and \emph{only} one) way to emulate the combined Poincar\'e and Weyl covariance of~\cref{WGTDer}, namely
\begin{align}
	\SCovD{_\mu}\varphi\equiv\Big(\PD{_\mu}+\tfrac{1}{2}\Big[&\FieldATilde{^{ij}_\mu}-\tfrac{2}{3}\FieldB{^{[i|}_\mu}\FieldH{^{|j]}^\nu}\big(3\FieldV{_\nu}
	\nonumber\\&\ \ \ 
	-2\FieldH{_k^\lambda}\PD{_{[\lambda|}}\FieldB{^k_{|\nu]}}\big)\Big]\Sigma{_{ij}}+w\FieldV{_\mu}\Big)\varphi.\label{EconomicalAction}
\end{align}
Although unique,\footnote{The extra terms in the square brackets in~\cref{EconomicalAction} have been constructed so as to emulate the inhomogeneous Lorentz transformation of the (missing) trace vector~$\FieldH{_j^\nu}\FieldA{^{ij}_\nu}$, without simultaneously generating an (unwanted) inhomogeneous Weyl transformation. There is only one such construction. One might ask whether a similar trick can be used to replace the axial vector~$\tensor{\epsilon}{^k_{ijl}}\FieldH{_k^\nu}\FieldA{^{ij}_\nu}$. The answer is of course affirmative, but the construction in this case does not need (and must not contain)~$\FieldV{_\mu}$ because the axial vector and purely tensor parts of the Ricci rotation coefficients are already Weyl-covariant: only the trace vector part has an inhomogeneous Weyl transformation, which requires correcting via~$\FieldV{_\mu}$ in the manner proposed. By carefully propagating this observation through to our main result in~\cref{eWGTAction}, we conclude that local scale invariance always leads to dynamical \emph{vector} torsion, and not \emph{axial vector} torsion.}we will soon show that~\cref{EconomicalAction} is just one notational way to write the eWGT covariant derivative from~\cite{Lasenby:2015dba}. Because eWGT shares the symmetries of WGT, it must also share the Yang--Mills-type action in~\cref{WGTAction}. The difference is that we must replace~$\FieldA{^{ij}_\mu}$, as when going from~\cref{WGTDer} to~\cref{EconomicalAction}. After this change, we notice from~\cref{TracelessTorsion} that~$\MAGTTilde{_\mu}\equiv 0$ identically. Thus,~$\Gam{3}$ in~\cref{WGTAction} drops out of eWGT.

\paragraph*{Equivalence of PGT and eWGT} With the field reparameterisation~$\FieldV{_\mu}\mapsto\FieldVV{_\mu}+\frac{2}{3}\FieldH{_k^\lambda}\PD{_{[\lambda|}}\FieldB{^k_{|\mu]}}$ we find~\cref{EconomicalAction} is 
\begin{align}
	\SCovD{_\mu}\varphi\equiv\Big(\PD{_\mu}+\tfrac{1}{2}\Big[&\FieldATilde{^{ij}_\mu}-2\FieldB{^{[i|}_\mu}\FieldH{^{|j]}^\nu}\FieldVV{_\nu}\Big]\Sigma{_{ij}}
	\nonumber\\&\ \ \ 
	+w\Big[\FieldVV{_\mu}+\tfrac{2}{3}\FieldH{_k^\lambda}\PD{_{[\lambda|}}\FieldB{^k_{|\mu]}}\Big]\Big)\varphi.\label{TrimmedAction}
\end{align}
In this~$\big\{\FieldB{^i_\mu},\FieldATilde{^{ij}_\mu},\FieldVV{_\mu},\phi\big\}$ formulation~$\FieldVV{_\mu}$ is a `Lorentz' boson, not a `Weyl' boson, because~$\FieldVV{_\mu}\mapsto\FieldVV{_\mu}-\frac{1}{3}\FieldH{^\lambda_l}\FieldB{^m_\mu}\FieldLambda{_k^l}\PD{_\lambda}\FieldLambda{^k_m}$. Relabelling~$\FieldA{^{ij}_\mu}\equiv\FieldATilde{^{ij}_\mu}-2\FieldB{^{[i|}_\mu}\FieldH{^{|j]}^\nu}\FieldVV{_\nu}$, where~$\FieldA{^{ij}_\mu}\equiv\FieldA{^{[ij]}_\mu}$ carries 24 d.o.f, ensures the PGT transformation of~$\FieldA{^{ij}_\mu}$. In this~$\big\{\FieldB{^i_\mu},\FieldA{^{ij}_\mu},\phi\big\}$ formulation~\cref{TrimmedAction} becomes
\begin{align}
	\SCovD{_\mu}\varphi&\equiv\Big(\PD{_\mu}+\tfrac{1}{2}\FieldA{^{ij}_\mu}\Sigma{_{ij}}
	+\tfrac{1}{3}w\FieldH{_k^\lambda}\Big[2\PD{_{[\lambda|}}\FieldB{^k_{|\mu]}}+3\FieldB{_l_\mu}\FieldA{^{kl}_\lambda}\Big]\Big)\varphi
	\nonumber\\&
	\equiv\Big(\PD{_\mu}+\tfrac{1}{2}\FieldA{^{ij}_\mu}\Sigma{_{ij}}
	+\tfrac{1}{3}w\MAGT{_\mu}\Big)\varphi,
\label{PreAction}
\end{align}
where we used~\cref{Torsion} in the last equality. New fields should not be introduced where they are not needed: in hindsight it was obvious from~\cref{TorsionTransform} that PGT contained a vector~$\MAGT{_\mu}/3$ which already performed the function of~$\FieldV{_\mu}$. This fact was observed by Obukhov in~\cite{Obukhov:1982zn}, and the covariant derivative in~\cref{PreAction} was first written down in~\cite{Lasenby:2015dba} (see also~\cite{Karananas:2021gco}). One could interpret~\cref{PreAction} as a hint that PGT descends from a scale-invariant theory. Explicit scales emerge in the~$\big\{\FieldBHat{^i_\mu},\FieldAHat{^{ij}_\mu}\big\}$ formulation, with scale-invariant variables~$\FieldB{^i_\mu}\mapsto\phi_0\FieldBHat{^i_\mu}/\phi$ and~$\FieldA{^{ij}_\mu}\mapsto\FieldAHat{^{ij}_\mu}$. By dropping the hats on variables, we find that the PGT Lagrangian couplings parameterise exactly the same physics as the following combinations of eWGT couplings (with  $\alpha_1,\ldots,\alpha_6$ identical)
\begin{equation}\label{Equivalences}
	\begin{gathered}
		\MPl{}^2\equiv \zeta\phi_0^2,\quad \MPl{}^2\Lambda\equiv\lambda\phi_0^4,\quad \Bet{1}\equiv\Gam{1}\phi_0^2,\\
		\Bet{2}\equiv\Gam{2}\phi_0^2,\quad \Bet{3}\equiv\big(\nu-6(2\Gam{1}+\Gam{2})\big)\phi_0^2/18.
	\end{gathered}
\end{equation}
Although the~$\Gam{3}$ coupling of WGT in~\cref{WGTAction} is missing from the eWGT action, its loss is exactly compensated for by~$\nu$ so that~$\Bet{3}$ emerges in~\cref{Equivalences} as an independent coupling in PGT. The couplings~$\chi$ and~$\xi$ do not appear in~\cref{Equivalences} --- these parameterise terms involving~$\MAGH{_{\mu\nu}}$, which is the Faraday tensor for~$\MAGT{_\mu}$ in the scale-invariant variables of PGT. Thus, eWGT motivates a correction to Yang--Mills-type PGT:
\begin{align}\label{eWGTAction}
	S_{\text{eWGT}}\equiv S_{\text{PGT}}
	+\frac{1}{3}\int\mathrm{d}^4x\sqrt{-g}\Big[&
	\chi\MAGF{_{[\mu\nu]}}\PD{^{[\mu}}\MAGT{^{\nu]}}
	\nonumber\\&
	-\frac{\xi}{12}\PD{_{[\mu}}\MAGT{_{\nu]}}\PD{^{[\mu}}\MAGT{^{\nu]}}
	\Big].
\end{align}
This is our central result: the relationship between~\cref{eWGTAction} and~\cref{PGTAction} is analogous to the relationship between~\cref{FullTheory} and~\cref{EinsteinKleinGordon}. Put another way, the traditional PGT action unfairly omits two specific terms, which are naturally motivated by the unique, scale-invariant embedding. This embedding allows us to claim that all PGT models are ultimately conformal, irrespective of the explicit mass scales in their spectra. We confirm in the supplemental material~\cite{Barker:2024} that eWGT, as formulated in~\cref{TracelessTorsion,WGTAction,EconomicalAction}, and the extension of PGT in~\cref{eWGTAction,PGTAction,Torsion,Curvature}, have completely equivalent spectra. In the expressions for masses and pole residues,~\cref{Equivalences} maps the dimensionful couplings to their dimensionless counterparts.\footnote{As a final remark, we return to the values of the dimensionless couplings. If one retains the forms of the coefficients on the RHS of~\cref{Equivalences} in the eWGT action, it is natural first to pull out a constant factor of $\phi_0^2$, which has the same dimensions as $\MPl{}^2$. 
When concerned with physics at the length-scale~$\LPhy{}$, it is helpful to work in units of~$\LPhy{}$. This can be done by taking~$\phi_0\LPhy{}\equiv 1$ as a gauge choice in the remainder of the action, which is equivalent to the standard practice of setting $\phi_0 = 1$ provided one works in length units of $\LPhy{}$. If one instead works in some other length units of $\LPhyprime{}$, then one should rescale $\phi_0$ by $\LPhyprime{}/\LPhy{}$. Having set $\phi_0=1$, however, it is the dimensionless couplings that should `run' with the length units used. This `running' is intended in a more prosaic sense than the action of beta functions in the quantum theory. Any classical field equations can, in some system of units, be integrated from a fixed numerical range of initial data and for a fixed numerical interval. The resulting solutions may look very different, depending on the numerical values of the couplings in the equations. These different solutions describe phenomena on different physical scales, and the values of the couplings are tied to these scales. To give a concrete example, the mass in the Klein--Gordon equation can be neglected numerically when the dynamics are probed at distances much shorter than the Compton wavelength. In our case, by observing the powers of $\phi_0$ associated with each dimensionless coupling constant (after factorising out $\phi^2_0$ as discussed above), it is straightforward to show that $\zeta$, $\nu$, and the $\beta$s do not change with scale (and so presumably should have values $\sim 1$), whereas the $\alpha$s, $\xi$ and $\chi$ scale as $(\LPhyprime{}/\LPhy{})^{-2}$, while $\lambda$ scales as $(\LPhyprime{}/\LPhy{})^2$. On adopting the convention~$\LPl{}\MPl{}\equiv 1$, it is worth noting from~\cref{Equivalences} that $\lambda\sim\SI{1e-122}{}(\LPhy{}/{\LPl{}})^2 \sim 1$ at the scale~$\LPhy{}\sim\SI{10}{\giga\parsec}$, which is close to the Hubble horizon.}

\paragraph*{Implications for the PGT spectrum} Since we have made a substantial correction to the PGT Lagrangian, it is important to update the well-known PGT particle spectrum in~\cite{Hayashi:1980qp}. To do this, we take~$\lambda= \Lambda = 0$ and focus on the linearisation near Minkowski spacetime. As shown in~\cref{ParticleSpectra} and~\cref{ParticleSpectrographGeneralCase}, the propagator of~$S_{\text{eWGT}}$ in~\cref{eWGTAction} contains a \emph{quartic} pole in the parity-odd spin-one sector
\begin{align}
	\label{MasterConstraint}
	&2\Big[\big(2\Alp{2}+4\Alp{4}+\Alp{5}\big)\xi-\chi^2\Big]k^4
	+3\Big[
		96\Alp{2}\big(2\Bet{1}+\Bet{2}+\Bet{3}\big)
		\nonumber\\&
		+8\big[24\Alp{4}\big(2\Bet{1}+\Bet{2}+\Bet{3}\big)+6\Alp{5}\big(2\Bet{1}+\Bet{2}+\Bet{3}\big)
		\nonumber\\&
		+\chi\big(4\Bet{1}+2\Bet{2}-\MPl{}^2\big)\big]+\big(4\Bet{1}+2\Bet{2}-\MPl{}^2\big)\xi
	\Big]k^2
		\nonumber\\&
	+72\big(4\Bet{1}+2\Bet{2}-\MPl{}^2\big)\big(2\Bet{1}+\Bet{2}+3\Bet{3}+\MPl{}^2\big).
\end{align}
The pole takes the familiar~$k^2-M^2$ form when the~$k^4$ coefficient is removed from~\cref{MasterConstraint}. For example, the branch~$2\Alp{2}+4\Alp{4}+\Alp{5}=\chi=0$ contains the Einstein--Proca theory (EPT)~$S_{\text{EPT}}\equiv\int\mathrm{d}^4x\sqrt{-g}\big[-\frac{1}{2}\MPl{}^2\rR{}-\frac{1}{4}\PD{_{[\mu}}\MAGT{_{\nu]}}\PD{^{[\mu}}\MAGT{^{\nu]}}+\frac{1}{2}M^2\MAGT{_\mu}\MAGT{^\mu}+\text{matter}\big]$ in~$\big\{\FieldG{_{\mu\nu}},\MAGT{^\alpha_{\mu}_{\nu}}\big\}$ variables. EPT is evidently not strongly coupled, so this is the much-sought-after \emph{vector torsion}. The more general branch~$\Alp{5}=\chi^2/\xi-2\left(\Alp{2}+2\Alp{4}\right)$ leads to the new mass spectrum in~\crefrange{Mass0p}{Mass2m}, as shown in~\cref{ParticleSpectrographGeneralCaseNoQuartic}. It is also interesting to instead remove the \emph{constant} term in~\cref{MasterConstraint}. This produces a~$k^2\big(k^2-M^2\big)$ pole, which affects the massless spectrum. As shown in~\cref{ParticleSpectrographBranch1Conservative}, we find an example case~$\Alp{2}=\Alp{3}-4\Alp{4}+\Alp{6}=4\Alp{4}+\Alp{5}=4\Bet{1}+2\Bet{2}-\MPl{}^2=0$ which propagates (without ghosts or tachyons) the Einstein graviton, one heavy scalar, one heavy pseudoscalar, and a \emph{one extra massless scalar}. To understand where this scalar comes from, we strip away all the uninvolved operators to show in~\cref{ParticleSpectrographSwitchonPGT} that the model 
\begin{equation}\label{SwitchonPGT}
	S=\int\mathrm{d}^4x\sqrt{-g}\big[\chi\MAGF{_{[\mu\nu]}}\PD{^{[\mu}}\MAGT{^{\nu]}}+\Bet{3}\MAGT{_{\mu}}\MAGT{^{\mu}}\big],	
\end{equation}
propagates, for~$\chi\neq 0$, one massless scalar with the no-ghost condition~$\Bet{3}>0$. For~$\chi=0$, the spectrum is empty. In fact~\cref{SwitchonPGT} can be reduced even further by removing~$\FieldB{^i_\mu}$, and keeping only the axial-free part of~$\FieldA{^{ij}_\mu}$. This part has the index symmetries of a Curtright field~$\Curtright{_{\mu\nu\sigma}}$, i.e.~$\Curtright{_{\mu\nu\sigma}}\equiv\Curtright{_{[\mu\nu]\sigma}}$ with~$\Curtright{_{[\mu\nu\sigma]}}\equiv 0$~\cite{Curtright:1980yk}, and~\cref{SwitchonPGT} reduces (linearly) to
\begin{equation}\label{NewCurtright}
	S=\int\mathrm{d}^4x\Big[
	\Bet{3}\Curtright{^\mu}\Curtright{_{\mu}}
	+\frac{\chi}{12}\Curtright{_{\mu\nu}}
	\Big(
			\Curtright{^{\mu\nu}}
			-\frac{4}{3}\PD{_\sigma}\Curtright{^{\mu(\nu\sigma)}}
	\Big)
	\Big],
\end{equation}
with trace~$\Curtright{_\mu}\equiv\Curtright{^\nu_{\mu\nu}}$ and field strength~$\Curtright{_{\mu\nu}}\equiv2\PD{_{[\mu}}\Curtright{_{\nu]}}$. As shown in~\cref{ParticleSpectrographSwitchonNoAxialVector},~\cref{NewCurtright,SwitchonPGT} have the same spectra. Finally, in the branch~$\xi=\chi=0$ we recover the known spectrum of~$S_{\text{PGT}}$ in~\cref{PGTAction}, as first presented in~\cite{Hayashi:1980qp}.

\paragraph*{Closing remarks} Poincar\'e gauge theory (PGT) strongly motivates spacetime torsion~\cite{Utiyama:1956sy,Kibble:1961ba,Sciama:1962}, but the consensus since the turn of the millennium has been that PGT prohibits vector torsion from propagating~\cite{Hecht:1996np,Chen:1998ad,Yo:1999ex,Yo:2001sy,Yo:2006qs,Shie:2008ms,Chen:2009at,Baekler:2010fr,Ho:2011qn,Ho:2011xf,Ho:2015ulu,Tseng:2018feo,Zhang:2019mhd,Zhang:2019xek,MaldonadoTorralba:2020mbh,delaCruzDombriz:2021nrg}. Separately, the full conformal group is a more appealing gauge group than the Poincar\'e group. With this context, our letter achieved three objectives:
\begin{enumerate}
	\item Starting just with the inflaton mass, scale-invariant embedding leads to the slow-roll plateau in~\cref{Potential,PotentialFig}. This is an intriguing, stand-alone result.
	\item We showed that PGT has a unique, locally scale-invariant embedding as \emph{extended} Weyl gauge theory (eWGT)~\cite{Lasenby:2015dba,Hobson:2020bms,Hobson:2020doi,Hobson:2021iwy,Hobson:2023rdw}. The embedding is unique if one is to achieve scale invariance via the introduction of a \emph{minimum} number of new gauge fields, beyond those already present in PGT. To reach eWGT, one need only introduce a compensator field. The compensator is purely gauge, so that the embedding theory is completely indistinguishable from PGT after gauge-fixing. In other words, PGT already descends from a conformal theory, without needing further work. By contrast, to reach the traditional Weyl gauge theory (WGT) embedding one must introduce both a compensator and a Weyl gauge boson. WGT actions may be written down which propagate this Weyl boson \emph{even after gauge fixing} resulting in phenomena that cannot be described by PGT alone. 
	\item The eWGT embedding of PGT means that the natural Yang--Mills-type PGT action is missing the two terms in~\cref{eWGTAction}. The new terms provide the first compelling argument for propagating vector torsion. The updated mass spectrum of PGT is shown in~\crefrange{Mass0p}{Mass2m}. The new terms reveal that the Curtright field in~\cref{NewCurtright} is embedded in the PGT dynamics.
\end{enumerate}
Further work is needed to develop our inflaton model in the full eWGT-PGT framework. One key question is whether the model is susceptible to the Weyl anomaly~\cite{Capper:1974ic}. Also, the implications of the new PGT terms for strong coupling should be investigated~\cite{Barker:2022kdk}. Finally, the Yang--Mills-type actions in~\cref{PGTAction,WGTAction} are restricted to parity-even terms for simplicity: the parity-odd extensions should be considered. The mass spectrum of parity-violating PGT was found in~\cite{Karananas:2014pxa}, and confirmed in~\cite{Blagojevic:2018dpz}, however the massless spectra and unitarity conditions of the various critical cases of the theory have not been thoroughly explored, nor have more than a handful of such cases been identified to date.\footnote{This is due to the fact that, in the parity-violating case, the formulae for the squares of the masses are \emph{irrational} functions of the couplings: the residues of such poles are comparatively more cumbersome to calculate.} As a result, the capacity for parity-violating PGT to propagate a vector from Yang--Mills-type terms has not been fully explored linearly, and the question of strong coupling in such models has not, to our knowledge, been addressed at all.

\vspace*{-0.3cm}
\begin{acknowledgements}

We are grateful for several useful discussions with Georgios Karananas, Carlo Marzo, Prabhoda Chandra Sarjapur and Sebastian Zell, and for the feedback of the anonymous referee.

WB is grateful for the kind hospitality of Leiden University and the Lorentz Institute, and the support of Girton College, Cambridge. Y-CL acknowledges support from the Ministry of Education of Taiwan and the Cambridge Commonwealth, European \& International Trust via a Taiwan Cambridge Scholarship, and from Corpus Christi College, Cambridge.

This work used the DiRAC Data Intensive service (CSD3 \href{www.csd3.cam.ac.uk}{www.csd3.cam.ac.uk}) at the University of Cambridge, managed by the University of Cambridge University Information Services on behalf of the STFC DiRAC HPC Facility (\href{www.dirac.ac.uk}{www.dirac.ac.uk}). The DiRAC component of CSD3 at Cambridge was funded by BEIS, UKRI and STFC capital funding and STFC operations grants. DiRAC is part of the UKRI Digital Research Infrastructure.

This work also used the Newton server, to which access was provided by Will Handley, and funded through an ERC grant.

\end{acknowledgements}

\bibliography{NotINSPIRE,Manuscript}

\begin{thebibliography}{196}%
\makeatletter
\providecommand \@ifxundefined [1]{%
 \@ifx{#1\undefined}
}%
\providecommand \@ifnum [1]{%
 \ifnum #1\expandafter \@firstoftwo
 \else \expandafter \@secondoftwo
 \fi
}%
\providecommand \@ifx [1]{%
 \ifx #1\expandafter \@firstoftwo
 \else \expandafter \@secondoftwo
 \fi
}%
\providecommand \natexlab [1]{#1}%
\providecommand \enquote  [1]{``#1''}%
\providecommand \bibnamefont  [1]{#1}%
\providecommand \bibfnamefont [1]{#1}%
\providecommand \citenamefont [1]{#1}%
\providecommand \href@noop [0]{\@secondoftwo}%
\providecommand \href [0]{\begingroup \@sanitize@url \@href}%
\providecommand \@href[1]{\@@startlink{#1}\@@href}%
\providecommand \@@href[1]{\endgroup#1\@@endlink}%
\providecommand \@sanitize@url [0]{\catcode `\\12\catcode `\$12\catcode
  `\&12\catcode `\#12\catcode `\^12\catcode `\_12\catcode `\%12\relax}%
\providecommand \@@startlink[1]{}%
\providecommand \@@endlink[0]{}%
\providecommand \url  [0]{\begingroup\@sanitize@url \@url }%
\providecommand \@url [1]{\endgroup\@href {#1}{\urlprefix }}%
\providecommand \urlprefix  [0]{URL }%
\providecommand \Eprint [0]{\href }%
\providecommand \doibase [0]{https://doi.org/}%
\providecommand \selectlanguage [0]{\@gobble}%
\providecommand \bibinfo  [0]{\@secondoftwo}%
\providecommand \bibfield  [0]{\@secondoftwo}%
\providecommand \translation [1]{[#1]}%
\providecommand \BibitemOpen [0]{}%
\providecommand \bibitemStop [0]{}%
\providecommand \bibitemNoStop [0]{.\EOS\space}%
\providecommand \EOS [0]{\spacefactor3000\relax}%
\providecommand \BibitemShut  [1]{\csname bibitem#1\endcsname}%
\let\auto@bib@innerbib\@empty
\bibitem [{\citenamefont {Einstein}(1915)}]{Einstein:1915}%
  \BibitemOpen
  \bibfield  {author} {\bibinfo {author} {\bibfnamefont {A.}~\bibnamefont
  {Einstein}},\ }\bibfield  {title} {\bibinfo {title} {{Die Feldgleichungen der
  Gravitation}},\ }\href@noop {} {\bibfield  {journal} {\bibinfo  {journal}
  {Sitzungsber. Preuss. Akad. Wiss}\ }\textbf {\bibinfo {volume} {18}},\
  \bibinfo {pages} {844} (\bibinfo {year} {1915})}\BibitemShut {NoStop}%
\bibitem [{\citenamefont {Utiyama}(1956)}]{Utiyama:1956sy}%
  \BibitemOpen
  \bibfield  {author} {\bibinfo {author} {\bibfnamefont {R.}~\bibnamefont
  {Utiyama}},\ }\bibfield  {title} {\bibinfo {title} {{Invariant theoretical
  interpretation of interaction}},\ }\href
  {https://doi.org/10.1103/PhysRev.101.1597} {\bibfield  {journal} {\bibinfo
  {journal} {Phys. Rev.}\ }\textbf {\bibinfo {volume} {101}},\ \bibinfo {pages}
  {1597} (\bibinfo {year} {1956})}\BibitemShut {NoStop}%
\bibitem [{\citenamefont {Kibble}(1961)}]{Kibble:1961ba}%
  \BibitemOpen
  \bibfield  {author} {\bibinfo {author} {\bibfnamefont {T.~W.~B.}\
  \bibnamefont {Kibble}},\ }\bibfield  {title} {\bibinfo {title} {{Lorentz
  invariance and the gravitational field}},\ }\href
  {https://doi.org/10.1063/1.1703702} {\bibfield  {journal} {\bibinfo
  {journal} {J. Math. Phys.}\ }\textbf {\bibinfo {volume} {2}},\ \bibinfo
  {pages} {212} (\bibinfo {year} {1961})}\BibitemShut {NoStop}%
\bibitem [{\citenamefont {Sciama}(1962)}]{Sciama:1962}%
  \BibitemOpen
  \bibfield  {author} {\bibinfo {author} {\bibfnamefont {D.~W.}\ \bibnamefont
  {Sciama}},\ }\bibfield  {title} {\bibinfo {title} {On the analogy between
  charge and spin in general relativity},\ }in\ \href@noop {} {\emph {\bibinfo
  {booktitle} {Recent developments in general relativity}}}\ (\bibinfo
  {publisher} {Pergamon Press},\ \bibinfo {address} {Oxford},\ \bibinfo {year}
  {1962})\ p.\ \bibinfo {pages} {415}\BibitemShut {NoStop}%
\bibitem [{\citenamefont {Blagojevic}(2002)}]{Blagojevic:2002}%
  \BibitemOpen
  \bibfield  {author} {\bibinfo {author} {\bibfnamefont {M.}~\bibnamefont
  {Blagojevic}},\ }\href@noop {} {\emph {\bibinfo {title} {{Gravitation and
  gauge symmetries}}}}\ (\bibinfo  {publisher} {CRC Press},\ \bibinfo {address}
  {Bristol and Philadelphia},\ \bibinfo {year} {2002})\BibitemShut {NoStop}%
\bibitem [{\citenamefont {Cartan}(1922)}]{Cartan:1922}%
  \BibitemOpen
  \bibfield  {author} {\bibinfo {author} {\bibfnamefont {{\'E}.}~\bibnamefont
  {Cartan}},\ }\bibfield  {title} {\bibinfo {title} {{Sur une
  g{\'e}n{\'e}ralisation de la notion de courbure de Riemann et les espaces
  {\`a} torsion}},\ }\href@noop {} {\bibfield  {journal} {\bibinfo  {journal}
  {Comptes Rendus, Ac. Sc. Paris}\ }\textbf {\bibinfo {volume} {174}},\
  \bibinfo {pages} {593} (\bibinfo {year} {1922})}\BibitemShut {NoStop}%
\bibitem [{\citenamefont {Cartan}(1923)}]{Cartan:1923}%
  \BibitemOpen
  \bibfield  {author} {\bibinfo {author} {\bibfnamefont {{\'E}.}~\bibnamefont
  {Cartan}},\ }\bibfield  {title} {\bibinfo {title} {{Sur les vari{\'e}t{\'e}s
  {\`a} connexion affine et la th{\'e}orie de la relativit{\'e}
  g{\'e}n{\'e}ralis{\'e}e (premi{\`e}re partie)}},\ }in\ \href@noop {} {\emph
  {\bibinfo {booktitle} {Annales scientifiques de l'{\'E}cole normale
  sup{\'e}rieure}}},\ Vol.~\bibinfo {volume} {40}\ (\bibinfo  {publisher}
  {Gauthier-Villars},\ \bibinfo {year} {1923})\ pp.\ \bibinfo {pages}
  {325--412}\BibitemShut {NoStop}%
\bibitem [{\citenamefont {Cartan}(1924)}]{Cartan:1924}%
  \BibitemOpen
  \bibfield  {author} {\bibinfo {author} {\bibfnamefont {{\'E}.}~\bibnamefont
  {Cartan}},\ }\bibfield  {title} {\bibinfo {title} {{Sur les vari{\'e}t{\'e}s
  {\`a} connexion affine, et la th{\'e}orie de la relativit{\'e}
  g{\'e}n{\'e}ralis{\'e}e (premi{\`e}re partie)(suite)}},\ }in\ \href@noop {}
  {\emph {\bibinfo {booktitle} {Annales scientifiques de l'{\'E}cole Normale
  Sup{\'e}rieure}}},\ Vol.~\bibinfo {volume} {41}\ (\bibinfo  {publisher}
  {Gauthier-Villars},\ \bibinfo {year} {1924})\ pp.\ \bibinfo {pages}
  {1--25}\BibitemShut {NoStop}%
\bibitem [{\citenamefont {Cartan}(1925)}]{Cartan:1925}%
  \BibitemOpen
  \bibfield  {author} {\bibinfo {author} {\bibfnamefont {{\'E}.}~\bibnamefont
  {Cartan}},\ }\bibfield  {title} {\bibinfo {title} {{Sur les vari{\'e}t{\'e}s
  {\`a} connexion affine, et la th{\'e}orie de la relativit{\'e}
  g{\'e}n{\'e}ralis{\'e}e (deuxi{\`e}me partie)}},\ }in\ \href@noop {} {\emph
  {\bibinfo {booktitle} {Annales scientifiques de l'{\'E}cole normale
  sup{\'e}rieure}}},\ Vol.~\bibinfo {volume} {42}\ (\bibinfo  {publisher}
  {Gauthier-Villars},\ \bibinfo {year} {1925})\ pp.\ \bibinfo {pages}
  {17--88}\BibitemShut {NoStop}%
\bibitem [{\citenamefont {Einstein}(1925)}]{Einstein:1925}%
  \BibitemOpen
  \bibfield  {author} {\bibinfo {author} {\bibfnamefont {A.}~\bibnamefont
  {Einstein}},\ }\bibfield  {title} {\bibinfo {title} {{Einheitliche
  Feldtheorie von Gravitation und Elektrizit\"at}},\ }\href@noop {} {\bibfield
  {journal} {\bibinfo  {journal} {Sitzungsber. Preuss. Akad. Wiss}\ }\textbf
  {\bibinfo {volume} {22}},\ \bibinfo {pages} {414} (\bibinfo {year}
  {1925})}\BibitemShut {NoStop}%
\bibitem [{\citenamefont {Einstein}(1928{\natexlab{a}})}]{Einstein:1928}%
  \BibitemOpen
  \bibfield  {author} {\bibinfo {author} {\bibfnamefont {A.}~\bibnamefont
  {Einstein}},\ }\bibfield  {title} {\bibinfo {title} {{Riemanngeometrie mit
  Aufrechterhaltung des Begriffes des Fern-Parallelismus}},\ }\href@noop {}
  {\bibfield  {journal} {\bibinfo  {journal} {Sitzungsber. Preuss. Akad. Wiss}\
  }\textbf {\bibinfo {volume} {17}},\ \bibinfo {pages} {217} (\bibinfo {year}
  {1928}{\natexlab{a}})}\BibitemShut {NoStop}%
\bibitem [{\citenamefont {Einstein}(1928{\natexlab{b}})}]{Einstein:19282}%
  \BibitemOpen
  \bibfield  {author} {\bibinfo {author} {\bibfnamefont {A.}~\bibnamefont
  {Einstein}},\ }\bibfield  {title} {\bibinfo {title} {{Neue M{\"o}glichkeit
  f{\"u}r eine einheitliche Feldtheorie von Gravitation und
  Elektrizit{\"a}t}},\ }\href@noop {} {\bibfield  {journal} {\bibinfo
  {journal} {Sitzungsber. Preuss. Akad. Wiss}\ }\textbf {\bibinfo {volume}
  {18}},\ \bibinfo {pages} {224} (\bibinfo {year}
  {1928}{\natexlab{b}})}\BibitemShut {NoStop}%
\bibitem [{\citenamefont {Rodichev}(1961)}]{Rodichev:1961}%
  \BibitemOpen
  \bibfield  {author} {\bibinfo {author} {\bibfnamefont {V.~I.}\ \bibnamefont
  {Rodichev}},\ }\bibfield  {title} {\bibinfo {title} {{Twisted Space and
  Nonlinear Field Equations}},\ }\href@noop {} {\bibfield  {journal} {\bibinfo
  {journal} {Zhur. Eksptl'. i Teoret. Fiz.}\ }\textbf {\bibinfo {volume}
  {40}},\ \bibinfo {pages} {1029} (\bibinfo {year} {1961})}\BibitemShut
  {NoStop}%
\bibitem [{\citenamefont {Freidel}\ \emph {et~al.}(2005)\citenamefont
  {Freidel}, \citenamefont {Minic},\ and\ \citenamefont
  {Takeuchi}}]{Freidel:2005sn}%
  \BibitemOpen
  \bibfield  {author} {\bibinfo {author} {\bibfnamefont {L.}~\bibnamefont
  {Freidel}}, \bibinfo {author} {\bibfnamefont {D.}~\bibnamefont {Minic}},\
  and\ \bibinfo {author} {\bibfnamefont {T.}~\bibnamefont {Takeuchi}},\
  }\bibfield  {title} {\bibinfo {title} {{Quantum gravity, torsion, parity
  violation and all that}},\ }\href
  {https://doi.org/10.1103/PhysRevD.72.104002} {\bibfield  {journal} {\bibinfo
  {journal} {Phys. Rev. D}\ }\textbf {\bibinfo {volume} {72}},\ \bibinfo
  {pages} {104002} (\bibinfo {year} {2005})},\ \Eprint
  {https://arxiv.org/abs/hep-th/0507253} {arXiv:hep-th/0507253} \BibitemShut
  {NoStop}%
\bibitem [{\citenamefont {Alexandrov}(2008)}]{Alexandrov:2008iy}%
  \BibitemOpen
  \bibfield  {author} {\bibinfo {author} {\bibfnamefont {S.}~\bibnamefont
  {Alexandrov}},\ }\bibfield  {title} {\bibinfo {title} {{Immirzi parameter and
  fermions with non-minimal coupling}},\ }\href
  {https://doi.org/10.1088/0264-9381/25/14/145012} {\bibfield  {journal}
  {\bibinfo  {journal} {Class. Quant. Grav.}\ }\textbf {\bibinfo {volume}
  {25}},\ \bibinfo {pages} {145012} (\bibinfo {year} {2008})},\ \Eprint
  {https://arxiv.org/abs/0802.1221} {arXiv:0802.1221 [gr-qc]} \BibitemShut
  {NoStop}%
\bibitem [{\citenamefont {Shaposhnikov}\ \emph {et~al.}(2020)\citenamefont
  {Shaposhnikov}, \citenamefont {Shkerin}, \citenamefont {Timiryasov},\ and\
  \citenamefont {Zell}}]{Shaposhnikov:2020frq}%
  \BibitemOpen
  \bibfield  {author} {\bibinfo {author} {\bibfnamefont {M.}~\bibnamefont
  {Shaposhnikov}}, \bibinfo {author} {\bibfnamefont {A.}~\bibnamefont
  {Shkerin}}, \bibinfo {author} {\bibfnamefont {I.}~\bibnamefont
  {Timiryasov}},\ and\ \bibinfo {author} {\bibfnamefont {S.}~\bibnamefont
  {Zell}},\ }\bibfield  {title} {\bibinfo {title} {{Einstein-Cartan gravity,
  matter, and scale-invariant generalization~}},\ }\href
  {https://doi.org/10.1007/JHEP08(2021)162} {\bibfield  {journal} {\bibinfo
  {journal} {JHEP}\ }\textbf {\bibinfo {volume} {10}},\ \bibinfo {pages}
  {177}},\ \Eprint {https://arxiv.org/abs/2007.16158} {arXiv:2007.16158
  [hep-th]} \BibitemShut {NoStop}%
\bibitem [{\citenamefont {Karananas}\ \emph
  {et~al.}(2021{\natexlab{a}})\citenamefont {Karananas}, \citenamefont
  {Shaposhnikov}, \citenamefont {Shkerin},\ and\ \citenamefont
  {Zell}}]{Karananas:2021zkl}%
  \BibitemOpen
  \bibfield  {author} {\bibinfo {author} {\bibfnamefont {G.~K.}\ \bibnamefont
  {Karananas}}, \bibinfo {author} {\bibfnamefont {M.}~\bibnamefont
  {Shaposhnikov}}, \bibinfo {author} {\bibfnamefont {A.}~\bibnamefont
  {Shkerin}},\ and\ \bibinfo {author} {\bibfnamefont {S.}~\bibnamefont
  {Zell}},\ }\bibfield  {title} {\bibinfo {title} {{Matter matters in
  Einstein-Cartan gravity}},\ }\href
  {https://doi.org/10.1103/PhysRevD.104.064036} {\bibfield  {journal} {\bibinfo
   {journal} {Phys. Rev. D}\ }\textbf {\bibinfo {volume} {104}},\ \bibinfo
  {pages} {064036} (\bibinfo {year} {2021}{\natexlab{a}})},\ \Eprint
  {https://arxiv.org/abs/2106.13811} {arXiv:2106.13811 [hep-th]} \BibitemShut
  {NoStop}%
\bibitem [{\citenamefont {Rigouzzo}\ and\ \citenamefont
  {Zell}(2023)}]{Rigouzzo:2023sbb}%
  \BibitemOpen
  \bibfield  {author} {\bibinfo {author} {\bibfnamefont {C.}~\bibnamefont
  {Rigouzzo}}\ and\ \bibinfo {author} {\bibfnamefont {S.}~\bibnamefont
  {Zell}},\ }\bibfield  {title} {\bibinfo {title} {{Coupling metric-affine
  gravity to the standard model and dark matter fermions}},\ }\href
  {https://doi.org/10.1103/PhysRevD.108.124067} {\bibfield  {journal} {\bibinfo
   {journal} {Phys. Rev. D}\ }\textbf {\bibinfo {volume} {108}},\ \bibinfo
  {pages} {124067} (\bibinfo {year} {2023})},\ \Eprint
  {https://arxiv.org/abs/2306.13134} {arXiv:2306.13134 [gr-qc]} \BibitemShut
  {NoStop}%
\bibitem [{\citenamefont {Bauer}\ and\ \citenamefont
  {Demir}(2008)}]{Bauer:2008zj}%
  \BibitemOpen
  \bibfield  {author} {\bibinfo {author} {\bibfnamefont {F.}~\bibnamefont
  {Bauer}}\ and\ \bibinfo {author} {\bibfnamefont {D.~A.}\ \bibnamefont
  {Demir}},\ }\bibfield  {title} {\bibinfo {title} {{Inflation with Non-Minimal
  Coupling: Metric versus Palatini Formulations}},\ }\href
  {https://doi.org/10.1016/j.physletb.2008.06.014} {\bibfield  {journal}
  {\bibinfo  {journal} {Phys. Lett. B}\ }\textbf {\bibinfo {volume} {665}},\
  \bibinfo {pages} {222} (\bibinfo {year} {2008})},\ \Eprint
  {https://arxiv.org/abs/0803.2664} {arXiv:0803.2664 [hep-ph]} \BibitemShut
  {NoStop}%
\bibitem [{\citenamefont {Poplawski}(2011)}]{Poplawski:2011xf}%
  \BibitemOpen
  \bibfield  {author} {\bibinfo {author} {\bibfnamefont {N.~J.}\ \bibnamefont
  {Poplawski}},\ }\bibfield  {title} {\bibinfo {title} {{Matter-antimatter
  asymmetry and dark matter from torsion}},\ }\href
  {https://doi.org/10.1103/PhysRevD.83.084033} {\bibfield  {journal} {\bibinfo
  {journal} {Phys. Rev. D}\ }\textbf {\bibinfo {volume} {83}},\ \bibinfo
  {pages} {084033} (\bibinfo {year} {2011})},\ \Eprint
  {https://arxiv.org/abs/1101.4012} {arXiv:1101.4012 [gr-qc]} \BibitemShut
  {NoStop}%
\bibitem [{\citenamefont {Diakonov}\ \emph {et~al.}(2011)\citenamefont
  {Diakonov}, \citenamefont {Tumanov},\ and\ \citenamefont
  {Vladimirov}}]{Diakonov:2011fs}%
  \BibitemOpen
  \bibfield  {author} {\bibinfo {author} {\bibfnamefont {D.}~\bibnamefont
  {Diakonov}}, \bibinfo {author} {\bibfnamefont {A.~G.}\ \bibnamefont
  {Tumanov}},\ and\ \bibinfo {author} {\bibfnamefont {A.~A.}\ \bibnamefont
  {Vladimirov}},\ }\bibfield  {title} {\bibinfo {title} {{Low-energy General
  Relativity with torsion: A Systematic derivative expansion}},\ }\href
  {https://doi.org/10.1103/PhysRevD.84.124042} {\bibfield  {journal} {\bibinfo
  {journal} {Phys. Rev. D}\ }\textbf {\bibinfo {volume} {84}},\ \bibinfo
  {pages} {124042} (\bibinfo {year} {2011})},\ \Eprint
  {https://arxiv.org/abs/1104.2432} {arXiv:1104.2432 [hep-th]} \BibitemShut
  {NoStop}%
\bibitem [{\citenamefont {Khriplovich}(2012)}]{Khriplovich:2012xg}%
  \BibitemOpen
  \bibfield  {author} {\bibinfo {author} {\bibfnamefont {I.~B.}\ \bibnamefont
  {Khriplovich}},\ }\bibfield  {title} {\bibinfo {title} {{Gravitational
  four-fermion interaction on the Planck scale}},\ }\href
  {https://doi.org/10.1016/j.physletb.2012.01.072} {\bibfield  {journal}
  {\bibinfo  {journal} {Phys. Lett. B}\ }\textbf {\bibinfo {volume} {709}},\
  \bibinfo {pages} {111} (\bibinfo {year} {2012})},\ \Eprint
  {https://arxiv.org/abs/1201.4226} {arXiv:1201.4226 [gr-qc]} \BibitemShut
  {NoStop}%
\bibitem [{\citenamefont {Magueijo}\ \emph {et~al.}(2013)\citenamefont
  {Magueijo}, \citenamefont {Zlosnik},\ and\ \citenamefont
  {Kibble}}]{Magueijo:2012ug}%
  \BibitemOpen
  \bibfield  {author} {\bibinfo {author} {\bibfnamefont {J.~a.}\ \bibnamefont
  {Magueijo}}, \bibinfo {author} {\bibfnamefont {T.~G.}\ \bibnamefont
  {Zlosnik}},\ and\ \bibinfo {author} {\bibfnamefont {T.~W.~B.}\ \bibnamefont
  {Kibble}},\ }\bibfield  {title} {\bibinfo {title} {{Cosmology with a spin}},\
  }\href {https://doi.org/10.1103/PhysRevD.87.063504} {\bibfield  {journal}
  {\bibinfo  {journal} {Phys. Rev. D}\ }\textbf {\bibinfo {volume} {87}},\
  \bibinfo {pages} {063504} (\bibinfo {year} {2013})},\ \Eprint
  {https://arxiv.org/abs/1212.0585} {arXiv:1212.0585 [astro-ph.CO]}
  \BibitemShut {NoStop}%
\bibitem [{\citenamefont {Khriplovich}\ and\ \citenamefont
  {Rudenko}(2013)}]{Khriplovich:2013tqa}%
  \BibitemOpen
  \bibfield  {author} {\bibinfo {author} {\bibfnamefont {I.~B.}\ \bibnamefont
  {Khriplovich}}\ and\ \bibinfo {author} {\bibfnamefont {A.~S.}\ \bibnamefont
  {Rudenko}},\ }\bibfield  {title} {\bibinfo {title} {{Gravitational
  four-fermion interaction and dynamics of the early Universe}},\ }\href
  {https://doi.org/10.1007/JHEP11(2013)174} {\bibfield  {journal} {\bibinfo
  {journal} {JHEP}\ }\textbf {\bibinfo {volume} {11}},\ \bibinfo {pages}
  {174}},\ \Eprint {https://arxiv.org/abs/1303.1348} {arXiv:1303.1348
  [astro-ph.CO]} \BibitemShut {NoStop}%
\bibitem [{\citenamefont {Markkanen}\ \emph {et~al.}(2018)\citenamefont
  {Markkanen}, \citenamefont {Tenkanen}, \citenamefont {Vaskonen},\ and\
  \citenamefont {Veerm\"ae}}]{Markkanen:2017tun}%
  \BibitemOpen
  \bibfield  {author} {\bibinfo {author} {\bibfnamefont {T.}~\bibnamefont
  {Markkanen}}, \bibinfo {author} {\bibfnamefont {T.}~\bibnamefont {Tenkanen}},
  \bibinfo {author} {\bibfnamefont {V.}~\bibnamefont {Vaskonen}},\ and\
  \bibinfo {author} {\bibfnamefont {H.}~\bibnamefont {Veerm\"ae}},\ }\bibfield
  {title} {\bibinfo {title} {{Quantum corrections to quartic inflation with a
  non-minimal coupling: metric vs. Palatini}},\ }\href
  {https://doi.org/10.1088/1475-7516/2018/03/029} {\bibfield  {journal}
  {\bibinfo  {journal} {JCAP}\ }\textbf {\bibinfo {volume} {03}},\ \bibinfo
  {pages} {029}},\ \Eprint {https://arxiv.org/abs/1712.04874} {arXiv:1712.04874
  [gr-qc]} \BibitemShut {NoStop}%
\bibitem [{\citenamefont {Carrilho}\ \emph {et~al.}(2018)\citenamefont
  {Carrilho}, \citenamefont {Mulryne}, \citenamefont {Ronayne},\ and\
  \citenamefont {Tenkanen}}]{Carrilho:2018ffi}%
  \BibitemOpen
  \bibfield  {author} {\bibinfo {author} {\bibfnamefont {P.}~\bibnamefont
  {Carrilho}}, \bibinfo {author} {\bibfnamefont {D.}~\bibnamefont {Mulryne}},
  \bibinfo {author} {\bibfnamefont {J.}~\bibnamefont {Ronayne}},\ and\ \bibinfo
  {author} {\bibfnamefont {T.}~\bibnamefont {Tenkanen}},\ }\bibfield  {title}
  {\bibinfo {title} {{Attractor Behaviour in Multifield Inflation}},\ }\href
  {https://doi.org/10.1088/1475-7516/2018/06/032} {\bibfield  {journal}
  {\bibinfo  {journal} {JCAP}\ }\textbf {\bibinfo {volume} {06}},\ \bibinfo
  {pages} {032}},\ \Eprint {https://arxiv.org/abs/1804.10489} {arXiv:1804.10489
  [astro-ph.CO]} \BibitemShut {NoStop}%
\bibitem [{\citenamefont {Enckell}\ \emph {et~al.}(2019)\citenamefont
  {Enckell}, \citenamefont {Enqvist}, \citenamefont {Rasanen},\ and\
  \citenamefont {Wahlman}}]{Enckell:2018hmo}%
  \BibitemOpen
  \bibfield  {author} {\bibinfo {author} {\bibfnamefont {V.-M.}\ \bibnamefont
  {Enckell}}, \bibinfo {author} {\bibfnamefont {K.}~\bibnamefont {Enqvist}},
  \bibinfo {author} {\bibfnamefont {S.}~\bibnamefont {Rasanen}},\ and\ \bibinfo
  {author} {\bibfnamefont {L.-P.}\ \bibnamefont {Wahlman}},\ }\bibfield
  {title} {\bibinfo {title} {{Inflation with $R^2$ term in the Palatini
  formalism}},\ }\href {https://doi.org/10.1088/1475-7516/2019/02/022}
  {\bibfield  {journal} {\bibinfo  {journal} {JCAP}\ }\textbf {\bibinfo
  {volume} {02}},\ \bibinfo {pages} {022}},\ \Eprint
  {https://arxiv.org/abs/1810.05536} {arXiv:1810.05536 [gr-qc]} \BibitemShut
  {NoStop}%
\bibitem [{\citenamefont {Rasanen}\ and\ \citenamefont
  {Tomberg}(2019)}]{Rasanen:2018fom}%
  \BibitemOpen
  \bibfield  {author} {\bibinfo {author} {\bibfnamefont {S.}~\bibnamefont
  {Rasanen}}\ and\ \bibinfo {author} {\bibfnamefont {E.}~\bibnamefont
  {Tomberg}},\ }\bibfield  {title} {\bibinfo {title} {{Planck scale black hole
  dark matter from Higgs inflation}},\ }\href
  {https://doi.org/10.1088/1475-7516/2019/01/038} {\bibfield  {journal}
  {\bibinfo  {journal} {JCAP}\ }\textbf {\bibinfo {volume} {01}},\ \bibinfo
  {pages} {038}},\ \Eprint {https://arxiv.org/abs/1810.12608} {arXiv:1810.12608
  [astro-ph.CO]} \BibitemShut {NoStop}%
\bibitem [{\citenamefont {Beltr\'an~Jim\'enez}\ and\ \citenamefont
  {Maldonado~Torralba}(2020{\natexlab{a}})}]{BeltranJimenez:2019hrm}%
  \BibitemOpen
  \bibfield  {author} {\bibinfo {author} {\bibfnamefont {J.}~\bibnamefont
  {Beltr\'an~Jim\'enez}}\ and\ \bibinfo {author} {\bibfnamefont {F.~J.}\
  \bibnamefont {Maldonado~Torralba}},\ }\bibfield  {title} {\bibinfo {title}
  {{Revisiting the stability of quadratic Poincar\'e gauge gravity}},\ }\href
  {https://doi.org/10.1140/epjc/s10052-020-8163-8} {\bibfield  {journal}
  {\bibinfo  {journal} {Eur. Phys. J. C}\ }\textbf {\bibinfo {volume} {80}},\
  \bibinfo {pages} {611} (\bibinfo {year} {2020}{\natexlab{a}})},\ \Eprint
  {https://arxiv.org/abs/1910.07506} {arXiv:1910.07506 [gr-qc]} \BibitemShut
  {NoStop}%
\bibitem [{\citenamefont {Rubio}\ and\ \citenamefont
  {Tomberg}(2019)}]{Rubio:2019ypq}%
  \BibitemOpen
  \bibfield  {author} {\bibinfo {author} {\bibfnamefont {J.}~\bibnamefont
  {Rubio}}\ and\ \bibinfo {author} {\bibfnamefont {E.~S.}\ \bibnamefont
  {Tomberg}},\ }\bibfield  {title} {\bibinfo {title} {{Preheating in Palatini
  Higgs inflation}},\ }\href {https://doi.org/10.1088/1475-7516/2019/04/021}
  {\bibfield  {journal} {\bibinfo  {journal} {JCAP}\ }\textbf {\bibinfo
  {volume} {04}},\ \bibinfo {pages} {021}},\ \Eprint
  {https://arxiv.org/abs/1902.10148} {arXiv:1902.10148 [hep-ph]} \BibitemShut
  {NoStop}%
\bibitem [{\citenamefont {Shaposhnikov}\ \emph
  {et~al.}(2021{\natexlab{a}})\citenamefont {Shaposhnikov}, \citenamefont
  {Shkerin},\ and\ \citenamefont {Zell}}]{Shaposhnikov:2020geh}%
  \BibitemOpen
  \bibfield  {author} {\bibinfo {author} {\bibfnamefont {M.}~\bibnamefont
  {Shaposhnikov}}, \bibinfo {author} {\bibfnamefont {A.}~\bibnamefont
  {Shkerin}},\ and\ \bibinfo {author} {\bibfnamefont {S.}~\bibnamefont
  {Zell}},\ }\bibfield  {title} {\bibinfo {title} {{Standard Model Meets
  Gravity: Electroweak Symmetry Breaking and Inflation}},\ }\href
  {https://doi.org/10.1103/PhysRevD.103.033006} {\bibfield  {journal} {\bibinfo
   {journal} {Phys. Rev. D}\ }\textbf {\bibinfo {volume} {103}},\ \bibinfo
  {pages} {033006} (\bibinfo {year} {2021}{\natexlab{a}})},\ \Eprint
  {https://arxiv.org/abs/2001.09088} {arXiv:2001.09088 [hep-th]} \BibitemShut
  {NoStop}%
\bibitem [{\citenamefont {Karananas}\ \emph {et~al.}(2020)\citenamefont
  {Karananas}, \citenamefont {Michel},\ and\ \citenamefont
  {Rubio}}]{Karananas:2020qkp}%
  \BibitemOpen
  \bibfield  {author} {\bibinfo {author} {\bibfnamefont {G.~K.}\ \bibnamefont
  {Karananas}}, \bibinfo {author} {\bibfnamefont {M.}~\bibnamefont {Michel}},\
  and\ \bibinfo {author} {\bibfnamefont {J.}~\bibnamefont {Rubio}},\ }\bibfield
   {title} {\bibinfo {title} {{One residue to rule them all: Electroweak
  symmetry breaking, inflation and field-space geometry}},\ }\href
  {https://doi.org/10.1016/j.physletb.2020.135876} {\bibfield  {journal}
  {\bibinfo  {journal} {Phys. Lett. B}\ }\textbf {\bibinfo {volume} {811}},\
  \bibinfo {pages} {135876} (\bibinfo {year} {2020})},\ \Eprint
  {https://arxiv.org/abs/2006.11290} {arXiv:2006.11290 [hep-th]} \BibitemShut
  {NoStop}%
\bibitem [{\citenamefont {L\r{a}ngvik}\ \emph {et~al.}(2021)\citenamefont
  {L\r{a}ngvik}, \citenamefont {Ojanper\"a}, \citenamefont {Raatikainen},\ and\
  \citenamefont {Rasanen}}]{Langvik:2020nrs}%
  \BibitemOpen
  \bibfield  {author} {\bibinfo {author} {\bibfnamefont {M.}~\bibnamefont
  {L\r{a}ngvik}}, \bibinfo {author} {\bibfnamefont {J.-M.}\ \bibnamefont
  {Ojanper\"a}}, \bibinfo {author} {\bibfnamefont {S.}~\bibnamefont
  {Raatikainen}},\ and\ \bibinfo {author} {\bibfnamefont {S.}~\bibnamefont
  {Rasanen}},\ }\bibfield  {title} {\bibinfo {title} {{Higgs inflation with the
  Holst and the Nieh\textendash{}Yan term}},\ }\href
  {https://doi.org/10.1103/PhysRevD.103.083514} {\bibfield  {journal} {\bibinfo
   {journal} {Phys. Rev. D}\ }\textbf {\bibinfo {volume} {103}},\ \bibinfo
  {pages} {083514} (\bibinfo {year} {2021})},\ \Eprint
  {https://arxiv.org/abs/2007.12595} {arXiv:2007.12595 [astro-ph.CO]}
  \BibitemShut {NoStop}%
\bibitem [{\citenamefont {Shaposhnikov}\ \emph
  {et~al.}(2021{\natexlab{b}})\citenamefont {Shaposhnikov}, \citenamefont
  {Shkerin}, \citenamefont {Timiryasov},\ and\ \citenamefont
  {Zell}}]{Shaposhnikov:2020gts}%
  \BibitemOpen
  \bibfield  {author} {\bibinfo {author} {\bibfnamefont {M.}~\bibnamefont
  {Shaposhnikov}}, \bibinfo {author} {\bibfnamefont {A.}~\bibnamefont
  {Shkerin}}, \bibinfo {author} {\bibfnamefont {I.}~\bibnamefont
  {Timiryasov}},\ and\ \bibinfo {author} {\bibfnamefont {S.}~\bibnamefont
  {Zell}},\ }\bibfield  {title} {\bibinfo {title} {{Higgs inflation in
  Einstein-Cartan gravity}},\ }\href
  {https://doi.org/10.1088/1475-7516/2021/10/E01} {\bibfield  {journal}
  {\bibinfo  {journal} {JCAP}\ }\textbf {\bibinfo {volume} {02}},\ \bibinfo
  {pages} {008}},\ \bibinfo {note} {[Erratum: JCAP 10, E01 (2021)]},\ \Eprint
  {https://arxiv.org/abs/2007.14978} {arXiv:2007.14978 [hep-ph]} \BibitemShut
  {NoStop}%
\bibitem [{\citenamefont {Mikura}\ \emph {et~al.}(2020)\citenamefont {Mikura},
  \citenamefont {Tada},\ and\ \citenamefont {Yokoyama}}]{Mikura:2020qhc}%
  \BibitemOpen
  \bibfield  {author} {\bibinfo {author} {\bibfnamefont {Y.}~\bibnamefont
  {Mikura}}, \bibinfo {author} {\bibfnamefont {Y.}~\bibnamefont {Tada}},\ and\
  \bibinfo {author} {\bibfnamefont {S.}~\bibnamefont {Yokoyama}},\ }\bibfield
  {title} {\bibinfo {title} {{Conformal inflation in the metric-affine
  geometry}},\ }\href {https://doi.org/10.1209/0295-5075/132/39001} {\bibfield
  {journal} {\bibinfo  {journal} {EPL}\ }\textbf {\bibinfo {volume} {132}},\
  \bibinfo {pages} {39001} (\bibinfo {year} {2020})},\ \Eprint
  {https://arxiv.org/abs/2008.00628} {arXiv:2008.00628 [hep-th]} \BibitemShut
  {NoStop}%
\bibitem [{\citenamefont {Shaposhnikov}\ \emph
  {et~al.}(2021{\natexlab{c}})\citenamefont {Shaposhnikov}, \citenamefont
  {Shkerin}, \citenamefont {Timiryasov},\ and\ \citenamefont
  {Zell}}]{Shaposhnikov:2020aen}%
  \BibitemOpen
  \bibfield  {author} {\bibinfo {author} {\bibfnamefont {M.}~\bibnamefont
  {Shaposhnikov}}, \bibinfo {author} {\bibfnamefont {A.}~\bibnamefont
  {Shkerin}}, \bibinfo {author} {\bibfnamefont {I.}~\bibnamefont
  {Timiryasov}},\ and\ \bibinfo {author} {\bibfnamefont {S.}~\bibnamefont
  {Zell}},\ }\bibfield  {title} {\bibinfo {title} {{Einstein-Cartan Portal to
  Dark Matter}},\ }\href {https://doi.org/10.1103/PhysRevLett.127.169901}
  {\bibfield  {journal} {\bibinfo  {journal} {Phys. Rev. Lett.}\ }\textbf
  {\bibinfo {volume} {126}},\ \bibinfo {pages} {161301} (\bibinfo {year}
  {2021}{\natexlab{c}})},\ \bibinfo {note} {[Erratum: Phys.Rev.Lett. 127,
  169901 (2021)]},\ \Eprint {https://arxiv.org/abs/2008.11686}
  {arXiv:2008.11686 [hep-ph]} \BibitemShut {NoStop}%
\bibitem [{\citenamefont {Kubota}\ \emph {et~al.}(2021)\citenamefont {Kubota},
  \citenamefont {Oda}, \citenamefont {Shimada},\ and\ \citenamefont
  {Yamaguchi}}]{Kubota:2020ehu}%
  \BibitemOpen
  \bibfield  {author} {\bibinfo {author} {\bibfnamefont {M.}~\bibnamefont
  {Kubota}}, \bibinfo {author} {\bibfnamefont {K.-Y.}\ \bibnamefont {Oda}},
  \bibinfo {author} {\bibfnamefont {K.}~\bibnamefont {Shimada}},\ and\ \bibinfo
  {author} {\bibfnamefont {M.}~\bibnamefont {Yamaguchi}},\ }\bibfield  {title}
  {\bibinfo {title} {{Cosmological Perturbations in Palatini Formalism}},\
  }\href {https://doi.org/10.1088/1475-7516/2021/03/006} {\bibfield  {journal}
  {\bibinfo  {journal} {JCAP}\ }\textbf {\bibinfo {volume} {03}},\ \bibinfo
  {pages} {006}},\ \Eprint {https://arxiv.org/abs/2010.07867} {arXiv:2010.07867
  [hep-th]} \BibitemShut {NoStop}%
\bibitem [{\citenamefont {Enckell}\ \emph {et~al.}(2021)\citenamefont
  {Enckell}, \citenamefont {Nurmi}, \citenamefont {R\"as\"anen},\ and\
  \citenamefont {Tomberg}}]{Enckell:2020lvn}%
  \BibitemOpen
  \bibfield  {author} {\bibinfo {author} {\bibfnamefont {V.-M.}\ \bibnamefont
  {Enckell}}, \bibinfo {author} {\bibfnamefont {S.}~\bibnamefont {Nurmi}},
  \bibinfo {author} {\bibfnamefont {S.}~\bibnamefont {R\"as\"anen}},\ and\
  \bibinfo {author} {\bibfnamefont {E.}~\bibnamefont {Tomberg}},\ }\bibfield
  {title} {\bibinfo {title} {{Critical point Higgs inflation in the Palatini
  formulation}},\ }\href {https://doi.org/10.1007/JHEP04(2021)059} {\bibfield
  {journal} {\bibinfo  {journal} {JHEP}\ }\textbf {\bibinfo {volume} {04}},\
  \bibinfo {pages} {059}},\ \Eprint {https://arxiv.org/abs/2012.03660}
  {arXiv:2012.03660 [astro-ph.CO]} \BibitemShut {NoStop}%
\bibitem [{\citenamefont {Iosifidis}\ and\ \citenamefont
  {Ravera}(2021)}]{Iosifidis:2021iuw}%
  \BibitemOpen
  \bibfield  {author} {\bibinfo {author} {\bibfnamefont {D.}~\bibnamefont
  {Iosifidis}}\ and\ \bibinfo {author} {\bibfnamefont {L.}~\bibnamefont
  {Ravera}},\ }\bibfield  {title} {\bibinfo {title} {{The cosmology of
  quadratic torsionful gravity}},\ }\href
  {https://doi.org/10.1140/epjc/s10052-021-09532-8} {\bibfield  {journal}
  {\bibinfo  {journal} {Eur. Phys. J. C}\ }\textbf {\bibinfo {volume} {81}},\
  \bibinfo {pages} {736} (\bibinfo {year} {2021})},\ \Eprint
  {https://arxiv.org/abs/2101.10339} {arXiv:2101.10339 [gr-qc]} \BibitemShut
  {NoStop}%
\bibitem [{\citenamefont {Bombacigno}\ \emph {et~al.}(2021)\citenamefont
  {Bombacigno}, \citenamefont {Boudet}, \citenamefont {Olmo},\ and\
  \citenamefont {Montani}}]{Bombacigno:2021bpk}%
  \BibitemOpen
  \bibfield  {author} {\bibinfo {author} {\bibfnamefont {F.}~\bibnamefont
  {Bombacigno}}, \bibinfo {author} {\bibfnamefont {S.}~\bibnamefont {Boudet}},
  \bibinfo {author} {\bibfnamefont {G.~J.}\ \bibnamefont {Olmo}},\ and\
  \bibinfo {author} {\bibfnamefont {G.}~\bibnamefont {Montani}},\ }\bibfield
  {title} {\bibinfo {title} {{Big bounce and future time singularity resolution
  in Bianchi I cosmologies: The projective invariant Nieh-Yan case}},\ }\href
  {https://doi.org/10.1103/PhysRevD.103.124031} {\bibfield  {journal} {\bibinfo
   {journal} {Phys. Rev. D}\ }\textbf {\bibinfo {volume} {103}},\ \bibinfo
  {pages} {124031} (\bibinfo {year} {2021})},\ \Eprint
  {https://arxiv.org/abs/2105.06870} {arXiv:2105.06870 [gr-qc]} \BibitemShut
  {NoStop}%
\bibitem [{\citenamefont {Racioppi}\ \emph {et~al.}(2022)\citenamefont
  {Racioppi}, \citenamefont {Rajasalu},\ and\ \citenamefont
  {Selke}}]{Racioppi:2021ynx}%
  \BibitemOpen
  \bibfield  {author} {\bibinfo {author} {\bibfnamefont {A.}~\bibnamefont
  {Racioppi}}, \bibinfo {author} {\bibfnamefont {J.}~\bibnamefont {Rajasalu}},\
  and\ \bibinfo {author} {\bibfnamefont {K.}~\bibnamefont {Selke}},\ }\bibfield
   {title} {\bibinfo {title} {{Multiple point criticality principle and
  Coleman-Weinberg inflation}},\ }\href
  {https://doi.org/10.1007/JHEP06(2022)107} {\bibfield  {journal} {\bibinfo
  {journal} {JHEP}\ }\textbf {\bibinfo {volume} {06}},\ \bibinfo {pages}
  {107}},\ \Eprint {https://arxiv.org/abs/2109.03238} {arXiv:2109.03238
  [astro-ph.CO]} \BibitemShut {NoStop}%
\bibitem [{\citenamefont {Cheong}\ \emph {et~al.}(2022)\citenamefont {Cheong},
  \citenamefont {Lee},\ and\ \citenamefont {Park}}]{Cheong:2021kyc}%
  \BibitemOpen
  \bibfield  {author} {\bibinfo {author} {\bibfnamefont {D.~Y.}\ \bibnamefont
  {Cheong}}, \bibinfo {author} {\bibfnamefont {S.~M.}\ \bibnamefont {Lee}},\
  and\ \bibinfo {author} {\bibfnamefont {S.~C.}\ \bibnamefont {Park}},\
  }\bibfield  {title} {\bibinfo {title} {{Reheating in models with non-minimal
  coupling in metric and~Palatini formalisms}},\ }\href
  {https://doi.org/10.1088/1475-7516/2022/02/029} {\bibfield  {journal}
  {\bibinfo  {journal} {JCAP}\ }\textbf {\bibinfo {volume} {02}}\bibfield
  {number} {\bibinfo  {number} { (02)},\ \bibinfo {pages} {029}},\ }\Eprint
  {https://arxiv.org/abs/2111.00825} {arXiv:2111.00825 [hep-ph]} \BibitemShut
  {NoStop}%
\bibitem [{\citenamefont {Dioguardi}\ \emph {et~al.}(2022)\citenamefont
  {Dioguardi}, \citenamefont {Racioppi},\ and\ \citenamefont
  {Tomberg}}]{Dioguardi:2021fmr}%
  \BibitemOpen
  \bibfield  {author} {\bibinfo {author} {\bibfnamefont {C.}~\bibnamefont
  {Dioguardi}}, \bibinfo {author} {\bibfnamefont {A.}~\bibnamefont
  {Racioppi}},\ and\ \bibinfo {author} {\bibfnamefont {E.}~\bibnamefont
  {Tomberg}},\ }\bibfield  {title} {\bibinfo {title} {{Slow-roll inflation in
  Palatini F(R) gravity}},\ }\href {https://doi.org/10.1007/JHEP06(2022)106}
  {\bibfield  {journal} {\bibinfo  {journal} {JHEP}\ }\textbf {\bibinfo
  {volume} {06}},\ \bibinfo {pages} {106}},\ \Eprint
  {https://arxiv.org/abs/2112.12149} {arXiv:2112.12149 [gr-qc]} \BibitemShut
  {NoStop}%
\bibitem [{\citenamefont {Piani}\ and\ \citenamefont
  {Rubio}(2022)}]{Piani:2022gon}%
  \BibitemOpen
  \bibfield  {author} {\bibinfo {author} {\bibfnamefont {M.}~\bibnamefont
  {Piani}}\ and\ \bibinfo {author} {\bibfnamefont {J.}~\bibnamefont {Rubio}},\
  }\bibfield  {title} {\bibinfo {title} {{Higgs-Dilaton inflation in
  Einstein-Cartan gravity}},\ }\href
  {https://doi.org/10.1088/1475-7516/2022/05/009} {\bibfield  {journal}
  {\bibinfo  {journal} {JCAP}\ }\textbf {\bibinfo {volume} {05}}\bibfield
  {number} {\bibinfo  {number} { (05)},\ \bibinfo {pages} {009}},\ }\Eprint
  {https://arxiv.org/abs/2202.04665} {arXiv:2202.04665 [gr-qc]} \BibitemShut
  {NoStop}%
\bibitem [{\citenamefont {Dux}\ \emph {et~al.}(2022)\citenamefont {Dux},
  \citenamefont {Florio}, \citenamefont {Klari\'c}, \citenamefont {Shkerin},\
  and\ \citenamefont {Timiryasov}}]{Dux:2022kuk}%
  \BibitemOpen
  \bibfield  {author} {\bibinfo {author} {\bibfnamefont {F.}~\bibnamefont
  {Dux}}, \bibinfo {author} {\bibfnamefont {A.}~\bibnamefont {Florio}},
  \bibinfo {author} {\bibfnamefont {J.}~\bibnamefont {Klari\'c}}, \bibinfo
  {author} {\bibfnamefont {A.}~\bibnamefont {Shkerin}},\ and\ \bibinfo {author}
  {\bibfnamefont {I.}~\bibnamefont {Timiryasov}},\ }\bibfield  {title}
  {\bibinfo {title} {{Preheating in Palatini Higgs inflation on the lattice}},\
  }\href {https://doi.org/10.1088/1475-7516/2022/09/015} {\bibfield  {journal}
  {\bibinfo  {journal} {JCAP}\ }\textbf {\bibinfo {volume} {09}},\ \bibinfo
  {pages} {015}},\ \Eprint {https://arxiv.org/abs/2203.13286} {arXiv:2203.13286
  [hep-ph]} \BibitemShut {NoStop}%
\bibitem [{\citenamefont {Rigouzzo}\ and\ \citenamefont
  {Zell}(2022)}]{Rigouzzo:2022yan}%
  \BibitemOpen
  \bibfield  {author} {\bibinfo {author} {\bibfnamefont {C.}~\bibnamefont
  {Rigouzzo}}\ and\ \bibinfo {author} {\bibfnamefont {S.}~\bibnamefont
  {Zell}},\ }\bibfield  {title} {\bibinfo {title} {{Coupling metric-affine
  gravity to a Higgs-like scalar field}},\ }\href
  {https://doi.org/10.1103/PhysRevD.106.024015} {\bibfield  {journal} {\bibinfo
   {journal} {Phys. Rev. D}\ }\textbf {\bibinfo {volume} {106}},\ \bibinfo
  {pages} {024015} (\bibinfo {year} {2022})},\ \Eprint
  {https://arxiv.org/abs/2204.03003} {arXiv:2204.03003 [hep-th]} \BibitemShut
  {NoStop}%
\bibitem [{\citenamefont {Pradisi}\ and\ \citenamefont
  {Salvio}(2022)}]{Pradisi:2022nmh}%
  \BibitemOpen
  \bibfield  {author} {\bibinfo {author} {\bibfnamefont {G.}~\bibnamefont
  {Pradisi}}\ and\ \bibinfo {author} {\bibfnamefont {A.}~\bibnamefont
  {Salvio}},\ }\bibfield  {title} {\bibinfo {title} {{(In)equivalence of
  metric-affine and metric effective field theories}},\ }\href
  {https://doi.org/10.1140/epjc/s10052-022-10825-9} {\bibfield  {journal}
  {\bibinfo  {journal} {Eur. Phys. J. C}\ }\textbf {\bibinfo {volume} {82}},\
  \bibinfo {pages} {840} (\bibinfo {year} {2022})},\ \Eprint
  {https://arxiv.org/abs/2206.15041} {arXiv:2206.15041 [hep-th]} \BibitemShut
  {NoStop}%
\bibitem [{\citenamefont {Salvio}(2022)}]{Salvio:2022suk}%
  \BibitemOpen
  \bibfield  {author} {\bibinfo {author} {\bibfnamefont {A.}~\bibnamefont
  {Salvio}},\ }\bibfield  {title} {\bibinfo {title} {{Inflating and reheating
  the Universe with an independent affine connection}},\ }\href
  {https://doi.org/10.1103/PhysRevD.106.103510} {\bibfield  {journal} {\bibinfo
   {journal} {Phys. Rev. D}\ }\textbf {\bibinfo {volume} {106}},\ \bibinfo
  {pages} {103510} (\bibinfo {year} {2022})},\ \Eprint
  {https://arxiv.org/abs/2207.08830} {arXiv:2207.08830 [hep-ph]} \BibitemShut
  {NoStop}%
\bibitem [{\citenamefont {Rasanen}\ and\ \citenamefont
  {Verbin}(2022)}]{Rasanen:2022ijc}%
  \BibitemOpen
  \bibfield  {author} {\bibinfo {author} {\bibfnamefont {S.}~\bibnamefont
  {Rasanen}}\ and\ \bibinfo {author} {\bibfnamefont {Y.}~\bibnamefont
  {Verbin}},\ }\bibfield  {title} {\bibinfo {title} {{Palatini formulation for
  gauge theory: implications for slow-roll inflation}}\ }\href
  {https://doi.org/10.21105/astro.2211.15584} {10.21105/astro.2211.15584}
  (\bibinfo {year} {2022}),\ \Eprint {https://arxiv.org/abs/2211.15584}
  {arXiv:2211.15584 [astro-ph.CO]} \BibitemShut {NoStop}%
\bibitem [{\citenamefont {Gialamas}\ and\ \citenamefont
  {Tamvakis}(2023)}]{Gialamas:2022xtt}%
  \BibitemOpen
  \bibfield  {author} {\bibinfo {author} {\bibfnamefont {I.~D.}\ \bibnamefont
  {Gialamas}}\ and\ \bibinfo {author} {\bibfnamefont {K.}~\bibnamefont
  {Tamvakis}},\ }\bibfield  {title} {\bibinfo {title} {{Inflation in
  metric-affine quadratic gravity}},\ }\href
  {https://doi.org/10.1088/1475-7516/2023/03/042} {\bibfield  {journal}
  {\bibinfo  {journal} {JCAP}\ }\textbf {\bibinfo {volume} {03}},\ \bibinfo
  {pages} {042}},\ \Eprint {https://arxiv.org/abs/2212.09896} {arXiv:2212.09896
  [gr-qc]} \BibitemShut {NoStop}%
\bibitem [{\citenamefont {Gialamas}\ and\ \citenamefont
  {Veerm\"ae}(2023)}]{Gialamas:2023emn}%
  \BibitemOpen
  \bibfield  {author} {\bibinfo {author} {\bibfnamefont {I.~D.}\ \bibnamefont
  {Gialamas}}\ and\ \bibinfo {author} {\bibfnamefont {H.}~\bibnamefont
  {Veerm\"ae}},\ }\bibfield  {title} {\bibinfo {title} {{Electroweak vacuum
  decay in metric-affine gravity}},\ }\href
  {https://doi.org/10.1016/j.physletb.2023.138109} {\bibfield  {journal}
  {\bibinfo  {journal} {Phys. Lett. B}\ }\textbf {\bibinfo {volume} {844}},\
  \bibinfo {pages} {138109} (\bibinfo {year} {2023})},\ \Eprint
  {https://arxiv.org/abs/2305.07693} {arXiv:2305.07693 [hep-th]} \BibitemShut
  {NoStop}%
\bibitem [{\citenamefont {Gialamas}\ \emph {et~al.}(2023)\citenamefont
  {Gialamas}, \citenamefont {Karam}, \citenamefont {Pappas},\ and\
  \citenamefont {Tomberg}}]{Gialamas:2023flv}%
  \BibitemOpen
  \bibfield  {author} {\bibinfo {author} {\bibfnamefont {I.~D.}\ \bibnamefont
  {Gialamas}}, \bibinfo {author} {\bibfnamefont {A.}~\bibnamefont {Karam}},
  \bibinfo {author} {\bibfnamefont {T.~D.}\ \bibnamefont {Pappas}},\ and\
  \bibinfo {author} {\bibfnamefont {E.}~\bibnamefont {Tomberg}},\ }\bibfield
  {title} {\bibinfo {title} {{Implications of Palatini gravity for inflation
  and beyond}},\ }\href {https://doi.org/10.1142/S0219887823300076} {\bibfield
  {journal} {\bibinfo  {journal} {Int. J. Geom. Meth. Mod. Phys.}\ }\textbf
  {\bibinfo {volume} {20}},\ \bibinfo {pages} {2330007} (\bibinfo {year}
  {2023})},\ \Eprint {https://arxiv.org/abs/2303.14148} {arXiv:2303.14148
  [gr-qc]} \BibitemShut {NoStop}%
\bibitem [{\citenamefont {Piani}\ and\ \citenamefont
  {Rubio}(2023)}]{Piani:2023aof}%
  \BibitemOpen
  \bibfield  {author} {\bibinfo {author} {\bibfnamefont {M.}~\bibnamefont
  {Piani}}\ and\ \bibinfo {author} {\bibfnamefont {J.}~\bibnamefont {Rubio}},\
  }\bibfield  {title} {\bibinfo {title} {{Preheating in Einstein-Cartan Higgs
  Inflation: oscillon formation}},\ }\href
  {https://doi.org/10.1088/1475-7516/2023/12/002} {\bibfield  {journal}
  {\bibinfo  {journal} {JCAP}\ }\textbf {\bibinfo {volume} {12}},\ \bibinfo
  {pages} {002}},\ \Eprint {https://arxiv.org/abs/2304.13056} {arXiv:2304.13056
  [hep-ph]} \BibitemShut {NoStop}%
\bibitem [{\citenamefont {Poisson}\ \emph {et~al.}(2024)\citenamefont
  {Poisson}, \citenamefont {Timiryasov},\ and\ \citenamefont
  {Zell}}]{Poisson:2023tja}%
  \BibitemOpen
  \bibfield  {author} {\bibinfo {author} {\bibfnamefont {A.}~\bibnamefont
  {Poisson}}, \bibinfo {author} {\bibfnamefont {I.}~\bibnamefont
  {Timiryasov}},\ and\ \bibinfo {author} {\bibfnamefont {S.}~\bibnamefont
  {Zell}},\ }\bibfield  {title} {\bibinfo {title} {{Critical points in Palatini
  Higgs inflation with small non-minimal coupling}},\ }\href
  {https://doi.org/10.1007/JHEP03(2024)130} {\bibfield  {journal} {\bibinfo
  {journal} {JHEP}\ }\textbf {\bibinfo {volume} {03}},\ \bibinfo {pages}
  {130}},\ \Eprint {https://arxiv.org/abs/2306.03893} {arXiv:2306.03893
  [hep-ph]} \BibitemShut {NoStop}%
\bibitem [{\citenamefont {Barker}\ and\ \citenamefont
  {Zell}(2024{\natexlab{a}})}]{Barker:2023fem}%
  \BibitemOpen
  \bibfield  {author} {\bibinfo {author} {\bibfnamefont {W.}~\bibnamefont
  {Barker}}\ and\ \bibinfo {author} {\bibfnamefont {S.}~\bibnamefont {Zell}},\
  }\bibfield  {title} {\bibinfo {title} {{Einstein-Proca theory from the
  Einstein-Cartan formulation}},\ }\href
  {https://doi.org/10.1103/PhysRevD.109.024007} {\bibfield  {journal} {\bibinfo
   {journal} {Phys. Rev. D}\ }\textbf {\bibinfo {volume} {109}},\ \bibinfo
  {pages} {024007} (\bibinfo {year} {2024}{\natexlab{a}})},\ \Eprint
  {https://arxiv.org/abs/2306.14953} {arXiv:2306.14953 [hep-th]} \BibitemShut
  {NoStop}%
\bibitem [{\citenamefont {Karananas}\ \emph {et~al.}(2023)\citenamefont
  {Karananas}, \citenamefont {Shaposhnikov},\ and\ \citenamefont
  {Zell}}]{Karananas:2023zgg}%
  \BibitemOpen
  \bibfield  {author} {\bibinfo {author} {\bibfnamefont {G.~K.}\ \bibnamefont
  {Karananas}}, \bibinfo {author} {\bibfnamefont {M.}~\bibnamefont
  {Shaposhnikov}},\ and\ \bibinfo {author} {\bibfnamefont {S.}~\bibnamefont
  {Zell}},\ }\bibfield  {title} {\bibinfo {title} {{Scale invariant
  Einstein-Cartan gravity and flat space conformal symmetry}},\ }\href
  {https://doi.org/10.1007/JHEP11(2023)171} {\bibfield  {journal} {\bibinfo
  {journal} {JHEP}\ }\textbf {\bibinfo {volume} {11}},\ \bibinfo {pages}
  {171}},\ \Eprint {https://arxiv.org/abs/2307.11151} {arXiv:2307.11151
  [hep-th]} \BibitemShut {NoStop}%
\bibitem [{\citenamefont {Martini}\ \emph {et~al.}(2024)\citenamefont
  {Martini}, \citenamefont {Paci},\ and\ \citenamefont
  {Sauro}}]{Martini:2023apm}%
  \BibitemOpen
  \bibfield  {author} {\bibinfo {author} {\bibfnamefont {R.}~\bibnamefont
  {Martini}}, \bibinfo {author} {\bibfnamefont {G.}~\bibnamefont {Paci}},\ and\
  \bibinfo {author} {\bibfnamefont {D.}~\bibnamefont {Sauro}},\ }\bibfield
  {title} {\bibinfo {title} {{Radiative corrections to the R and R$^{2}$
  invariants from torsion fluctuations on maximally symmetric spaces}},\ }\href
  {https://doi.org/10.1007/JHEP12(2024)138} {\bibfield  {journal} {\bibinfo
  {journal} {JHEP}\ }\textbf {\bibinfo {volume} {12}},\ \bibinfo {pages}
  {138}},\ \Eprint {https://arxiv.org/abs/2312.16681} {arXiv:2312.16681
  [gr-qc]} \BibitemShut {NoStop}%
\bibitem [{\citenamefont {He}\ \emph {et~al.}(2024)\citenamefont {He},
  \citenamefont {Hong},\ and\ \citenamefont {Mukaida}}]{He:2024wqv}%
  \BibitemOpen
  \bibfield  {author} {\bibinfo {author} {\bibfnamefont {M.}~\bibnamefont
  {He}}, \bibinfo {author} {\bibfnamefont {M.}~\bibnamefont {Hong}},\ and\
  \bibinfo {author} {\bibfnamefont {K.}~\bibnamefont {Mukaida}},\ }\bibfield
  {title} {\bibinfo {title} {{Starobinsky inflation and beyond in
  Einstein-Cartan gravity}},\ }\href
  {https://doi.org/10.1088/1475-7516/2024/05/107} {\bibfield  {journal}
  {\bibinfo  {journal} {JCAP}\ }\textbf {\bibinfo {volume} {05}},\ \bibinfo
  {pages} {107}},\ \Eprint {https://arxiv.org/abs/2402.05358} {arXiv:2402.05358
  [gr-qc]} \BibitemShut {NoStop}%
\bibitem [{\citenamefont {Hayashi}\ and\ \citenamefont
  {Nakano}(1967)}]{Hayashi:1967se}%
  \BibitemOpen
  \bibfield  {author} {\bibinfo {author} {\bibfnamefont {K.}~\bibnamefont
  {Hayashi}}\ and\ \bibinfo {author} {\bibfnamefont {T.}~\bibnamefont
  {Nakano}},\ }\bibfield  {title} {\bibinfo {title} {{Extended translation
  invariance and associated gauge fields}},\ }\href
  {https://doi.org/10.1143/PTP.38.491} {\bibfield  {journal} {\bibinfo
  {journal} {Prog. Theor. Phys.}\ }\textbf {\bibinfo {volume} {38}},\ \bibinfo
  {pages} {491} (\bibinfo {year} {1967})}\BibitemShut {NoStop}%
\bibitem [{\citenamefont {Hayashi}\ and\ \citenamefont
  {Shirafuji}(1980)}]{Hayashi:1980qp}%
  \BibitemOpen
  \bibfield  {author} {\bibinfo {author} {\bibfnamefont {K.}~\bibnamefont
  {Hayashi}}\ and\ \bibinfo {author} {\bibfnamefont {T.}~\bibnamefont
  {Shirafuji}},\ }\bibfield  {title} {\bibinfo {title} {{Gravity From Poincare
  Gauge Theory of the Fundamental Particles. 4. Mass and Energy of Particle
  Spectrum}},\ }\href {https://doi.org/10.1143/PTP.64.2222} {\bibfield
  {journal} {\bibinfo  {journal} {Prog. Theor. Phys.}\ }\textbf {\bibinfo
  {volume} {64}},\ \bibinfo {pages} {2222} (\bibinfo {year}
  {1980})}\BibitemShut {NoStop}%
\bibitem [{\citenamefont {Yo}\ and\ \citenamefont {Nester}(1999)}]{Yo:1999ex}%
  \BibitemOpen
  \bibfield  {author} {\bibinfo {author} {\bibfnamefont {H.-j.}\ \bibnamefont
  {Yo}}\ and\ \bibinfo {author} {\bibfnamefont {J.~M.}\ \bibnamefont
  {Nester}},\ }\bibfield  {title} {\bibinfo {title} {{Hamiltonian analysis of
  Poincare gauge theory scalar modes}},\ }\href
  {https://doi.org/10.1142/S021827189900033X} {\bibfield  {journal} {\bibinfo
  {journal} {Int. J. Mod. Phys. D}\ }\textbf {\bibinfo {volume} {8}},\ \bibinfo
  {pages} {459} (\bibinfo {year} {1999})},\ \Eprint
  {https://arxiv.org/abs/gr-qc/9902032} {arXiv:gr-qc/9902032} \BibitemShut
  {NoStop}%
\bibitem [{\citenamefont {Yo}\ and\ \citenamefont {Nester}(2002)}]{Yo:2001sy}%
  \BibitemOpen
  \bibfield  {author} {\bibinfo {author} {\bibfnamefont {H.-J.}\ \bibnamefont
  {Yo}}\ and\ \bibinfo {author} {\bibfnamefont {J.~M.}\ \bibnamefont
  {Nester}},\ }\bibfield  {title} {\bibinfo {title} {{Hamiltonian analysis of
  Poincare gauge theory: Higher spin modes}},\ }\href
  {https://doi.org/10.1142/S0218271802001998} {\bibfield  {journal} {\bibinfo
  {journal} {Int. J. Mod. Phys. D}\ }\textbf {\bibinfo {volume} {11}},\
  \bibinfo {pages} {747} (\bibinfo {year} {2002})},\ \Eprint
  {https://arxiv.org/abs/gr-qc/0112030} {arXiv:gr-qc/0112030} \BibitemShut
  {NoStop}%
\bibitem [{\citenamefont {Puetzfeld}(2005)}]{Puetzfeld:2004yg}%
  \BibitemOpen
  \bibfield  {author} {\bibinfo {author} {\bibfnamefont {D.}~\bibnamefont
  {Puetzfeld}},\ }\bibfield  {title} {\bibinfo {title} {{Status of
  non-Riemannian cosmology}},\ }\href
  {https://doi.org/10.1016/j.newar.2005.01.022} {\bibfield  {journal} {\bibinfo
   {journal} {New Astron. Rev.}\ }\textbf {\bibinfo {volume} {49}},\ \bibinfo
  {pages} {59} (\bibinfo {year} {2005})},\ \Eprint
  {https://arxiv.org/abs/gr-qc/0404119} {arXiv:gr-qc/0404119} \BibitemShut
  {NoStop}%
\bibitem [{\citenamefont {Karananas}(2015)}]{Karananas:2014pxa}%
  \BibitemOpen
  \bibfield  {author} {\bibinfo {author} {\bibfnamefont {G.~K.}\ \bibnamefont
  {Karananas}},\ }\bibfield  {title} {\bibinfo {title} {{The particle spectrum
  of parity-violating Poincar\'e gravitational theory}},\ }\href
  {https://doi.org/10.1088/0264-9381/32/5/055012} {\bibfield  {journal}
  {\bibinfo  {journal} {Class. Quant. Grav.}\ }\textbf {\bibinfo {volume}
  {32}},\ \bibinfo {pages} {055012} (\bibinfo {year} {2015})},\ \Eprint
  {https://arxiv.org/abs/1411.5613} {arXiv:1411.5613 [gr-qc]} \BibitemShut
  {NoStop}%
\bibitem [{\citenamefont {Blagojevi\'c}\ and\ \citenamefont
  {Cvetkovi\'c}(2018)}]{Blagojevic:2018dpz}%
  \BibitemOpen
  \bibfield  {author} {\bibinfo {author} {\bibfnamefont {M.}~\bibnamefont
  {Blagojevi\'c}}\ and\ \bibinfo {author} {\bibfnamefont {B.}~\bibnamefont
  {Cvetkovi\'c}},\ }\bibfield  {title} {\bibinfo {title} {{General Poincar\'e
  gauge theory: Hamiltonian structure and particle spectrum}},\ }\href
  {https://doi.org/10.1103/PhysRevD.98.024014} {\bibfield  {journal} {\bibinfo
  {journal} {Phys. Rev. D}\ }\textbf {\bibinfo {volume} {98}},\ \bibinfo
  {pages} {024014} (\bibinfo {year} {2018})},\ \Eprint
  {https://arxiv.org/abs/1804.05556} {arXiv:1804.05556 [gr-qc]} \BibitemShut
  {NoStop}%
\bibitem [{\citenamefont {Sezgin}(1981)}]{Sezgin:1981xs}%
  \BibitemOpen
  \bibfield  {author} {\bibinfo {author} {\bibfnamefont {E.}~\bibnamefont
  {Sezgin}},\ }\bibfield  {title} {\bibinfo {title} {{Class of Ghost Free
  Gravity Lagrangians With Massive or Massless Propagating Torsion}},\ }\href
  {https://doi.org/10.1103/PhysRevD.24.1677} {\bibfield  {journal} {\bibinfo
  {journal} {Phys. Rev. D}\ }\textbf {\bibinfo {volume} {24}},\ \bibinfo
  {pages} {1677} (\bibinfo {year} {1981})}\BibitemShut {NoStop}%
\bibitem [{\citenamefont {Blagojevic}\ and\ \citenamefont
  {Nikolic}(1983)}]{Blagojevic:1983zz}%
  \BibitemOpen
  \bibfield  {author} {\bibinfo {author} {\bibfnamefont {M.}~\bibnamefont
  {Blagojevic}}\ and\ \bibinfo {author} {\bibfnamefont {I.~A.}\ \bibnamefont
  {Nikolic}},\ }\bibfield  {title} {\bibinfo {title} {{Hamiltonian dynamics of
  Poincare gauge theory: General structure in the time gauge}},\ }\href
  {https://doi.org/10.1103/PhysRevD.28.2455} {\bibfield  {journal} {\bibinfo
  {journal} {Phys. Rev. D}\ }\textbf {\bibinfo {volume} {28}},\ \bibinfo
  {pages} {2455} (\bibinfo {year} {1983})}\BibitemShut {NoStop}%
\bibitem [{\citenamefont {Blagojevic}\ and\ \citenamefont
  {Vasilic}(1987)}]{Blagojevic:1986dm}%
  \BibitemOpen
  \bibfield  {author} {\bibinfo {author} {\bibfnamefont {M.}~\bibnamefont
  {Blagojevic}}\ and\ \bibinfo {author} {\bibfnamefont {M.}~\bibnamefont
  {Vasilic}},\ }\bibfield  {title} {\bibinfo {title} {{EXTRA GAUGE SYMMETRIES
  IN A WEAK FIELD APPROXIMATION OF AN R + T**2 + R**2 THEORY OF GRAVITY}},\
  }\href {https://doi.org/10.1103/PhysRevD.35.3748} {\bibfield  {journal}
  {\bibinfo  {journal} {Phys. Rev. D}\ }\textbf {\bibinfo {volume} {35}},\
  \bibinfo {pages} {3748} (\bibinfo {year} {1987})}\BibitemShut {NoStop}%
\bibitem [{\citenamefont {Kuhfuss}\ and\ \citenamefont
  {Nitsch}(1986)}]{Kuhfuss:1986rb}%
  \BibitemOpen
  \bibfield  {author} {\bibinfo {author} {\bibfnamefont {R.}~\bibnamefont
  {Kuhfuss}}\ and\ \bibinfo {author} {\bibfnamefont {J.}~\bibnamefont
  {Nitsch}},\ }\bibfield  {title} {\bibinfo {title} {{Propagating Modes in
  Gauge Field Theories of Gravity}},\ }\href
  {https://doi.org/10.1007/BF00763447} {\bibfield  {journal} {\bibinfo
  {journal} {Gen. Rel. Grav.}\ }\textbf {\bibinfo {volume} {18}},\ \bibinfo
  {pages} {1207} (\bibinfo {year} {1986})}\BibitemShut {NoStop}%
\bibitem [{\citenamefont {Yo}\ and\ \citenamefont {Nester}(2007)}]{Yo:2006qs}%
  \BibitemOpen
  \bibfield  {author} {\bibinfo {author} {\bibfnamefont {H.-J.}\ \bibnamefont
  {Yo}}\ and\ \bibinfo {author} {\bibfnamefont {J.~M.}\ \bibnamefont
  {Nester}},\ }\bibfield  {title} {\bibinfo {title} {{Dynamic Scalar Torsion
  and an Oscillating Universe}},\ }\href
  {https://doi.org/10.1142/S0217732307025303} {\bibfield  {journal} {\bibinfo
  {journal} {Mod. Phys. Lett. A}\ }\textbf {\bibinfo {volume} {22}},\ \bibinfo
  {pages} {2057} (\bibinfo {year} {2007})},\ \Eprint
  {https://arxiv.org/abs/astro-ph/0612738} {arXiv:astro-ph/0612738}
  \BibitemShut {NoStop}%
\bibitem [{\citenamefont {Shie}\ \emph {et~al.}(2008)\citenamefont {Shie},
  \citenamefont {Nester},\ and\ \citenamefont {Yo}}]{Shie:2008ms}%
  \BibitemOpen
  \bibfield  {author} {\bibinfo {author} {\bibfnamefont {K.-F.}\ \bibnamefont
  {Shie}}, \bibinfo {author} {\bibfnamefont {J.~M.}\ \bibnamefont {Nester}},\
  and\ \bibinfo {author} {\bibfnamefont {H.-J.}\ \bibnamefont {Yo}},\
  }\bibfield  {title} {\bibinfo {title} {{Torsion Cosmology and the
  Accelerating Universe}},\ }\href {https://doi.org/10.1103/PhysRevD.78.023522}
  {\bibfield  {journal} {\bibinfo  {journal} {Phys. Rev. D}\ }\textbf {\bibinfo
  {volume} {78}},\ \bibinfo {pages} {023522} (\bibinfo {year} {2008})},\
  \Eprint {https://arxiv.org/abs/0805.3834} {arXiv:0805.3834 [gr-qc]}
  \BibitemShut {NoStop}%
\bibitem [{\citenamefont {Nair}\ \emph {et~al.}(2009)\citenamefont {Nair},
  \citenamefont {Randjbar-Daemi},\ and\ \citenamefont {Rubakov}}]{Nair:2008yh}%
  \BibitemOpen
  \bibfield  {author} {\bibinfo {author} {\bibfnamefont {V.~P.}\ \bibnamefont
  {Nair}}, \bibinfo {author} {\bibfnamefont {S.}~\bibnamefont
  {Randjbar-Daemi}},\ and\ \bibinfo {author} {\bibfnamefont {V.}~\bibnamefont
  {Rubakov}},\ }\bibfield  {title} {\bibinfo {title} {{Massive Spin-2 fields of
  Geometric Origin in Curved Spacetimes}},\ }\href
  {https://doi.org/10.1103/PhysRevD.80.104031} {\bibfield  {journal} {\bibinfo
  {journal} {Phys. Rev. D}\ }\textbf {\bibinfo {volume} {80}},\ \bibinfo
  {pages} {104031} (\bibinfo {year} {2009})},\ \Eprint
  {https://arxiv.org/abs/0811.3781} {arXiv:0811.3781 [hep-th]} \BibitemShut
  {NoStop}%
\bibitem [{\citenamefont {Nikiforova}\ \emph {et~al.}(2009)\citenamefont
  {Nikiforova}, \citenamefont {Randjbar-Daemi},\ and\ \citenamefont
  {Rubakov}}]{Nikiforova:2009qr}%
  \BibitemOpen
  \bibfield  {author} {\bibinfo {author} {\bibfnamefont {V.}~\bibnamefont
  {Nikiforova}}, \bibinfo {author} {\bibfnamefont {S.}~\bibnamefont
  {Randjbar-Daemi}},\ and\ \bibinfo {author} {\bibfnamefont {V.}~\bibnamefont
  {Rubakov}},\ }\bibfield  {title} {\bibinfo {title} {{Infrared Modified
  Gravity with Dynamical Torsion}},\ }\href
  {https://doi.org/10.1103/PhysRevD.80.124050} {\bibfield  {journal} {\bibinfo
  {journal} {Phys. Rev. D}\ }\textbf {\bibinfo {volume} {80}},\ \bibinfo
  {pages} {124050} (\bibinfo {year} {2009})},\ \Eprint
  {https://arxiv.org/abs/0905.3732} {arXiv:0905.3732 [hep-th]} \BibitemShut
  {NoStop}%
\bibitem [{\citenamefont {Chen}\ \emph {et~al.}(2009)\citenamefont {Chen},
  \citenamefont {Ho}, \citenamefont {Nester}, \citenamefont {Wang},\ and\
  \citenamefont {Yo}}]{Chen:2009at}%
  \BibitemOpen
  \bibfield  {author} {\bibinfo {author} {\bibfnamefont {H.}~\bibnamefont
  {Chen}}, \bibinfo {author} {\bibfnamefont {F.-H.}\ \bibnamefont {Ho}},
  \bibinfo {author} {\bibfnamefont {J.~M.}\ \bibnamefont {Nester}}, \bibinfo
  {author} {\bibfnamefont {C.-H.}\ \bibnamefont {Wang}},\ and\ \bibinfo
  {author} {\bibfnamefont {H.-J.}\ \bibnamefont {Yo}},\ }\bibfield  {title}
  {\bibinfo {title} {{Cosmological dynamics with propagating Lorentz connection
  modes of spin zero}},\ }\href {https://doi.org/10.1088/1475-7516/2009/10/027}
  {\bibfield  {journal} {\bibinfo  {journal} {JCAP}\ }\textbf {\bibinfo
  {volume} {10}},\ \bibinfo {pages} {027}},\ \Eprint
  {https://arxiv.org/abs/0908.3323} {arXiv:0908.3323 [gr-qc]} \BibitemShut
  {NoStop}%
\bibitem [{\citenamefont {Ni}(2010)}]{Ni:2009fg}%
  \BibitemOpen
  \bibfield  {author} {\bibinfo {author} {\bibfnamefont {W.-T.}\ \bibnamefont
  {Ni}},\ }\bibfield  {title} {\bibinfo {title} {{Searches for the role of spin
  and polarization in gravity}},\ }\href
  {https://doi.org/10.1088/0034-4885/73/5/056901} {\bibfield  {journal}
  {\bibinfo  {journal} {Rept. Prog. Phys.}\ }\textbf {\bibinfo {volume} {73}},\
  \bibinfo {pages} {056901} (\bibinfo {year} {2010})},\ \Eprint
  {https://arxiv.org/abs/0912.5057} {arXiv:0912.5057 [gr-qc]} \BibitemShut
  {NoStop}%
\bibitem [{\citenamefont {Baekler}\ \emph {et~al.}(2011)\citenamefont
  {Baekler}, \citenamefont {Hehl},\ and\ \citenamefont
  {Nester}}]{Baekler:2010fr}%
  \BibitemOpen
  \bibfield  {author} {\bibinfo {author} {\bibfnamefont {P.}~\bibnamefont
  {Baekler}}, \bibinfo {author} {\bibfnamefont {F.~W.}\ \bibnamefont {Hehl}},\
  and\ \bibinfo {author} {\bibfnamefont {J.~M.}\ \bibnamefont {Nester}},\
  }\bibfield  {title} {\bibinfo {title} {{Poincare gauge theory of gravity:
  Friedman cosmology with even and odd parity modes. Analytic part}},\ }\href
  {https://doi.org/10.1103/PhysRevD.83.024001} {\bibfield  {journal} {\bibinfo
  {journal} {Phys. Rev. D}\ }\textbf {\bibinfo {volume} {83}},\ \bibinfo
  {pages} {024001} (\bibinfo {year} {2011})},\ \Eprint
  {https://arxiv.org/abs/1009.5112} {arXiv:1009.5112 [gr-qc]} \BibitemShut
  {NoStop}%
\bibitem [{\citenamefont {Ho}\ and\ \citenamefont {Nester}(2011)}]{Ho:2011qn}%
  \BibitemOpen
  \bibfield  {author} {\bibinfo {author} {\bibfnamefont {F.-H.}\ \bibnamefont
  {Ho}}\ and\ \bibinfo {author} {\bibfnamefont {J.~M.}\ \bibnamefont
  {Nester}},\ }\bibfield  {title} {\bibinfo {title} {{Poincar\'e gauge theory
  with even and odd parity dynamic connection modes: isotropic Bianchi
  cosmological models}},\ }\href
  {https://doi.org/10.1088/1742-6596/330/1/012005} {\bibfield  {journal}
  {\bibinfo  {journal} {J. Phys. Conf. Ser.}\ }\textbf {\bibinfo {volume}
  {330}},\ \bibinfo {pages} {012005} (\bibinfo {year} {2011})},\ \Eprint
  {https://arxiv.org/abs/1105.5001} {arXiv:1105.5001 [gr-qc]} \BibitemShut
  {NoStop}%
\bibitem [{\citenamefont {Ho}\ and\ \citenamefont {Nester}(2012)}]{Ho:2011xf}%
  \BibitemOpen
  \bibfield  {author} {\bibinfo {author} {\bibfnamefont {F.-H.}\ \bibnamefont
  {Ho}}\ and\ \bibinfo {author} {\bibfnamefont {J.~M.}\ \bibnamefont
  {Nester}},\ }\bibfield  {title} {\bibinfo {title} {{Poincar\'e Gauge Theory
  With Coupled Even And Odd Parity Dynamic Spin-0 Modes: Dynamic Equations For
  Isotropic Bianchi Cosmologies}},\ }\href
  {https://doi.org/10.1002/andp.201100101} {\bibfield  {journal} {\bibinfo
  {journal} {Annalen Phys.}\ }\textbf {\bibinfo {volume} {524}},\ \bibinfo
  {pages} {97} (\bibinfo {year} {2012})},\ \Eprint
  {https://arxiv.org/abs/1106.0711} {arXiv:1106.0711 [gr-qc]} \BibitemShut
  {NoStop}%
\bibitem [{\citenamefont {Ong}\ \emph {et~al.}(2013)\citenamefont {Ong},
  \citenamefont {Izumi}, \citenamefont {Nester},\ and\ \citenamefont
  {Chen}}]{Ong:2013qja}%
  \BibitemOpen
  \bibfield  {author} {\bibinfo {author} {\bibfnamefont {Y.~C.}\ \bibnamefont
  {Ong}}, \bibinfo {author} {\bibfnamefont {K.}~\bibnamefont {Izumi}}, \bibinfo
  {author} {\bibfnamefont {J.~M.}\ \bibnamefont {Nester}},\ and\ \bibinfo
  {author} {\bibfnamefont {P.}~\bibnamefont {Chen}},\ }\bibfield  {title}
  {\bibinfo {title} {{Problems with Propagation and Time Evolution in f(T)
  Gravity}},\ }\href {https://doi.org/10.1103/PhysRevD.88.024019} {\bibfield
  {journal} {\bibinfo  {journal} {Phys. Rev. D}\ }\textbf {\bibinfo {volume}
  {88}},\ \bibinfo {pages} {024019} (\bibinfo {year} {2013})},\ \Eprint
  {https://arxiv.org/abs/1303.0993} {arXiv:1303.0993 [gr-qc]} \BibitemShut
  {NoStop}%
\bibitem [{\citenamefont {Puetzfeld}\ and\ \citenamefont
  {Obukhov}(2014)}]{Puetzfeld:2014sja}%
  \BibitemOpen
  \bibfield  {author} {\bibinfo {author} {\bibfnamefont {D.}~\bibnamefont
  {Puetzfeld}}\ and\ \bibinfo {author} {\bibfnamefont {Y.~N.}\ \bibnamefont
  {Obukhov}},\ }\bibfield  {title} {\bibinfo {title} {{Prospects of detecting
  spacetime torsion}},\ }\href {https://doi.org/10.1142/S0218271814420048}
  {\bibfield  {journal} {\bibinfo  {journal} {Int. J. Mod. Phys. D}\ }\textbf
  {\bibinfo {volume} {23}},\ \bibinfo {pages} {1442004} (\bibinfo {year}
  {2014})},\ \Eprint {https://arxiv.org/abs/1405.4137} {arXiv:1405.4137
  [gr-qc]} \BibitemShut {NoStop}%
\bibitem [{\citenamefont {Ni}(2016)}]{Ni:2015poa}%
  \BibitemOpen
  \bibfield  {author} {\bibinfo {author} {\bibfnamefont {W.-T.}\ \bibnamefont
  {Ni}},\ }\bibfield  {title} {\bibinfo {title} {{Searches for the role of spin
  and polarization in gravity: a five-year update}},\ }\href
  {https://doi.org/10.1142/S2010194516600107} {\bibfield  {journal} {\bibinfo
  {journal} {Int. J. Mod. Phys. Conf. Ser.}\ }\textbf {\bibinfo {volume}
  {40}},\ \bibinfo {pages} {1660010} (\bibinfo {year} {2016})},\ \Eprint
  {https://arxiv.org/abs/1501.07696} {arXiv:1501.07696 [hep-ph]} \BibitemShut
  {NoStop}%
\bibitem [{\citenamefont {Ho}\ \emph {et~al.}(2015)\citenamefont {Ho},
  \citenamefont {Chen}, \citenamefont {Nester},\ and\ \citenamefont
  {Yo}}]{Ho:2015ulu}%
  \BibitemOpen
  \bibfield  {author} {\bibinfo {author} {\bibfnamefont {F.-H.}\ \bibnamefont
  {Ho}}, \bibinfo {author} {\bibfnamefont {H.}~\bibnamefont {Chen}}, \bibinfo
  {author} {\bibfnamefont {J.~M.}\ \bibnamefont {Nester}},\ and\ \bibinfo
  {author} {\bibfnamefont {H.-J.}\ \bibnamefont {Yo}},\ }\bibfield  {title}
  {\bibinfo {title} {{General Poincar\'e Gauge Theory Cosmology}},\ }\href
  {https://doi.org/10.6122/CJP.20151014} {\bibfield  {journal} {\bibinfo
  {journal} {Chin. J. Phys.}\ }\textbf {\bibinfo {volume} {53}},\ \bibinfo
  {pages} {110109} (\bibinfo {year} {2015})},\ \Eprint
  {https://arxiv.org/abs/1512.01202} {arXiv:1512.01202 [gr-qc]} \BibitemShut
  {NoStop}%
\bibitem [{\citenamefont {Karananas}(2016)}]{Karananas:2016ltn}%
  \BibitemOpen
  \bibfield  {author} {\bibinfo {author} {\bibfnamefont {G.~K.}\ \bibnamefont
  {Karananas}},\ }\emph {\bibinfo {title} {{Poincar\'e, Scale and Conformal
  Symmetries Gauge Perspective and Cosmological Ramifications}}},\ \href
  {https://doi.org/10.5075/epfl-thesis-7173} {Ph.D. thesis},\ \bibinfo
  {school} {Ecole Polytechnique, Lausanne} (\bibinfo {year} {2016}),\ \Eprint
  {https://arxiv.org/abs/1608.08451} {arXiv:1608.08451 [hep-th]} \BibitemShut
  {NoStop}%
\bibitem [{\citenamefont {Obukhov}(2017)}]{Obukhov:2017pxa}%
  \BibitemOpen
  \bibfield  {author} {\bibinfo {author} {\bibfnamefont {Y.~N.}\ \bibnamefont
  {Obukhov}},\ }\bibfield  {title} {\bibinfo {title} {{Gravitational waves in
  Poincar\'e gauge gravity theory}},\ }\href
  {https://doi.org/10.1103/PhysRevD.95.084028} {\bibfield  {journal} {\bibinfo
  {journal} {Phys. Rev. D}\ }\textbf {\bibinfo {volume} {95}},\ \bibinfo
  {pages} {084028} (\bibinfo {year} {2017})},\ \Eprint
  {https://arxiv.org/abs/1702.05185} {arXiv:1702.05185 [gr-qc]} \BibitemShut
  {NoStop}%
\bibitem [{\citenamefont {Blagojevi\'c}\ \emph {et~al.}(2017)\citenamefont
  {Blagojevi\'c}, \citenamefont {Cvetkovi\'c},\ and\ \citenamefont
  {Obukhov}}]{Blagojevic:2017ssv}%
  \BibitemOpen
  \bibfield  {author} {\bibinfo {author} {\bibfnamefont {M.}~\bibnamefont
  {Blagojevi\'c}}, \bibinfo {author} {\bibfnamefont {B.}~\bibnamefont
  {Cvetkovi\'c}},\ and\ \bibinfo {author} {\bibfnamefont {Y.~N.}\ \bibnamefont
  {Obukhov}},\ }\bibfield  {title} {\bibinfo {title} {{Generalized plane waves
  in Poincar\'e gauge theory of gravity}},\ }\href
  {https://doi.org/10.1103/PhysRevD.96.064031} {\bibfield  {journal} {\bibinfo
  {journal} {Phys. Rev. D}\ }\textbf {\bibinfo {volume} {96}},\ \bibinfo
  {pages} {064031} (\bibinfo {year} {2017})},\ \Eprint
  {https://arxiv.org/abs/1708.08766} {arXiv:1708.08766 [gr-qc]} \BibitemShut
  {NoStop}%
\bibitem [{\citenamefont {Tseng}(2018)}]{Tseng:2018feo}%
  \BibitemOpen
  \bibfield  {author} {\bibinfo {author} {\bibfnamefont {H.-H.}\ \bibnamefont
  {Tseng}},\ }\emph {\bibinfo {title} {{Gravitational Theories with
  Torsion}}},\ \href@noop {} {Ph.D. thesis},\ \bibinfo  {school} {Taiwan, Natl.
  Tsing Hua U.} (\bibinfo {year} {2018}),\ \Eprint
  {https://arxiv.org/abs/1812.00314} {arXiv:1812.00314 [gr-qc]} \BibitemShut
  {NoStop}%
\bibitem [{\citenamefont {Lin}\ \emph {et~al.}(2019)\citenamefont {Lin},
  \citenamefont {Hobson},\ and\ \citenamefont {Lasenby}}]{Lin:2018awc}%
  \BibitemOpen
  \bibfield  {author} {\bibinfo {author} {\bibfnamefont {Y.-C.}\ \bibnamefont
  {Lin}}, \bibinfo {author} {\bibfnamefont {M.~P.}\ \bibnamefont {Hobson}},\
  and\ \bibinfo {author} {\bibfnamefont {A.~N.}\ \bibnamefont {Lasenby}},\
  }\bibfield  {title} {\bibinfo {title} {{Ghost and tachyon free Poincar\'e
  gauge theories: A systematic approach}},\ }\href
  {https://doi.org/10.1103/PhysRevD.99.064001} {\bibfield  {journal} {\bibinfo
  {journal} {Phys. Rev. D}\ }\textbf {\bibinfo {volume} {99}},\ \bibinfo
  {pages} {064001} (\bibinfo {year} {2019})},\ \Eprint
  {https://arxiv.org/abs/1812.02675} {arXiv:1812.02675 [gr-qc]} \BibitemShut
  {NoStop}%
\bibitem [{\citenamefont {Beltr\'an~Jim\'enez}\ and\ \citenamefont
  {Delhom}(2019)}]{BeltranJimenez:2019acz}%
  \BibitemOpen
  \bibfield  {author} {\bibinfo {author} {\bibfnamefont {J.}~\bibnamefont
  {Beltr\'an~Jim\'enez}}\ and\ \bibinfo {author} {\bibfnamefont
  {A.}~\bibnamefont {Delhom}},\ }\bibfield  {title} {\bibinfo {title} {{Ghosts
  in metric-affine higher order curvature gravity}},\ }\href
  {https://doi.org/10.1140/epjc/s10052-019-7149-x} {\bibfield  {journal}
  {\bibinfo  {journal} {Eur. Phys. J. C}\ }\textbf {\bibinfo {volume} {79}},\
  \bibinfo {pages} {656} (\bibinfo {year} {2019})},\ \Eprint
  {https://arxiv.org/abs/1901.08988} {arXiv:1901.08988 [gr-qc]} \BibitemShut
  {NoStop}%
\bibitem [{\citenamefont {Zhang}\ and\ \citenamefont
  {Xu}(2019)}]{Zhang:2019mhd}%
  \BibitemOpen
  \bibfield  {author} {\bibinfo {author} {\bibfnamefont {H.}~\bibnamefont
  {Zhang}}\ and\ \bibinfo {author} {\bibfnamefont {L.}~\bibnamefont {Xu}},\
  }\bibfield  {title} {\bibinfo {title} {{Late-time acceleration and inflation
  in a Poincar\'e gauge cosmological model}},\ }\href
  {https://doi.org/10.1088/1475-7516/2019/09/050} {\bibfield  {journal}
  {\bibinfo  {journal} {JCAP}\ }\textbf {\bibinfo {volume} {09}},\ \bibinfo
  {pages} {050}},\ \Eprint {https://arxiv.org/abs/1904.03545} {arXiv:1904.03545
  [gr-qc]} \BibitemShut {NoStop}%
\bibitem [{\citenamefont {Aoki}\ and\ \citenamefont
  {Shimada}(2019)}]{Aoki:2019rvi}%
  \BibitemOpen
  \bibfield  {author} {\bibinfo {author} {\bibfnamefont {K.}~\bibnamefont
  {Aoki}}\ and\ \bibinfo {author} {\bibfnamefont {K.}~\bibnamefont {Shimada}},\
  }\bibfield  {title} {\bibinfo {title} {{Scalar-metric-affine theories: Can we
  get ghost-free theories from symmetry?}},\ }\href
  {https://doi.org/10.1103/PhysRevD.100.044037} {\bibfield  {journal} {\bibinfo
   {journal} {Phys. Rev. D}\ }\textbf {\bibinfo {volume} {100}},\ \bibinfo
  {pages} {044037} (\bibinfo {year} {2019})},\ \Eprint
  {https://arxiv.org/abs/1904.10175} {arXiv:1904.10175 [hep-th]} \BibitemShut
  {NoStop}%
\bibitem [{\citenamefont {Zhang}\ and\ \citenamefont
  {Xu}(2020)}]{Zhang:2019xek}%
  \BibitemOpen
  \bibfield  {author} {\bibinfo {author} {\bibfnamefont {H.}~\bibnamefont
  {Zhang}}\ and\ \bibinfo {author} {\bibfnamefont {L.}~\bibnamefont {Xu}},\
  }\bibfield  {title} {\bibinfo {title} {{Inflation in the parity-conserving
  Poincar\'e gauge cosmology}},\ }\href
  {https://doi.org/10.1088/1475-7516/2020/10/003} {\bibfield  {journal}
  {\bibinfo  {journal} {JCAP}\ }\textbf {\bibinfo {volume} {10}},\ \bibinfo
  {pages} {003}},\ \Eprint {https://arxiv.org/abs/1906.04340} {arXiv:1906.04340
  [gr-qc]} \BibitemShut {NoStop}%
\bibitem [{\citenamefont {Beltr\'an~Jim\'enez}\ and\ \citenamefont
  {Maldonado~Torralba}(2020{\natexlab{b}})}]{Jimenez:2019qjc}%
  \BibitemOpen
  \bibfield  {author} {\bibinfo {author} {\bibfnamefont {J.}~\bibnamefont
  {Beltr\'an~Jim\'enez}}\ and\ \bibinfo {author} {\bibfnamefont {F.~J.}\
  \bibnamefont {Maldonado~Torralba}},\ }\bibfield  {title} {\bibinfo {title}
  {{Revisiting the stability of quadratic Poincar\'e gauge gravity}},\ }\href
  {https://doi.org/10.1140/epjc/s10052-020-8163-8} {\bibfield  {journal}
  {\bibinfo  {journal} {Eur. Phys. J. C}\ }\textbf {\bibinfo {volume} {80}},\
  \bibinfo {pages} {611} (\bibinfo {year} {2020}{\natexlab{b}})},\ \Eprint
  {https://arxiv.org/abs/1910.07506} {arXiv:1910.07506 [gr-qc]} \BibitemShut
  {NoStop}%
\bibitem [{\citenamefont {Lin}\ \emph {et~al.}(2020)\citenamefont {Lin},
  \citenamefont {Hobson},\ and\ \citenamefont {Lasenby}}]{Lin:2019ugq}%
  \BibitemOpen
  \bibfield  {author} {\bibinfo {author} {\bibfnamefont {Y.-C.}\ \bibnamefont
  {Lin}}, \bibinfo {author} {\bibfnamefont {M.~P.}\ \bibnamefont {Hobson}},\
  and\ \bibinfo {author} {\bibfnamefont {A.~N.}\ \bibnamefont {Lasenby}},\
  }\bibfield  {title} {\bibinfo {title} {{Power-counting renormalizable,
  ghost-and-tachyon-free Poincar\'e gauge theories}},\ }\href
  {https://doi.org/10.1103/PhysRevD.101.064038} {\bibfield  {journal} {\bibinfo
   {journal} {Phys. Rev. D}\ }\textbf {\bibinfo {volume} {101}},\ \bibinfo
  {pages} {064038} (\bibinfo {year} {2020})},\ \Eprint
  {https://arxiv.org/abs/1910.14197} {arXiv:1910.14197 [gr-qc]} \BibitemShut
  {NoStop}%
\bibitem [{\citenamefont {Percacci}\ and\ \citenamefont
  {Sezgin}(2020{\natexlab{a}})}]{Percacci:2019hxn}%
  \BibitemOpen
  \bibfield  {author} {\bibinfo {author} {\bibfnamefont {R.}~\bibnamefont
  {Percacci}}\ and\ \bibinfo {author} {\bibfnamefont {E.}~\bibnamefont
  {Sezgin}},\ }\bibfield  {title} {\bibinfo {title} {{New class of ghost- and
  tachyon-free metric affine gravities}},\ }\href
  {https://doi.org/10.1103/PhysRevD.101.084040} {\bibfield  {journal} {\bibinfo
   {journal} {Phys. Rev. D}\ }\textbf {\bibinfo {volume} {101}},\ \bibinfo
  {pages} {084040} (\bibinfo {year} {2020}{\natexlab{a}})},\ \Eprint
  {https://arxiv.org/abs/1912.01023} {arXiv:1912.01023 [hep-th]} \BibitemShut
  {NoStop}%
\bibitem [{\citenamefont {Barker}\ \emph {et~al.}(2020)\citenamefont {Barker},
  \citenamefont {Lasenby}, \citenamefont {Hobson},\ and\ \citenamefont
  {Handley}}]{Barker:2020gcp}%
  \BibitemOpen
  \bibfield  {author} {\bibinfo {author} {\bibfnamefont {W.~E.~V.}\
  \bibnamefont {Barker}}, \bibinfo {author} {\bibfnamefont {A.~N.}\
  \bibnamefont {Lasenby}}, \bibinfo {author} {\bibfnamefont {M.~P.}\
  \bibnamefont {Hobson}},\ and\ \bibinfo {author} {\bibfnamefont {W.~J.}\
  \bibnamefont {Handley}},\ }\bibfield  {title} {\bibinfo {title} {{Systematic
  study of background cosmology in unitary Poincar\'e gauge theories with
  application to emergent dark radiation and $H_0$ tension}},\ }\href
  {https://doi.org/10.1103/PhysRevD.102.024048} {\bibfield  {journal} {\bibinfo
   {journal} {Phys. Rev. D}\ }\textbf {\bibinfo {volume} {102}},\ \bibinfo
  {pages} {024048} (\bibinfo {year} {2020})},\ \Eprint
  {https://arxiv.org/abs/2003.02690} {arXiv:2003.02690 [gr-qc]} \BibitemShut
  {NoStop}%
\bibitem [{\citenamefont {Beltr\'an~Jim\'enez}\ and\ \citenamefont
  {Delhom}(2020)}]{BeltranJimenez:2020sqf}%
  \BibitemOpen
  \bibfield  {author} {\bibinfo {author} {\bibfnamefont {J.}~\bibnamefont
  {Beltr\'an~Jim\'enez}}\ and\ \bibinfo {author} {\bibfnamefont
  {A.}~\bibnamefont {Delhom}},\ }\bibfield  {title} {\bibinfo {title}
  {{Instabilities in metric-affine theories of gravity with higher order
  curvature terms}},\ }\href {https://doi.org/10.1140/epjc/s10052-020-8143-z}
  {\bibfield  {journal} {\bibinfo  {journal} {Eur. Phys. J. C}\ }\textbf
  {\bibinfo {volume} {80}},\ \bibinfo {pages} {585} (\bibinfo {year} {2020})},\
  \Eprint {https://arxiv.org/abs/2004.11357} {arXiv:2004.11357 [gr-qc]}
  \BibitemShut {NoStop}%
\bibitem [{\citenamefont
  {Maldonado~Torralba}(2020)}]{MaldonadoTorralba:2020mbh}%
  \BibitemOpen
  \bibfield  {author} {\bibinfo {author} {\bibfnamefont {F.~J.}\ \bibnamefont
  {Maldonado~Torralba}},\ }\emph {\bibinfo {title} {{New effective theories of
  gravitation and their phenomenological consequences}}},\ \href
  {https://doi.org/10.33612/diss.143961423} {Ph.D. thesis},\ \bibinfo  {school}
  {Cape Town U., Dept. Math.} (\bibinfo {year} {2020}),\ \Eprint
  {https://arxiv.org/abs/2101.11523} {arXiv:2101.11523 [gr-qc]} \BibitemShut
  {NoStop}%
\bibitem [{\citenamefont {Barker}\ \emph {et~al.}(2021)\citenamefont {Barker},
  \citenamefont {Lasenby}, \citenamefont {Hobson},\ and\ \citenamefont
  {Handley}}]{Barker:2021oez}%
  \BibitemOpen
  \bibfield  {author} {\bibinfo {author} {\bibfnamefont {W.~E.~V.}\
  \bibnamefont {Barker}}, \bibinfo {author} {\bibfnamefont {A.~N.}\
  \bibnamefont {Lasenby}}, \bibinfo {author} {\bibfnamefont {M.~P.}\
  \bibnamefont {Hobson}},\ and\ \bibinfo {author} {\bibfnamefont {W.~J.}\
  \bibnamefont {Handley}},\ }\bibfield  {title} {\bibinfo {title} {{Nonlinear
  Hamiltonian analysis of new quadratic torsion theories: Cases with
  curvature-free constraints}},\ }\href
  {https://doi.org/10.1103/PhysRevD.104.084036} {\bibfield  {journal} {\bibinfo
   {journal} {Phys. Rev. D}\ }\textbf {\bibinfo {volume} {104}},\ \bibinfo
  {pages} {084036} (\bibinfo {year} {2021})},\ \Eprint
  {https://arxiv.org/abs/2101.02645} {arXiv:2101.02645 [gr-qc]} \BibitemShut
  {NoStop}%
\bibitem [{\citenamefont {Marzo}(2022{\natexlab{a}})}]{Marzo:2021esg}%
  \BibitemOpen
  \bibfield  {author} {\bibinfo {author} {\bibfnamefont {C.}~\bibnamefont
  {Marzo}},\ }\bibfield  {title} {\bibinfo {title} {{Ghost and tachyon free
  propagation up to spin 3 in Lorentz invariant field theories}},\ }\href
  {https://doi.org/10.1103/PhysRevD.105.065017} {\bibfield  {journal} {\bibinfo
   {journal} {Phys. Rev. D}\ }\textbf {\bibinfo {volume} {105}},\ \bibinfo
  {pages} {065017} (\bibinfo {year} {2022}{\natexlab{a}})},\ \Eprint
  {https://arxiv.org/abs/2108.11982} {arXiv:2108.11982 [hep-ph]} \BibitemShut
  {NoStop}%
\bibitem [{\citenamefont {Marzo}(2022{\natexlab{b}})}]{Marzo:2021iok}%
  \BibitemOpen
  \bibfield  {author} {\bibinfo {author} {\bibfnamefont {C.}~\bibnamefont
  {Marzo}},\ }\bibfield  {title} {\bibinfo {title} {{Radiatively stable ghost
  and tachyon freedom in metric affine gravity}},\ }\href
  {https://doi.org/10.1103/PhysRevD.106.024045} {\bibfield  {journal} {\bibinfo
   {journal} {Phys. Rev. D}\ }\textbf {\bibinfo {volume} {106}},\ \bibinfo
  {pages} {024045} (\bibinfo {year} {2022}{\natexlab{b}})},\ \Eprint
  {https://arxiv.org/abs/2110.14788} {arXiv:2110.14788 [hep-th]} \BibitemShut
  {NoStop}%
\bibitem [{\citenamefont {de~la Cruz~Dombriz}\ \emph
  {et~al.}(2022)\citenamefont {de~la Cruz~Dombriz}, \citenamefont
  {Maldonado~Torralba},\ and\ \citenamefont {Mota}}]{delaCruzDombriz:2021nrg}%
  \BibitemOpen
  \bibfield  {author} {\bibinfo {author} {\bibfnamefont {A.}~\bibnamefont
  {de~la Cruz~Dombriz}}, \bibinfo {author} {\bibfnamefont {F.~J.}\ \bibnamefont
  {Maldonado~Torralba}},\ and\ \bibinfo {author} {\bibfnamefont {D.~F.}\
  \bibnamefont {Mota}},\ }\bibfield  {title} {\bibinfo {title} {{Dark matter
  candidate from torsion}},\ }\href
  {https://doi.org/10.1016/j.physletb.2022.137488} {\bibfield  {journal}
  {\bibinfo  {journal} {Phys. Lett. B}\ }\textbf {\bibinfo {volume} {834}},\
  \bibinfo {pages} {137488} (\bibinfo {year} {2022})},\ \Eprint
  {https://arxiv.org/abs/2112.03957} {arXiv:2112.03957 [gr-qc]} \BibitemShut
  {NoStop}%
\bibitem [{\citenamefont {Baldazzi}\ \emph {et~al.}(2022)\citenamefont
  {Baldazzi}, \citenamefont {Melichev},\ and\ \citenamefont
  {Percacci}}]{Baldazzi:2021kaf}%
  \BibitemOpen
  \bibfield  {author} {\bibinfo {author} {\bibfnamefont {A.}~\bibnamefont
  {Baldazzi}}, \bibinfo {author} {\bibfnamefont {O.}~\bibnamefont {Melichev}},\
  and\ \bibinfo {author} {\bibfnamefont {R.}~\bibnamefont {Percacci}},\
  }\bibfield  {title} {\bibinfo {title} {{Metric-Affine Gravity as an effective
  field theory}},\ }\href {https://doi.org/10.1016/j.aop.2022.168757}
  {\bibfield  {journal} {\bibinfo  {journal} {Annals Phys.}\ }\textbf {\bibinfo
  {volume} {438}},\ \bibinfo {pages} {168757} (\bibinfo {year} {2022})},\
  \Eprint {https://arxiv.org/abs/2112.10193} {arXiv:2112.10193 [gr-qc]}
  \BibitemShut {NoStop}%
\bibitem [{\citenamefont {Annala}\ and\ \citenamefont
  {Rasanen}(2023)}]{Annala:2022gtl}%
  \BibitemOpen
  \bibfield  {author} {\bibinfo {author} {\bibfnamefont {J.}~\bibnamefont
  {Annala}}\ and\ \bibinfo {author} {\bibfnamefont {S.}~\bibnamefont
  {Rasanen}},\ }\bibfield  {title} {\bibinfo {title} {{Stability of
  non-degenerate Ricci-type Palatini theories}},\ }\href
  {https://doi.org/10.1088/1475-7516/2023/04/014} {\bibfield  {journal}
  {\bibinfo  {journal} {JCAP}\ }\textbf {\bibinfo {volume} {04}},\ \bibinfo
  {pages} {014}},\ \bibinfo {note} {[Erratum: JCAP 08, E02 (2023)]},\ \Eprint
  {https://arxiv.org/abs/2212.09820} {arXiv:2212.09820 [gr-qc]} \BibitemShut
  {NoStop}%
\bibitem [{\citenamefont {Mikura}\ \emph {et~al.}(2024)\citenamefont {Mikura},
  \citenamefont {Naso},\ and\ \citenamefont {Percacci}}]{Mikura:2023ruz}%
  \BibitemOpen
  \bibfield  {author} {\bibinfo {author} {\bibfnamefont {Y.}~\bibnamefont
  {Mikura}}, \bibinfo {author} {\bibfnamefont {V.}~\bibnamefont {Naso}},\ and\
  \bibinfo {author} {\bibfnamefont {R.}~\bibnamefont {Percacci}},\ }\bibfield
  {title} {\bibinfo {title} {{Some simple theories of gravity with propagating
  torsion}},\ }\href {https://doi.org/10.1103/PhysRevD.109.104071} {\bibfield
  {journal} {\bibinfo  {journal} {Phys. Rev. D}\ }\textbf {\bibinfo {volume}
  {109}},\ \bibinfo {pages} {104071} (\bibinfo {year} {2024})},\ \Eprint
  {https://arxiv.org/abs/2312.10249} {arXiv:2312.10249 [gr-qc]} \BibitemShut
  {NoStop}%
\bibitem [{\citenamefont {Mikura}\ and\ \citenamefont
  {Percacci}(2024)}]{Mikura:2024mji}%
  \BibitemOpen
  \bibfield  {author} {\bibinfo {author} {\bibfnamefont {Y.}~\bibnamefont
  {Mikura}}\ and\ \bibinfo {author} {\bibfnamefont {R.}~\bibnamefont
  {Percacci}},\ }\bibfield  {title} {\bibinfo {title} {{Some simple theories of
  gravity with propagating nonmetricity}},\ }\href@noop {} {\  (\bibinfo {year}
  {2024})},\ \Eprint {https://arxiv.org/abs/2401.10097} {arXiv:2401.10097
  [gr-qc]} \BibitemShut {NoStop}%
\bibitem [{\citenamefont {Barker}\ and\ \citenamefont
  {Marzo}(2024)}]{Barker:2024ydb}%
  \BibitemOpen
  \bibfield  {author} {\bibinfo {author} {\bibfnamefont {W.}~\bibnamefont
  {Barker}}\ and\ \bibinfo {author} {\bibfnamefont {C.}~\bibnamefont {Marzo}},\
  }\bibfield  {title} {\bibinfo {title} {{Particle spectra of general
  Ricci-type Palatini or metric-affine theories}},\ }\href
  {https://doi.org/10.1103/PhysRevD.109.104017} {\bibfield  {journal} {\bibinfo
   {journal} {Phys. Rev. D}\ }\textbf {\bibinfo {volume} {109}},\ \bibinfo
  {pages} {104017} (\bibinfo {year} {2024})},\ \Eprint
  {https://arxiv.org/abs/2402.07641} {arXiv:2402.07641 [hep-th]} \BibitemShut
  {NoStop}%
\bibitem [{\citenamefont {Karananas}\ \emph {et~al.}(2024)\citenamefont
  {Karananas}, \citenamefont {Shaposhnikov},\ and\ \citenamefont
  {Zell}}]{Karananas:2024xja}%
  \BibitemOpen
  \bibfield  {author} {\bibinfo {author} {\bibfnamefont {G.~K.}\ \bibnamefont
  {Karananas}}, \bibinfo {author} {\bibfnamefont {M.}~\bibnamefont
  {Shaposhnikov}},\ and\ \bibinfo {author} {\bibfnamefont {S.}~\bibnamefont
  {Zell}},\ }\bibfield  {title} {\bibinfo {title} {{Weyl-invariant
  Einstein-Cartan gravity: unifying the strong CP and hierarchy puzzles}},\
  }\href {https://doi.org/10.1007/JHEP11(2024)146} {\bibfield  {journal}
  {\bibinfo  {journal} {JHEP}\ }\textbf {\bibinfo {volume} {11}},\ \bibinfo
  {pages} {146}},\ \Eprint {https://arxiv.org/abs/2406.11956} {arXiv:2406.11956
  [hep-th]} \BibitemShut {NoStop}%
\bibitem [{\citenamefont {M{\o}ller}(1961)}]{Moller:1961}%
  \BibitemOpen
  \bibfield  {author} {\bibinfo {author} {\bibfnamefont {C.}~\bibnamefont
  {M{\o}ller}},\ }\bibfield  {title} {\bibinfo {title} {{Conservation Law and
  Absolute Parallelism in General Relativity}},\ }\href@noop {} {\bibfield
  {journal} {\bibinfo  {journal} {K. Dan. Vidensk. Selsk. Mat. Fys. Skr.}\
  }\textbf {\bibinfo {volume} {1}},\ \bibinfo {pages} {1} (\bibinfo {year}
  {1961})}\BibitemShut {NoStop}%
\bibitem [{\citenamefont {Pellegrini}\ and\ \citenamefont
  {Plebanski}(1963)}]{Pellegrini:1963}%
  \BibitemOpen
  \bibfield  {author} {\bibinfo {author} {\bibfnamefont {C.}~\bibnamefont
  {Pellegrini}}\ and\ \bibinfo {author} {\bibfnamefont {J.}~\bibnamefont
  {Plebanski}},\ }\bibfield  {title} {\bibinfo {title} {{Tetrad Fields and
  Gravitational Fields}},\ }\href@noop {} {\bibfield  {journal} {\bibinfo
  {journal} {K. Dan. Vidensk. Selsk. Mat. Fys. Skr.}\ }\textbf {\bibinfo
  {volume} {2}},\ \bibinfo {pages} {1} (\bibinfo {year} {1963})}\BibitemShut
  {NoStop}%
\bibitem [{\citenamefont {Cho}(1976)}]{Cho:1975dh}%
  \BibitemOpen
  \bibfield  {author} {\bibinfo {author} {\bibfnamefont {Y.~M.}\ \bibnamefont
  {Cho}},\ }\bibfield  {title} {\bibinfo {title} {{Einstein Lagrangian as the
  Translational Yang-Mills Lagrangian}},\ }\href
  {https://doi.org/10.1103/PhysRevD.14.2521} {\bibfield  {journal} {\bibinfo
  {journal} {Phys. Rev. D}\ }\textbf {\bibinfo {volume} {14}},\ \bibinfo
  {pages} {2521} (\bibinfo {year} {1976})}\BibitemShut {NoStop}%
\bibitem [{\citenamefont {Hayashi}\ and\ \citenamefont
  {Shirafuji}(1979)}]{Hayashi:1979qx}%
  \BibitemOpen
  \bibfield  {author} {\bibinfo {author} {\bibfnamefont {K.}~\bibnamefont
  {Hayashi}}\ and\ \bibinfo {author} {\bibfnamefont {T.}~\bibnamefont
  {Shirafuji}},\ }\bibfield  {title} {\bibinfo {title} {{New general
  relativity.}},\ }\href {https://doi.org/10.1103/PhysRevD.19.3524} {\bibfield
  {journal} {\bibinfo  {journal} {Phys. Rev. D}\ }\textbf {\bibinfo {volume}
  {19}},\ \bibinfo {pages} {3524} (\bibinfo {year} {1979})},\ \bibinfo {note}
  {[Addendum: Phys.Rev.D 24, 3312--3314 (1982)]}\BibitemShut {NoStop}%
\bibitem [{\citenamefont {Dimakis}(1989{\natexlab{a}})}]{Dimakis:1989az}%
  \BibitemOpen
  \bibfield  {author} {\bibinfo {author} {\bibfnamefont {A.}~\bibnamefont
  {Dimakis}},\ }\bibfield  {title} {\bibinfo {title} {{The Initial Value
  Problem of the Poincare Gauge Theory in Vacuum. 1: Second Order Formalism}},\
  }\href@noop {} {\bibfield  {journal} {\bibinfo  {journal} {Ann. Inst. H.
  Poincare Phys. Theor.}\ }\textbf {\bibinfo {volume} {51}},\ \bibinfo {pages}
  {371} (\bibinfo {year} {1989}{\natexlab{a}})}\BibitemShut {NoStop}%
\bibitem [{\citenamefont {Dimakis}(1989{\natexlab{b}})}]{Dimakis:1989ba}%
  \BibitemOpen
  \bibfield  {author} {\bibinfo {author} {\bibfnamefont {A.}~\bibnamefont
  {Dimakis}},\ }\bibfield  {title} {\bibinfo {title} {{THE INITIAL VALUE
  PROBLEM OF THE POINCARE GAUGE THEORY IN VACUUM. 1: FIRST ORDER FORMALISM}},\
  }\href@noop {} {\bibfield  {journal} {\bibinfo  {journal} {Ann. Inst. H.
  Poincare Phys. Theor.}\ }\textbf {\bibinfo {volume} {51}},\ \bibinfo {pages}
  {389} (\bibinfo {year} {1989}{\natexlab{b}})}\BibitemShut {NoStop}%
\bibitem [{\citenamefont {Lemke}(1990)}]{Lemke:1990su}%
  \BibitemOpen
  \bibfield  {author} {\bibinfo {author} {\bibfnamefont {J.}~\bibnamefont
  {Lemke}},\ }\bibfield  {title} {\bibinfo {title} {{Shock waves in the
  Poincare gauge theory of gravitation}},\ }\href
  {https://doi.org/10.1016/0375-9601(90)90789-Q} {\bibfield  {journal}
  {\bibinfo  {journal} {Phys. Lett. A}\ }\textbf {\bibinfo {volume} {143}},\
  \bibinfo {pages} {13} (\bibinfo {year} {1990})}\BibitemShut {NoStop}%
\bibitem [{\citenamefont {Hecht}\ \emph {et~al.}(1990)\citenamefont {Hecht},
  \citenamefont {Lemke},\ and\ \citenamefont {Wallner}}]{Hecht:1990wn}%
  \BibitemOpen
  \bibfield  {author} {\bibinfo {author} {\bibfnamefont {R.~D.}\ \bibnamefont
  {Hecht}}, \bibinfo {author} {\bibfnamefont {J.}~\bibnamefont {Lemke}},\ and\
  \bibinfo {author} {\bibfnamefont {R.~P.}\ \bibnamefont {Wallner}},\
  }\bibfield  {title} {\bibinfo {title} {{Tachyonic torsion shock waves in
  Poincare gauge theory}},\ }\href
  {https://doi.org/10.1016/0375-9601(90)90837-E} {\bibfield  {journal}
  {\bibinfo  {journal} {Phys. Lett. A}\ }\textbf {\bibinfo {volume} {151}},\
  \bibinfo {pages} {12} (\bibinfo {year} {1990})}\BibitemShut {NoStop}%
\bibitem [{\citenamefont {Hecht}\ \emph {et~al.}(1991)\citenamefont {Hecht},
  \citenamefont {Lemke},\ and\ \citenamefont {Wallner}}]{Hecht:1991jh}%
  \BibitemOpen
  \bibfield  {author} {\bibinfo {author} {\bibfnamefont {R.~D.}\ \bibnamefont
  {Hecht}}, \bibinfo {author} {\bibfnamefont {J.}~\bibnamefont {Lemke}},\ and\
  \bibinfo {author} {\bibfnamefont {R.~P.}\ \bibnamefont {Wallner}},\
  }\bibfield  {title} {\bibinfo {title} {{Can Poincare gauge theory be
  saved?}},\ }\href {https://doi.org/10.1103/PhysRevD.44.2442} {\bibfield
  {journal} {\bibinfo  {journal} {Phys. Rev. D}\ }\textbf {\bibinfo {volume}
  {44}},\ \bibinfo {pages} {2442} (\bibinfo {year} {1991})}\BibitemShut
  {NoStop}%
\bibitem [{\citenamefont {Afshordi}\ \emph {et~al.}(2007)\citenamefont
  {Afshordi}, \citenamefont {Chung},\ and\ \citenamefont
  {Geshnizjani}}]{Afshordi:2006ad}%
  \BibitemOpen
  \bibfield  {author} {\bibinfo {author} {\bibfnamefont {N.}~\bibnamefont
  {Afshordi}}, \bibinfo {author} {\bibfnamefont {D.~J.~H.}\ \bibnamefont
  {Chung}},\ and\ \bibinfo {author} {\bibfnamefont {G.}~\bibnamefont
  {Geshnizjani}},\ }\bibfield  {title} {\bibinfo {title} {{Cuscuton: A Causal
  Field Theory with an Infinite Speed of Sound}},\ }\href
  {https://doi.org/10.1103/PhysRevD.75.083513} {\bibfield  {journal} {\bibinfo
  {journal} {Phys. Rev. D}\ }\textbf {\bibinfo {volume} {75}},\ \bibinfo
  {pages} {083513} (\bibinfo {year} {2007})},\ \Eprint
  {https://arxiv.org/abs/hep-th/0609150} {arXiv:hep-th/0609150} \BibitemShut
  {NoStop}%
\bibitem [{\citenamefont {Magueijo}(2009)}]{Magueijo:2008sx}%
  \BibitemOpen
  \bibfield  {author} {\bibinfo {author} {\bibfnamefont {J.}~\bibnamefont
  {Magueijo}},\ }\bibfield  {title} {\bibinfo {title} {{Bimetric varying speed
  of light theories and primordial fluctuations}},\ }\href
  {https://doi.org/10.1103/PhysRevD.79.043525} {\bibfield  {journal} {\bibinfo
  {journal} {Phys. Rev. D}\ }\textbf {\bibinfo {volume} {79}},\ \bibinfo
  {pages} {043525} (\bibinfo {year} {2009})},\ \Eprint
  {https://arxiv.org/abs/0807.1689} {arXiv:0807.1689 [gr-qc]} \BibitemShut
  {NoStop}%
\bibitem [{\citenamefont {Charmousis}\ and\ \citenamefont
  {Padilla}(2008)}]{Charmousis:2008ce}%
  \BibitemOpen
  \bibfield  {author} {\bibinfo {author} {\bibfnamefont {C.}~\bibnamefont
  {Charmousis}}\ and\ \bibinfo {author} {\bibfnamefont {A.}~\bibnamefont
  {Padilla}},\ }\bibfield  {title} {\bibinfo {title} {{The Instability of Vacua
  in Gauss-Bonnet Gravity}},\ }\href
  {https://doi.org/10.1088/1126-6708/2008/12/038} {\bibfield  {journal}
  {\bibinfo  {journal} {JHEP}\ }\textbf {\bibinfo {volume} {12}},\ \bibinfo
  {pages} {038}},\ \Eprint {https://arxiv.org/abs/0807.2864} {arXiv:0807.2864
  [hep-th]} \BibitemShut {NoStop}%
\bibitem [{\citenamefont {Charmousis}\ \emph {et~al.}(2009)\citenamefont
  {Charmousis}, \citenamefont {Niz}, \citenamefont {Padilla},\ and\
  \citenamefont {Saffin}}]{Charmousis:2009tc}%
  \BibitemOpen
  \bibfield  {author} {\bibinfo {author} {\bibfnamefont {C.}~\bibnamefont
  {Charmousis}}, \bibinfo {author} {\bibfnamefont {G.}~\bibnamefont {Niz}},
  \bibinfo {author} {\bibfnamefont {A.}~\bibnamefont {Padilla}},\ and\ \bibinfo
  {author} {\bibfnamefont {P.~M.}\ \bibnamefont {Saffin}},\ }\bibfield  {title}
  {\bibinfo {title} {{Strong coupling in Horava gravity}},\ }\href
  {https://doi.org/10.1088/1126-6708/2009/08/070} {\bibfield  {journal}
  {\bibinfo  {journal} {JHEP}\ }\textbf {\bibinfo {volume} {08}},\ \bibinfo
  {pages} {070}},\ \Eprint {https://arxiv.org/abs/0905.2579} {arXiv:0905.2579
  [hep-th]} \BibitemShut {NoStop}%
\bibitem [{\citenamefont {Papazoglou}\ and\ \citenamefont
  {Sotiriou}(2010)}]{Papazoglou:2009fj}%
  \BibitemOpen
  \bibfield  {author} {\bibinfo {author} {\bibfnamefont {A.}~\bibnamefont
  {Papazoglou}}\ and\ \bibinfo {author} {\bibfnamefont {T.~P.}\ \bibnamefont
  {Sotiriou}},\ }\bibfield  {title} {\bibinfo {title} {{Strong coupling in
  extended Horava-Lifshitz gravity}},\ }\href
  {https://doi.org/10.1016/j.physletb.2010.01.054} {\bibfield  {journal}
  {\bibinfo  {journal} {Phys. Lett. B}\ }\textbf {\bibinfo {volume} {685}},\
  \bibinfo {pages} {197} (\bibinfo {year} {2010})},\ \Eprint
  {https://arxiv.org/abs/0911.1299} {arXiv:0911.1299 [hep-th]} \BibitemShut
  {NoStop}%
\bibitem [{\citenamefont {Baumann}\ \emph {et~al.}(2011)\citenamefont
  {Baumann}, \citenamefont {Senatore},\ and\ \citenamefont
  {Zaldarriaga}}]{Baumann:2011dt}%
  \BibitemOpen
  \bibfield  {author} {\bibinfo {author} {\bibfnamefont {D.}~\bibnamefont
  {Baumann}}, \bibinfo {author} {\bibfnamefont {L.}~\bibnamefont {Senatore}},\
  and\ \bibinfo {author} {\bibfnamefont {M.}~\bibnamefont {Zaldarriaga}},\
  }\bibfield  {title} {\bibinfo {title} {{Scale-Invariance and the Strong
  Coupling Problem}},\ }\href {https://doi.org/10.1088/1475-7516/2011/05/004}
  {\bibfield  {journal} {\bibinfo  {journal} {JCAP}\ }\textbf {\bibinfo
  {volume} {05}},\ \bibinfo {pages} {004}},\ \Eprint
  {https://arxiv.org/abs/1101.3320} {arXiv:1101.3320 [hep-th]} \BibitemShut
  {NoStop}%
\bibitem [{\citenamefont {D'Amico}\ \emph {et~al.}(2011)\citenamefont
  {D'Amico}, \citenamefont {de~Rham}, \citenamefont {Dubovsky}, \citenamefont
  {Gabadadze}, \citenamefont {Pirtskhalava},\ and\ \citenamefont
  {Tolley}}]{DAmico:2011eto}%
  \BibitemOpen
  \bibfield  {author} {\bibinfo {author} {\bibfnamefont {G.}~\bibnamefont
  {D'Amico}}, \bibinfo {author} {\bibfnamefont {C.}~\bibnamefont {de~Rham}},
  \bibinfo {author} {\bibfnamefont {S.}~\bibnamefont {Dubovsky}}, \bibinfo
  {author} {\bibfnamefont {G.}~\bibnamefont {Gabadadze}}, \bibinfo {author}
  {\bibfnamefont {D.}~\bibnamefont {Pirtskhalava}},\ and\ \bibinfo {author}
  {\bibfnamefont {A.~J.}\ \bibnamefont {Tolley}},\ }\bibfield  {title}
  {\bibinfo {title} {{Massive Cosmologies}},\ }\href
  {https://doi.org/10.1103/PhysRevD.84.124046} {\bibfield  {journal} {\bibinfo
  {journal} {Phys. Rev. D}\ }\textbf {\bibinfo {volume} {84}},\ \bibinfo
  {pages} {124046} (\bibinfo {year} {2011})},\ \Eprint
  {https://arxiv.org/abs/1108.5231} {arXiv:1108.5231 [hep-th]} \BibitemShut
  {NoStop}%
\bibitem [{\citenamefont {Gumrukcuoglu}\ \emph {et~al.}(2012)\citenamefont
  {Gumrukcuoglu}, \citenamefont {Lin},\ and\ \citenamefont
  {Mukohyama}}]{Gumrukcuoglu:2012aa}%
  \BibitemOpen
  \bibfield  {author} {\bibinfo {author} {\bibfnamefont {A.~E.}\ \bibnamefont
  {Gumrukcuoglu}}, \bibinfo {author} {\bibfnamefont {C.}~\bibnamefont {Lin}},\
  and\ \bibinfo {author} {\bibfnamefont {S.}~\bibnamefont {Mukohyama}},\
  }\bibfield  {title} {\bibinfo {title} {{Anisotropic
  Friedmann-Robertson-Walker universe from nonlinear massive gravity}},\ }\href
  {https://doi.org/10.1016/j.physletb.2012.09.049} {\bibfield  {journal}
  {\bibinfo  {journal} {Phys. Lett. B}\ }\textbf {\bibinfo {volume} {717}},\
  \bibinfo {pages} {295} (\bibinfo {year} {2012})},\ \Eprint
  {https://arxiv.org/abs/1206.2723} {arXiv:1206.2723 [hep-th]} \BibitemShut
  {NoStop}%
\bibitem [{\citenamefont {Wang}(2017)}]{Wang:2017brl}%
  \BibitemOpen
  \bibfield  {author} {\bibinfo {author} {\bibfnamefont {A.}~\bibnamefont
  {Wang}},\ }\bibfield  {title} {\bibinfo {title} {{Ho\v{r}ava gravity at a
  Lifshitz point: A progress report}},\ }\href
  {https://doi.org/10.1142/S0218271817300142} {\bibfield  {journal} {\bibinfo
  {journal} {Int. J. Mod. Phys. D}\ }\textbf {\bibinfo {volume} {26}},\
  \bibinfo {pages} {1730014} (\bibinfo {year} {2017})},\ \Eprint
  {https://arxiv.org/abs/1701.06087} {arXiv:1701.06087 [gr-qc]} \BibitemShut
  {NoStop}%
\bibitem [{\citenamefont {Mazuet}\ \emph {et~al.}(2017)\citenamefont {Mazuet},
  \citenamefont {Mukohyama},\ and\ \citenamefont {Volkov}}]{Mazuet:2017rgq}%
  \BibitemOpen
  \bibfield  {author} {\bibinfo {author} {\bibfnamefont {C.}~\bibnamefont
  {Mazuet}}, \bibinfo {author} {\bibfnamefont {S.}~\bibnamefont {Mukohyama}},\
  and\ \bibinfo {author} {\bibfnamefont {M.~S.}\ \bibnamefont {Volkov}},\
  }\bibfield  {title} {\bibinfo {title} {{Anisotropic deformations of spatially
  open cosmology in massive gravity theory}},\ }\href
  {https://doi.org/10.1088/1475-7516/2017/04/039} {\bibfield  {journal}
  {\bibinfo  {journal} {JCAP}\ }\textbf {\bibinfo {volume} {04}},\ \bibinfo
  {pages} {039}},\ \Eprint {https://arxiv.org/abs/1702.04205} {arXiv:1702.04205
  [hep-th]} \BibitemShut {NoStop}%
\bibitem [{\citenamefont {Beltr\'an~Jim\'enez}\ and\ \citenamefont
  {Jim\'enez-Cano}(2021)}]{BeltranJimenez:2020lee}%
  \BibitemOpen
  \bibfield  {author} {\bibinfo {author} {\bibfnamefont {J.}~\bibnamefont
  {Beltr\'an~Jim\'enez}}\ and\ \bibinfo {author} {\bibfnamefont
  {A.}~\bibnamefont {Jim\'enez-Cano}},\ }\bibfield  {title} {\bibinfo {title}
  {{On the strong coupling of Einsteinian Cubic Gravity and its
  generalisations}},\ }\href {https://doi.org/10.1088/1475-7516/2021/01/069}
  {\bibfield  {journal} {\bibinfo  {journal} {JCAP}\ }\textbf {\bibinfo
  {volume} {01}},\ \bibinfo {pages} {069}},\ \Eprint
  {https://arxiv.org/abs/2009.08197} {arXiv:2009.08197 [gr-qc]} \BibitemShut
  {NoStop}%
\bibitem [{\citenamefont {Jim\'enez~Cano}(2021)}]{JimenezCano:2021rlu}%
  \BibitemOpen
  \bibfield  {author} {\bibinfo {author} {\bibfnamefont {A.}~\bibnamefont
  {Jim\'enez~Cano}},\ }\emph {\bibinfo {title} {{Metric-affine Gauge theories
  of gravity. Foundations and new insights}}},\ \href@noop {} {Ph.D. thesis},\
  \bibinfo  {school} {Granada U., Theor. Phys. Astrophys.} (\bibinfo {year}
  {2021}),\ \Eprint {https://arxiv.org/abs/2201.12847} {arXiv:2201.12847
  [gr-qc]} \BibitemShut {NoStop}%
\bibitem [{\citenamefont {Barker}(2023{\natexlab{a}})}]{Barker:2022kdk}%
  \BibitemOpen
  \bibfield  {author} {\bibinfo {author} {\bibfnamefont {W.~E.~V.}\
  \bibnamefont {Barker}},\ }\bibfield  {title} {\bibinfo {title}
  {{Supercomputers against strong coupling in gravity with curvature and
  torsion}},\ }\href {https://doi.org/10.1140/epjc/s10052-023-11179-6}
  {\bibfield  {journal} {\bibinfo  {journal} {Eur. Phys. J. C}\ }\textbf
  {\bibinfo {volume} {83}},\ \bibinfo {pages} {228} (\bibinfo {year}
  {2023}{\natexlab{a}})},\ \Eprint {https://arxiv.org/abs/2206.00658}
  {arXiv:2206.00658 [gr-qc]} \BibitemShut {NoStop}%
\bibitem [{\citenamefont {Delhom}\ \emph {et~al.}(2022)\citenamefont {Delhom},
  \citenamefont {Jim\'enez-Cano},\ and\ \citenamefont
  {Maldonado~Torralba}}]{Delhom:2022vae}%
  \BibitemOpen
  \bibfield  {author} {\bibinfo {author} {\bibfnamefont {A.}~\bibnamefont
  {Delhom}}, \bibinfo {author} {\bibfnamefont {A.}~\bibnamefont
  {Jim\'enez-Cano}},\ and\ \bibinfo {author} {\bibfnamefont {F.~J.}\
  \bibnamefont {Maldonado~Torralba}},\ }\bibfield  {title} {\bibinfo {title}
  {{Instabilities in Field Theory: A Primer with Applications in Modified
  Gravity}}\ }(\bibinfo  {publisher} {Springer},\ \bibinfo {year} {2022})\
  \Eprint {https://arxiv.org/abs/2207.13431} {arXiv:2207.13431 [gr-qc]}
  \BibitemShut {NoStop}%
\bibitem [{\citenamefont {Karananas}(2025{\natexlab{a}})}]{Karananas:2024hoh}%
  \BibitemOpen
  \bibfield  {author} {\bibinfo {author} {\bibfnamefont {G.~K.}\ \bibnamefont
  {Karananas}},\ }\bibfield  {title} {\bibinfo {title} {{Particle content of
  (scalar\,curvature)2 gravities revisited}},\ }\href
  {https://doi.org/10.1103/PhysRevD.111.044068} {\bibfield  {journal} {\bibinfo
   {journal} {Phys. Rev. D}\ }\textbf {\bibinfo {volume} {111}},\ \bibinfo
  {pages} {044068} (\bibinfo {year} {2025}{\natexlab{a}})},\ \Eprint
  {https://arxiv.org/abs/2407.09598} {arXiv:2407.09598 [hep-th]} \BibitemShut
  {NoStop}%
\bibitem [{\citenamefont {Vainshtein}(1972)}]{Vainshtein:1972sx}%
  \BibitemOpen
  \bibfield  {author} {\bibinfo {author} {\bibfnamefont {A.~I.}\ \bibnamefont
  {Vainshtein}},\ }\bibfield  {title} {\bibinfo {title} {{To the problem of
  nonvanishing gravitation mass}},\ }\href
  {https://doi.org/10.1016/0370-2693(72)90147-5} {\bibfield  {journal}
  {\bibinfo  {journal} {Phys. Lett. B}\ }\textbf {\bibinfo {volume} {39}},\
  \bibinfo {pages} {393} (\bibinfo {year} {1972})}\BibitemShut {NoStop}%
\bibitem [{\citenamefont {Deffayet}\ \emph {et~al.}(2002)\citenamefont
  {Deffayet}, \citenamefont {Dvali}, \citenamefont {Gabadadze},\ and\
  \citenamefont {Vainshtein}}]{Deffayet:2001uk}%
  \BibitemOpen
  \bibfield  {author} {\bibinfo {author} {\bibfnamefont {C.}~\bibnamefont
  {Deffayet}}, \bibinfo {author} {\bibfnamefont {G.~R.}\ \bibnamefont {Dvali}},
  \bibinfo {author} {\bibfnamefont {G.}~\bibnamefont {Gabadadze}},\ and\
  \bibinfo {author} {\bibfnamefont {A.~I.}\ \bibnamefont {Vainshtein}},\
  }\bibfield  {title} {\bibinfo {title} {{Nonperturbative continuity in
  graviton mass versus perturbative discontinuity}},\ }\href
  {https://doi.org/10.1103/PhysRevD.65.044026} {\bibfield  {journal} {\bibinfo
  {journal} {Phys. Rev. D}\ }\textbf {\bibinfo {volume} {65}},\ \bibinfo
  {pages} {044026} (\bibinfo {year} {2002})},\ \Eprint
  {https://arxiv.org/abs/hep-th/0106001} {arXiv:hep-th/0106001} \BibitemShut
  {NoStop}%
\bibitem [{\citenamefont {Deffayet}\ and\ \citenamefont
  {Rombouts}(2005)}]{Deffayet:2005ys}%
  \BibitemOpen
  \bibfield  {author} {\bibinfo {author} {\bibfnamefont {C.}~\bibnamefont
  {Deffayet}}\ and\ \bibinfo {author} {\bibfnamefont {J.-W.}\ \bibnamefont
  {Rombouts}},\ }\bibfield  {title} {\bibinfo {title} {{Ghosts, strong coupling
  and accidental symmetries in massive gravity}},\ }\href
  {https://doi.org/10.1103/PhysRevD.72.044003} {\bibfield  {journal} {\bibinfo
  {journal} {Phys. Rev. D}\ }\textbf {\bibinfo {volume} {72}},\ \bibinfo
  {pages} {044003} (\bibinfo {year} {2005})},\ \Eprint
  {https://arxiv.org/abs/gr-qc/0505134} {arXiv:gr-qc/0505134} \BibitemShut
  {NoStop}%
\bibitem [{\citenamefont {de~Rham}(2014)}]{deRham:2014zqa}%
  \BibitemOpen
  \bibfield  {author} {\bibinfo {author} {\bibfnamefont {C.}~\bibnamefont
  {de~Rham}},\ }\bibfield  {title} {\bibinfo {title} {{Massive Gravity}},\
  }\href {https://doi.org/10.12942/lrr-2014-7} {\bibfield  {journal} {\bibinfo
  {journal} {Living Rev. Rel.}\ }\textbf {\bibinfo {volume} {17}},\ \bibinfo
  {pages} {7} (\bibinfo {year} {2014})},\ \Eprint
  {https://arxiv.org/abs/1401.4173} {arXiv:1401.4173 [hep-th]} \BibitemShut
  {NoStop}%
\bibitem [{\citenamefont {Deser}\ \emph {et~al.}(2014)\citenamefont {Deser},
  \citenamefont {Sandora}, \citenamefont {Waldron},\ and\ \citenamefont
  {Zahariade}}]{Deser:2014hga}%
  \BibitemOpen
  \bibfield  {author} {\bibinfo {author} {\bibfnamefont {S.}~\bibnamefont
  {Deser}}, \bibinfo {author} {\bibfnamefont {M.}~\bibnamefont {Sandora}},
  \bibinfo {author} {\bibfnamefont {A.}~\bibnamefont {Waldron}},\ and\ \bibinfo
  {author} {\bibfnamefont {G.}~\bibnamefont {Zahariade}},\ }\bibfield  {title}
  {\bibinfo {title} {{Covariant constraints for generic massive gravity and
  analysis of its characteristics}},\ }\href
  {https://doi.org/10.1103/PhysRevD.90.104043} {\bibfield  {journal} {\bibinfo
  {journal} {Phys. Rev. D}\ }\textbf {\bibinfo {volume} {90}},\ \bibinfo
  {pages} {104043} (\bibinfo {year} {2014})},\ \Eprint
  {https://arxiv.org/abs/1408.0561} {arXiv:1408.0561 [hep-th]} \BibitemShut
  {NoStop}%
\bibitem [{\citenamefont {Velo}\ and\ \citenamefont
  {Zwanziger}(1969)}]{Velo:1969txo}%
  \BibitemOpen
  \bibfield  {author} {\bibinfo {author} {\bibfnamefont {G.}~\bibnamefont
  {Velo}}\ and\ \bibinfo {author} {\bibfnamefont {D.}~\bibnamefont
  {Zwanziger}},\ }\bibfield  {title} {\bibinfo {title} {{Noncausality and other
  defects of interaction lagrangians for particles with spin one and higher}},\
  }\href {https://doi.org/10.1103/PhysRev.188.2218} {\bibfield  {journal}
  {\bibinfo  {journal} {Phys. Rev.}\ }\textbf {\bibinfo {volume} {188}},\
  \bibinfo {pages} {2218} (\bibinfo {year} {1969})}\BibitemShut {NoStop}%
\bibitem [{\citenamefont {Aragone}\ and\ \citenamefont
  {Deser}(1971)}]{Aragone:1971kh}%
  \BibitemOpen
  \bibfield  {author} {\bibinfo {author} {\bibfnamefont {C.}~\bibnamefont
  {Aragone}}\ and\ \bibinfo {author} {\bibfnamefont {S.}~\bibnamefont
  {Deser}},\ }\bibfield  {title} {\bibinfo {title} {{Constraints on
  gravitationally coupled tensor fields}},\ }\href
  {https://doi.org/10.1007/BF02813572} {\bibfield  {journal} {\bibinfo
  {journal} {Nuovo Cim. A}\ }\textbf {\bibinfo {volume} {3}},\ \bibinfo {pages}
  {709} (\bibinfo {year} {1971})}\BibitemShut {NoStop}%
\bibitem [{\citenamefont {Cheng}\ \emph {et~al.}(1988)\citenamefont {Cheng},
  \citenamefont {Chern},\ and\ \citenamefont {Nester}}]{Cheng:1988zg}%
  \BibitemOpen
  \bibfield  {author} {\bibinfo {author} {\bibfnamefont {W.-H.}\ \bibnamefont
  {Cheng}}, \bibinfo {author} {\bibfnamefont {D.-C.}\ \bibnamefont {Chern}},\
  and\ \bibinfo {author} {\bibfnamefont {J.~M.}\ \bibnamefont {Nester}},\
  }\bibfield  {title} {\bibinfo {title} {{Canonical Analysis of the One
  Parameter Teleparallel Theory}},\ }\href
  {https://doi.org/10.1103/PhysRevD.38.2656} {\bibfield  {journal} {\bibinfo
  {journal} {Phys. Rev. D}\ }\textbf {\bibinfo {volume} {38}},\ \bibinfo
  {pages} {2656} (\bibinfo {year} {1988})}\BibitemShut {NoStop}%
\bibitem [{\citenamefont {Hecht}\ \emph {et~al.}(1996)\citenamefont {Hecht},
  \citenamefont {Nester},\ and\ \citenamefont {Zhytnikov}}]{Hecht:1996np}%
  \BibitemOpen
  \bibfield  {author} {\bibinfo {author} {\bibfnamefont {R.~D.}\ \bibnamefont
  {Hecht}}, \bibinfo {author} {\bibfnamefont {J.~M.}\ \bibnamefont {Nester}},\
  and\ \bibinfo {author} {\bibfnamefont {V.~V.}\ \bibnamefont {Zhytnikov}},\
  }\bibfield  {title} {\bibinfo {title} {{Some Poincare gauge theory
  Lagrangians with well posed initial value problems}},\ }\href
  {https://doi.org/10.1016/0375-9601(96)00622-6} {\bibfield  {journal}
  {\bibinfo  {journal} {Phys. Lett. A}\ }\textbf {\bibinfo {volume} {222}},\
  \bibinfo {pages} {37} (\bibinfo {year} {1996})}\BibitemShut {NoStop}%
\bibitem [{\citenamefont {Chen}\ \emph {et~al.}(1998)\citenamefont {Chen},
  \citenamefont {Nester},\ and\ \citenamefont {Yo}}]{Chen:1998ad}%
  \BibitemOpen
  \bibfield  {author} {\bibinfo {author} {\bibfnamefont {H.}~\bibnamefont
  {Chen}}, \bibinfo {author} {\bibfnamefont {J.~M.}\ \bibnamefont {Nester}},\
  and\ \bibinfo {author} {\bibfnamefont {H.-J.}\ \bibnamefont {Yo}},\
  }\bibfield  {title} {\bibinfo {title} {{Acausal PGT modes and the nonlinear
  constraint effect}},\ }\href@noop {} {\bibfield  {journal} {\bibinfo
  {journal} {Acta Phys. Polon. B}\ }\textbf {\bibinfo {volume} {29}},\ \bibinfo
  {pages} {961} (\bibinfo {year} {1998})}\BibitemShut {NoStop}%
\bibitem [{\citenamefont {Blixt}\ \emph
  {et~al.}(2019{\natexlab{a}})\citenamefont {Blixt}, \citenamefont {Hohmann},\
  and\ \citenamefont {Pfeifer}}]{Blixt:2018znp}%
  \BibitemOpen
  \bibfield  {author} {\bibinfo {author} {\bibfnamefont {D.}~\bibnamefont
  {Blixt}}, \bibinfo {author} {\bibfnamefont {M.}~\bibnamefont {Hohmann}},\
  and\ \bibinfo {author} {\bibfnamefont {C.}~\bibnamefont {Pfeifer}},\
  }\bibfield  {title} {\bibinfo {title} {{Hamiltonian and primary constraints
  of new general relativity}},\ }\href
  {https://doi.org/10.1103/PhysRevD.99.084025} {\bibfield  {journal} {\bibinfo
  {journal} {Phys. Rev. D}\ }\textbf {\bibinfo {volume} {99}},\ \bibinfo
  {pages} {084025} (\bibinfo {year} {2019}{\natexlab{a}})},\ \Eprint
  {https://arxiv.org/abs/1811.11137} {arXiv:1811.11137 [gr-qc]} \BibitemShut
  {NoStop}%
\bibitem [{\citenamefont {Blixt}\ \emph
  {et~al.}(2019{\natexlab{b}})\citenamefont {Blixt}, \citenamefont {Hohmann},
  \citenamefont {Kr\v{s}\v{s}\'ak},\ and\ \citenamefont
  {Pfeifer}}]{Blixt:2019ene}%
  \BibitemOpen
  \bibfield  {author} {\bibinfo {author} {\bibfnamefont {D.}~\bibnamefont
  {Blixt}}, \bibinfo {author} {\bibfnamefont {M.}~\bibnamefont {Hohmann}},
  \bibinfo {author} {\bibfnamefont {M.}~\bibnamefont {Kr\v{s}\v{s}\'ak}},\ and\
  \bibinfo {author} {\bibfnamefont {C.}~\bibnamefont {Pfeifer}},\ }\bibfield
  {title} {\bibinfo {title} {{Hamiltonian analysis in new general
  relativity}},\ }in\ \href {https://doi.org/10.1142/9789811258251_0038} {\emph
  {\bibinfo {booktitle} {{15th Marcel Grossmann Meeting on Recent Developments
  in Theoretical and Experimental General Relativity, Astrophysics, and
  Relativistic Field Theories}}}}\ (\bibinfo {year} {2019})\ \Eprint
  {https://arxiv.org/abs/1905.11919} {arXiv:1905.11919 [gr-qc]} \BibitemShut
  {NoStop}%
\bibitem [{\citenamefont {Blixt}\ \emph {et~al.}(2021)\citenamefont {Blixt},
  \citenamefont {Guzm\'an}, \citenamefont {Hohmann},\ and\ \citenamefont
  {Pfeifer}}]{Blixt:2020ekl}%
  \BibitemOpen
  \bibfield  {author} {\bibinfo {author} {\bibfnamefont {D.}~\bibnamefont
  {Blixt}}, \bibinfo {author} {\bibfnamefont {M.-J.}\ \bibnamefont {Guzm\'an}},
  \bibinfo {author} {\bibfnamefont {M.}~\bibnamefont {Hohmann}},\ and\ \bibinfo
  {author} {\bibfnamefont {C.}~\bibnamefont {Pfeifer}},\ }\bibfield  {title}
  {\bibinfo {title} {{Review of the Hamiltonian analysis in teleparallel
  gravity}},\ }\href {https://doi.org/10.1142/S0219887821300051} {\bibfield
  {journal} {\bibinfo  {journal} {Int. J. Geom. Meth. Mod. Phys.}\ }\textbf
  {\bibinfo {volume} {18}},\ \bibinfo {pages} {2130005} (\bibinfo {year}
  {2021})},\ \Eprint {https://arxiv.org/abs/2012.09180} {arXiv:2012.09180
  [gr-qc]} \BibitemShut {NoStop}%
\bibitem [{\citenamefont {Krasnov}\ and\ \citenamefont
  {Mitsou}(2021)}]{Krasnov:2021zen}%
  \BibitemOpen
  \bibfield  {author} {\bibinfo {author} {\bibfnamefont {K.}~\bibnamefont
  {Krasnov}}\ and\ \bibinfo {author} {\bibfnamefont {E.}~\bibnamefont
  {Mitsou}},\ }\bibfield  {title} {\bibinfo {title} {{Pure Lorentz spin
  connection theories and uniqueness of general relativity}},\ }\href
  {https://doi.org/10.1088/1361-6382/ac25e3} {\bibfield  {journal} {\bibinfo
  {journal} {Class. Quant. Grav.}\ }\textbf {\bibinfo {volume} {38}},\ \bibinfo
  {pages} {205009} (\bibinfo {year} {2021})},\ \Eprint
  {https://arxiv.org/abs/2106.05803} {arXiv:2106.05803 [gr-qc]} \BibitemShut
  {NoStop}%
\bibitem [{\citenamefont {Bahamonde}\ \emph {et~al.}(2023)\citenamefont
  {Bahamonde}, \citenamefont {Dialektopoulos}, \citenamefont
  {Escamilla-Rivera}, \citenamefont {Farrugia}, \citenamefont {Gakis},
  \citenamefont {Hendry}, \citenamefont {Hohmann}, \citenamefont {Levi~Said},
  \citenamefont {Mifsud},\ and\ \citenamefont
  {Di~Valentino}}]{Bahamonde:2021gfp}%
  \BibitemOpen
  \bibfield  {author} {\bibinfo {author} {\bibfnamefont {S.}~\bibnamefont
  {Bahamonde}}, \bibinfo {author} {\bibfnamefont {K.~F.}\ \bibnamefont
  {Dialektopoulos}}, \bibinfo {author} {\bibfnamefont {C.}~\bibnamefont
  {Escamilla-Rivera}}, \bibinfo {author} {\bibfnamefont {G.}~\bibnamefont
  {Farrugia}}, \bibinfo {author} {\bibfnamefont {V.}~\bibnamefont {Gakis}},
  \bibinfo {author} {\bibfnamefont {M.}~\bibnamefont {Hendry}}, \bibinfo
  {author} {\bibfnamefont {M.}~\bibnamefont {Hohmann}}, \bibinfo {author}
  {\bibfnamefont {J.}~\bibnamefont {Levi~Said}}, \bibinfo {author}
  {\bibfnamefont {J.}~\bibnamefont {Mifsud}},\ and\ \bibinfo {author}
  {\bibfnamefont {E.}~\bibnamefont {Di~Valentino}},\ }\bibfield  {title}
  {\bibinfo {title} {{Teleparallel gravity: from theory to cosmology}},\ }\href
  {https://doi.org/10.1088/1361-6633/ac9cef} {\bibfield  {journal} {\bibinfo
  {journal} {Rept. Prog. Phys.}\ }\textbf {\bibinfo {volume} {86}},\ \bibinfo
  {pages} {026901} (\bibinfo {year} {2023})},\ \Eprint
  {https://arxiv.org/abs/2106.13793} {arXiv:2106.13793 [gr-qc]} \BibitemShut
  {NoStop}%
\bibitem [{\citenamefont {Blagojevi\'c}\ and\ \citenamefont
  {Cvetkovi\'c}(2013{\natexlab{a}})}]{Blagojevic:2013dea}%
  \BibitemOpen
  \bibfield  {author} {\bibinfo {author} {\bibfnamefont {M.}~\bibnamefont
  {Blagojevi\'c}}\ and\ \bibinfo {author} {\bibfnamefont {B.}~\bibnamefont
  {Cvetkovi\'c}},\ }\bibfield  {title} {\bibinfo {title} {{Three-dimensional
  gravity with propagating torsion: Hamiltonian structure of the scalar
  sector}},\ }\href {https://doi.org/10.1103/PhysRevD.88.104032} {\bibfield
  {journal} {\bibinfo  {journal} {Phys. Rev. D}\ }\textbf {\bibinfo {volume}
  {88}},\ \bibinfo {pages} {104032} (\bibinfo {year} {2013}{\natexlab{a}})},\
  \Eprint {https://arxiv.org/abs/1309.0411} {arXiv:1309.0411 [gr-qc]}
  \BibitemShut {NoStop}%
\bibitem [{\citenamefont {Blagojevi\'c}\ and\ \citenamefont
  {Cvetkovi\'c}(2013{\natexlab{b}})}]{Blagojevic:2013taa}%
  \BibitemOpen
  \bibfield  {author} {\bibinfo {author} {\bibfnamefont {M.}~\bibnamefont
  {Blagojevi\'c}}\ and\ \bibinfo {author} {\bibfnamefont {B.}~\bibnamefont
  {Cvetkovi\'c}},\ }\bibfield  {title} {\bibinfo {title} {{Poincar\'e gauge
  theory in 3D: canonical stability of the scalar sector}},\ }\href@noop {} {\
  (\bibinfo {year} {2013}{\natexlab{b}})},\ \Eprint
  {https://arxiv.org/abs/1310.8309} {arXiv:1310.8309 [gr-qc]} \BibitemShut
  {NoStop}%
\bibitem [{\citenamefont {Aoki}(2020)}]{Aoki:2020rae}%
  \BibitemOpen
  \bibfield  {author} {\bibinfo {author} {\bibfnamefont {K.}~\bibnamefont
  {Aoki}},\ }\bibfield  {title} {\bibinfo {title} {{Nonlinearly ghost-free
  higher curvature gravity}},\ }\href
  {https://doi.org/10.1103/PhysRevD.102.124049} {\bibfield  {journal} {\bibinfo
   {journal} {Phys. Rev. D}\ }\textbf {\bibinfo {volume} {102}},\ \bibinfo
  {pages} {124049} (\bibinfo {year} {2020})},\ \Eprint
  {https://arxiv.org/abs/2009.11739} {arXiv:2009.11739 [hep-th]} \BibitemShut
  {NoStop}%
\bibitem [{\citenamefont {Beltr\'an~Jim\'enez}\ \emph
  {et~al.}(2019)\citenamefont {Beltr\'an~Jim\'enez}, \citenamefont
  {Heisenberg},\ and\ \citenamefont {Koivisto}}]{BeltranJimenez:2019esp}%
  \BibitemOpen
  \bibfield  {author} {\bibinfo {author} {\bibfnamefont {J.}~\bibnamefont
  {Beltr\'an~Jim\'enez}}, \bibinfo {author} {\bibfnamefont {L.}~\bibnamefont
  {Heisenberg}},\ and\ \bibinfo {author} {\bibfnamefont {T.~S.}\ \bibnamefont
  {Koivisto}},\ }\bibfield  {title} {\bibinfo {title} {{The Geometrical Trinity
  of Gravity}},\ }\href {https://doi.org/10.3390/universe5070173} {\bibfield
  {journal} {\bibinfo  {journal} {Universe}\ }\textbf {\bibinfo {volume} {5}},\
  \bibinfo {pages} {173} (\bibinfo {year} {2019})},\ \Eprint
  {https://arxiv.org/abs/1903.06830} {arXiv:1903.06830 [hep-th]} \BibitemShut
  {NoStop}%
\bibitem [{\citenamefont {Percacci}\ and\ \citenamefont
  {Sezgin}(2020{\natexlab{b}})}]{Percacci:2020ddy}%
  \BibitemOpen
  \bibfield  {author} {\bibinfo {author} {\bibfnamefont {R.}~\bibnamefont
  {Percacci}}\ and\ \bibinfo {author} {\bibfnamefont {E.}~\bibnamefont
  {Sezgin}},\ }\bibfield  {title} {\bibinfo {title} {{New class of ghost- and
  tachyon-free metric affine gravities}},\ }\href
  {https://doi.org/10.1103/PhysRevD.101.084040} {\bibfield  {journal} {\bibinfo
   {journal} {Phys. Rev. D}\ }\textbf {\bibinfo {volume} {101}},\ \bibinfo
  {pages} {084040} (\bibinfo {year} {2020}{\natexlab{b}})},\ \Eprint
  {https://arxiv.org/abs/1912.01023} {arXiv:1912.01023 [hep-th]} \BibitemShut
  {NoStop}%
\bibitem [{\citenamefont {Piva}(2022)}]{Piva:2021nyj}%
  \BibitemOpen
  \bibfield  {author} {\bibinfo {author} {\bibfnamefont {M.}~\bibnamefont
  {Piva}},\ }\bibfield  {title} {\bibinfo {title} {{Massive higher-spin
  multiplets and asymptotic freedom in quantum gravity}},\ }\href
  {https://doi.org/10.1103/PhysRevD.105.045006} {\bibfield  {journal} {\bibinfo
   {journal} {Phys. Rev. D}\ }\textbf {\bibinfo {volume} {105}},\ \bibinfo
  {pages} {045006} (\bibinfo {year} {2022})},\ \Eprint
  {https://arxiv.org/abs/2110.09649} {arXiv:2110.09649 [hep-th]} \BibitemShut
  {NoStop}%
\bibitem [{\citenamefont {Iosifidis}\ \emph {et~al.}(2022)\citenamefont
  {Iosifidis}, \citenamefont {Myrzakulov}, \citenamefont {Ravera},
  \citenamefont {Yergaliyeva},\ and\ \citenamefont
  {Yerzhanov}}]{Iosifidis:2021xdx}%
  \BibitemOpen
  \bibfield  {author} {\bibinfo {author} {\bibfnamefont {D.}~\bibnamefont
  {Iosifidis}}, \bibinfo {author} {\bibfnamefont {R.}~\bibnamefont
  {Myrzakulov}}, \bibinfo {author} {\bibfnamefont {L.}~\bibnamefont {Ravera}},
  \bibinfo {author} {\bibfnamefont {G.}~\bibnamefont {Yergaliyeva}},\ and\
  \bibinfo {author} {\bibfnamefont {K.}~\bibnamefont {Yerzhanov}},\ }\bibfield
  {title} {\bibinfo {title} {{Metric-Affine Vector\textendash{}Tensor
  correspondence and implications in F(R,T,Q,T,D) gravity}},\ }\href
  {https://doi.org/10.1016/j.dark.2022.101094} {\bibfield  {journal} {\bibinfo
  {journal} {Phys. Dark Univ.}\ }\textbf {\bibinfo {volume} {37}},\ \bibinfo
  {pages} {101094} (\bibinfo {year} {2022})},\ \Eprint
  {https://arxiv.org/abs/2111.14214} {arXiv:2111.14214 [gr-qc]} \BibitemShut
  {NoStop}%
\bibitem [{\citenamefont {Jim\'enez-Cano}\ and\ \citenamefont
  {Maldonado~Torralba}(2022)}]{Jimenez-Cano:2022sds}%
  \BibitemOpen
  \bibfield  {author} {\bibinfo {author} {\bibfnamefont {A.}~\bibnamefont
  {Jim\'enez-Cano}}\ and\ \bibinfo {author} {\bibfnamefont {F.~J.}\
  \bibnamefont {Maldonado~Torralba}},\ }\bibfield  {title} {\bibinfo {title}
  {{Vector stability in quadratic metric-affine theories}},\ }\href
  {https://doi.org/10.1088/1475-7516/2022/09/044} {\bibfield  {journal}
  {\bibinfo  {journal} {JCAP}\ }\textbf {\bibinfo {volume} {09}},\ \bibinfo
  {pages} {044}},\ \Eprint {https://arxiv.org/abs/2205.05674} {arXiv:2205.05674
  [gr-qc]} \BibitemShut {NoStop}%
\bibitem [{\citenamefont {Iosifidis}\ and\ \citenamefont
  {Pallikaris}(2023)}]{Iosifidis:2023pvz}%
  \BibitemOpen
  \bibfield  {author} {\bibinfo {author} {\bibfnamefont {D.}~\bibnamefont
  {Iosifidis}}\ and\ \bibinfo {author} {\bibfnamefont {K.}~\bibnamefont
  {Pallikaris}},\ }\bibfield  {title} {\bibinfo {title} {{Describing
  metric-affine theories anew: alternative frameworks, examples and
  solutions}},\ }\href {https://doi.org/10.1088/1475-7516/2023/05/037}
  {\bibfield  {journal} {\bibinfo  {journal} {JCAP}\ }\textbf {\bibinfo
  {volume} {05}},\ \bibinfo {pages} {037}},\ \Eprint
  {https://arxiv.org/abs/2301.11364} {arXiv:2301.11364 [gr-qc]} \BibitemShut
  {NoStop}%
\bibitem [{\citenamefont {Lasenby}\ and\ \citenamefont
  {Hobson}(2016)}]{Lasenby:2015dba}%
  \BibitemOpen
  \bibfield  {author} {\bibinfo {author} {\bibfnamefont {A.}~\bibnamefont
  {Lasenby}}\ and\ \bibinfo {author} {\bibfnamefont {M.}~\bibnamefont
  {Hobson}},\ }\bibfield  {title} {\bibinfo {title} {{Scale-invariant gauge
  theories of gravity: theoretical foundations}},\ }\href
  {https://doi.org/10.1063/1.4963143} {\bibfield  {journal} {\bibinfo
  {journal} {J. Math. Phys.}\ }\textbf {\bibinfo {volume} {57}},\ \bibinfo
  {pages} {092505} (\bibinfo {year} {2016})},\ \Eprint
  {https://arxiv.org/abs/1510.06699} {arXiv:1510.06699 [gr-qc]} \BibitemShut
  {NoStop}%
\bibitem [{\citenamefont {Hobson}\ and\ \citenamefont
  {Lasenby}(2021)}]{Hobson:2020bms}%
  \BibitemOpen
  \bibfield  {author} {\bibinfo {author} {\bibfnamefont {M.~P.}\ \bibnamefont
  {Hobson}}\ and\ \bibinfo {author} {\bibfnamefont {A.~N.}\ \bibnamefont
  {Lasenby}},\ }\bibfield  {title} {\bibinfo {title} {{Fresh perspective on
  gauging the conformal group}},\ }\href
  {https://doi.org/10.1103/PhysRevD.103.104042} {\bibfield  {journal} {\bibinfo
   {journal} {Phys. Rev. D}\ }\textbf {\bibinfo {volume} {103}},\ \bibinfo
  {pages} {104042} (\bibinfo {year} {2021})},\ \Eprint
  {https://arxiv.org/abs/2008.09053} {arXiv:2008.09053 [gr-qc]} \BibitemShut
  {NoStop}%
\bibitem [{\citenamefont {Hobson}\ and\ \citenamefont
  {Lasenby}(2020)}]{Hobson:2020doi}%
  \BibitemOpen
  \bibfield  {author} {\bibinfo {author} {\bibfnamefont {M.~P.}\ \bibnamefont
  {Hobson}}\ and\ \bibinfo {author} {\bibfnamefont {A.~N.}\ \bibnamefont
  {Lasenby}},\ }\bibfield  {title} {\bibinfo {title} {{Weyl gauge theories of
  gravity do not predict a second clock effect}},\ }\href
  {https://doi.org/10.1103/PhysRevD.102.084040} {\bibfield  {journal} {\bibinfo
   {journal} {Phys. Rev. D}\ }\textbf {\bibinfo {volume} {102}},\ \bibinfo
  {pages} {084040} (\bibinfo {year} {2020})},\ \Eprint
  {https://arxiv.org/abs/2009.06407} {arXiv:2009.06407 [gr-qc]} \BibitemShut
  {NoStop}%
\bibitem [{\citenamefont {Hobson}\ and\ \citenamefont
  {Lasenby}(2022)}]{Hobson:2021iwy}%
  \BibitemOpen
  \bibfield  {author} {\bibinfo {author} {\bibfnamefont {M.}~\bibnamefont
  {Hobson}}\ and\ \bibinfo {author} {\bibfnamefont {A.}~\bibnamefont
  {Lasenby}},\ }\bibfield  {title} {\bibinfo {title} {{Note on the absence of
  the second clock effect in Weyl gauge theories of gravity}},\ }\href
  {https://doi.org/10.1103/PhysRevD.105.L021501} {\bibfield  {journal}
  {\bibinfo  {journal} {Phys. Rev. D}\ }\textbf {\bibinfo {volume} {105}},\
  \bibinfo {pages} {L021501} (\bibinfo {year} {2022})},\ \Eprint
  {https://arxiv.org/abs/2112.09967} {arXiv:2112.09967 [gr-qc]} \BibitemShut
  {NoStop}%
\bibitem [{\citenamefont {Hobson}\ \emph {et~al.}(2024)\citenamefont {Hobson},
  \citenamefont {Lasenby},\ and\ \citenamefont {Barker}}]{Hobson:2023rdw}%
  \BibitemOpen
  \bibfield  {author} {\bibinfo {author} {\bibfnamefont {M.}~\bibnamefont
  {Hobson}}, \bibinfo {author} {\bibfnamefont {A.}~\bibnamefont {Lasenby}},\
  and\ \bibinfo {author} {\bibfnamefont {W.~E.~V.}\ \bibnamefont {Barker}},\
  }\bibfield  {title} {\bibinfo {title} {{Manifestly covariant variational
  principle for gauge theories of gravity}},\ }\href
  {https://doi.org/10.1103/PhysRevD.109.024022} {\bibfield  {journal} {\bibinfo
   {journal} {Phys. Rev. D}\ }\textbf {\bibinfo {volume} {109}},\ \bibinfo
  {pages} {024022} (\bibinfo {year} {2024})},\ \Eprint
  {https://arxiv.org/abs/2309.14783} {arXiv:2309.14783 [gr-qc]} \BibitemShut
  {NoStop}%
\bibitem [{\citenamefont {Hobson}\ \emph {et~al.}(2006)\citenamefont {Hobson},
  \citenamefont {Efstathiou},\ and\ \citenamefont {Lasenby}}]{Hobson:2006se}%
  \BibitemOpen
  \bibfield  {author} {\bibinfo {author} {\bibfnamefont {M.~P.}\ \bibnamefont
  {Hobson}}, \bibinfo {author} {\bibfnamefont {G.~P.}\ \bibnamefont
  {Efstathiou}},\ and\ \bibinfo {author} {\bibfnamefont {A.~N.}\ \bibnamefont
  {Lasenby}},\ }\href@noop {} {\emph {\bibinfo {title} {{General relativity: An
  introduction for physicists}}}}\ (\bibinfo {year} {2006})\BibitemShut
  {NoStop}%
\bibitem [{\citenamefont {V\'azquez}\ \emph {et~al.}(2020)\citenamefont
  {V\'azquez}, \citenamefont {Padilla},\ and\ \citenamefont
  {Matos}}]{Vazquez:2018qdg}%
  \BibitemOpen
  \bibfield  {author} {\bibinfo {author} {\bibfnamefont {J.~A.}\ \bibnamefont
  {V\'azquez}}, \bibinfo {author} {\bibfnamefont {L.~E.}\ \bibnamefont
  {Padilla}},\ and\ \bibinfo {author} {\bibfnamefont {T.}~\bibnamefont
  {Matos}},\ }\bibfield  {title} {\bibinfo {title} {{Inflationary cosmology:
  from theory to observations}},\ }\href
  {https://doi.org/10.31349/RevMexFisE.17.73} {\bibfield  {journal} {\bibinfo
  {journal} {Rev. Mex. Fis. E}\ }\textbf {\bibinfo {volume} {17}},\ \bibinfo
  {pages} {73} (\bibinfo {year} {2020})},\ \Eprint
  {https://arxiv.org/abs/1810.09934} {arXiv:1810.09934 [astro-ph.CO]}
  \BibitemShut {NoStop}%
\bibitem [{\citenamefont {Ach\'ucarro}\ \emph {et~al.}(2022)\citenamefont
  {Ach\'ucarro} \emph {et~al.}}]{Achucarro:2022qrl}%
  \BibitemOpen
  \bibfield  {author} {\bibinfo {author} {\bibfnamefont {A.}~\bibnamefont
  {Ach\'ucarro}} \emph {et~al.},\ }\bibfield  {title} {\bibinfo {title}
  {{Inflation: Theory and Observations}},\ }\href@noop {} {\  (\bibinfo {year}
  {2022})},\ \Eprint {https://arxiv.org/abs/2203.08128} {arXiv:2203.08128
  [astro-ph.CO]} \BibitemShut {NoStop}%
\bibitem [{\citenamefont {Bezrukov}\ and\ \citenamefont
  {Shaposhnikov}(2008)}]{Bezrukov:2007ep}%
  \BibitemOpen
  \bibfield  {author} {\bibinfo {author} {\bibfnamefont {F.~L.}\ \bibnamefont
  {Bezrukov}}\ and\ \bibinfo {author} {\bibfnamefont {M.}~\bibnamefont
  {Shaposhnikov}},\ }\bibfield  {title} {\bibinfo {title} {{The Standard Model
  Higgs boson as the inflaton}},\ }\href
  {https://doi.org/10.1016/j.physletb.2007.11.072} {\bibfield  {journal}
  {\bibinfo  {journal} {Phys. Lett. B}\ }\textbf {\bibinfo {volume} {659}},\
  \bibinfo {pages} {703} (\bibinfo {year} {2008})},\ \Eprint
  {https://arxiv.org/abs/0710.3755} {arXiv:0710.3755 [hep-th]} \BibitemShut
  {NoStop}%
\bibitem [{\citenamefont {Allahverdi}\ \emph {et~al.}(2010)\citenamefont
  {Allahverdi}, \citenamefont {Brandenberger}, \citenamefont {Cyr-Racine},\
  and\ \citenamefont {Mazumdar}}]{Allahverdi:2010xz}%
  \BibitemOpen
  \bibfield  {author} {\bibinfo {author} {\bibfnamefont {R.}~\bibnamefont
  {Allahverdi}}, \bibinfo {author} {\bibfnamefont {R.}~\bibnamefont
  {Brandenberger}}, \bibinfo {author} {\bibfnamefont {F.-Y.}\ \bibnamefont
  {Cyr-Racine}},\ and\ \bibinfo {author} {\bibfnamefont {A.}~\bibnamefont
  {Mazumdar}},\ }\bibfield  {title} {\bibinfo {title} {{Reheating in
  Inflationary Cosmology: Theory and Applications}},\ }\href
  {https://doi.org/10.1146/annurev.nucl.012809.104511} {\bibfield  {journal}
  {\bibinfo  {journal} {Ann. Rev. Nucl. Part. Sci.}\ }\textbf {\bibinfo
  {volume} {60}},\ \bibinfo {pages} {27} (\bibinfo {year} {2010})},\ \Eprint
  {https://arxiv.org/abs/1001.2600} {arXiv:1001.2600 [hep-th]} \BibitemShut
  {NoStop}%
\bibitem [{\citenamefont {Ade}\ \emph {et~al.}(2014)\citenamefont {Ade} \emph
  {et~al.}}]{Planck:2013jfk}%
  \BibitemOpen
  \bibfield  {author} {\bibinfo {author} {\bibfnamefont {P.~A.~R.}\
  \bibnamefont {Ade}} \emph {et~al.} (\bibinfo {collaboration} {Planck}),\
  }\bibfield  {title} {\bibinfo {title} {{Planck 2013 results. XXII.
  Constraints on inflation}},\ }\href
  {https://doi.org/10.1051/0004-6361/201321569} {\bibfield  {journal} {\bibinfo
   {journal} {Astron. Astrophys.}\ }\textbf {\bibinfo {volume} {571}},\
  \bibinfo {pages} {A22} (\bibinfo {year} {2014})},\ \Eprint
  {https://arxiv.org/abs/1303.5082} {arXiv:1303.5082 [astro-ph.CO]}
  \BibitemShut {NoStop}%
\bibitem [{\citenamefont {Ade}\ \emph {et~al.}(2016)\citenamefont {Ade} \emph
  {et~al.}}]{Planck:2015sxf}%
  \BibitemOpen
  \bibfield  {author} {\bibinfo {author} {\bibfnamefont {P.~A.~R.}\
  \bibnamefont {Ade}} \emph {et~al.} (\bibinfo {collaboration} {Planck}),\
  }\bibfield  {title} {\bibinfo {title} {{Planck 2015 results. XX. Constraints
  on inflation}},\ }\href {https://doi.org/10.1051/0004-6361/201525898}
  {\bibfield  {journal} {\bibinfo  {journal} {Astron. Astrophys.}\ }\textbf
  {\bibinfo {volume} {594}},\ \bibinfo {pages} {A20} (\bibinfo {year}
  {2016})},\ \Eprint {https://arxiv.org/abs/1502.02114} {arXiv:1502.02114
  [astro-ph.CO]} \BibitemShut {NoStop}%
\bibitem [{\citenamefont {Akrami}\ \emph {et~al.}(2020)\citenamefont {Akrami}
  \emph {et~al.}}]{Planck:2018jri}%
  \BibitemOpen
  \bibfield  {author} {\bibinfo {author} {\bibfnamefont {Y.}~\bibnamefont
  {Akrami}} \emph {et~al.} (\bibinfo {collaboration} {Planck}),\ }\bibfield
  {title} {\bibinfo {title} {{Planck 2018 results. X. Constraints on
  inflation}},\ }\href {https://doi.org/10.1051/0004-6361/201833887} {\bibfield
   {journal} {\bibinfo  {journal} {Astron. Astrophys.}\ }\textbf {\bibinfo
  {volume} {641}},\ \bibinfo {pages} {A10} (\bibinfo {year} {2020})},\ \Eprint
  {https://arxiv.org/abs/1807.06211} {arXiv:1807.06211 [astro-ph.CO]}
  \BibitemShut {NoStop}%
\bibitem [{\citenamefont {Georgi}(1984)}]{Georgi:1984zwz}%
  \BibitemOpen
  \bibfield  {author} {\bibinfo {author} {\bibfnamefont {H.}~\bibnamefont
  {Georgi}},\ }\href@noop {} {\emph {\bibinfo {title} {{Weak Interactions and
  Modern Particle Theory}}}}\ (\bibinfo {year} {1984})\BibitemShut {NoStop}%
\bibitem [{\citenamefont {Donoghue}\ \emph {et~al.}(2014)\citenamefont
  {Donoghue}, \citenamefont {Golowich},\ and\ \citenamefont
  {Holstein}}]{Donoghue:1992dd}%
  \BibitemOpen
  \bibfield  {author} {\bibinfo {author} {\bibfnamefont {J.~F.}\ \bibnamefont
  {Donoghue}}, \bibinfo {author} {\bibfnamefont {E.}~\bibnamefont {Golowich}},\
  and\ \bibinfo {author} {\bibfnamefont {B.~R.}\ \bibnamefont {Holstein}},\
  }\href {https://doi.org/10.1017/CBO9780511524370} {\emph {\bibinfo {title}
  {{Dynamics of the standard model}}}},\ Vol.~\bibinfo {volume} {2}\ (\bibinfo
  {publisher} {CUP},\ \bibinfo {year} {2014})\BibitemShut {NoStop}%
\bibitem [{\citenamefont {Manohar}\ and\ \citenamefont
  {Wise}(2000)}]{Manohar:2000dt}%
  \BibitemOpen
  \bibfield  {author} {\bibinfo {author} {\bibfnamefont {A.~V.}\ \bibnamefont
  {Manohar}}\ and\ \bibinfo {author} {\bibfnamefont {M.~B.}\ \bibnamefont
  {Wise}},\ }\href {https://doi.org/10.1017/9781009402125} {\emph {\bibinfo
  {title} {{Heavy quark physics}}}},\ Vol.~\bibinfo {volume} {10}\ (\bibinfo
  {year} {2000})\BibitemShut {NoStop}%
\bibitem [{\citenamefont {Scherer}\ and\ \citenamefont
  {Schindler}(2012)}]{Scherer:2012xha}%
  \BibitemOpen
  \bibfield  {author} {\bibinfo {author} {\bibfnamefont {S.}~\bibnamefont
  {Scherer}}\ and\ \bibinfo {author} {\bibfnamefont {M.~R.}\ \bibnamefont
  {Schindler}},\ }\href {https://doi.org/10.1007/978-3-642-19254-8} {\emph
  {\bibinfo {title} {{A Primer for Chiral Perturbation Theory}}}},\ Vol.\
  \bibinfo {volume} {830}\ (\bibinfo {year} {2012})\BibitemShut {NoStop}%
\bibitem [{\citenamefont {Higgs}(1964)}]{Higgs:1964pj}%
  \BibitemOpen
  \bibfield  {author} {\bibinfo {author} {\bibfnamefont {P.~W.}\ \bibnamefont
  {Higgs}},\ }\bibfield  {title} {\bibinfo {title} {{Broken Symmetries and the
  Masses of Gauge Bosons}},\ }\href
  {https://doi.org/10.1103/PhysRevLett.13.508} {\bibfield  {journal} {\bibinfo
  {journal} {Phys. Rev. Lett.}\ }\textbf {\bibinfo {volume} {13}},\ \bibinfo
  {pages} {508} (\bibinfo {year} {1964})}\BibitemShut {NoStop}%
\bibitem [{\citenamefont {Guralnik}\ \emph {et~al.}(1964)\citenamefont
  {Guralnik}, \citenamefont {Hagen},\ and\ \citenamefont
  {Kibble}}]{Guralnik:1964eu}%
  \BibitemOpen
  \bibfield  {author} {\bibinfo {author} {\bibfnamefont {G.~S.}\ \bibnamefont
  {Guralnik}}, \bibinfo {author} {\bibfnamefont {C.~R.}\ \bibnamefont
  {Hagen}},\ and\ \bibinfo {author} {\bibfnamefont {T.~W.~B.}\ \bibnamefont
  {Kibble}},\ }\bibfield  {title} {\bibinfo {title} {{Global Conservation Laws
  and Massless Particles}},\ }\href
  {https://doi.org/10.1103/PhysRevLett.13.585} {\bibfield  {journal} {\bibinfo
  {journal} {Phys. Rev. Lett.}\ }\textbf {\bibinfo {volume} {13}},\ \bibinfo
  {pages} {585} (\bibinfo {year} {1964})}\BibitemShut {NoStop}%
\bibitem [{\citenamefont {Henneaux}\ and\ \citenamefont
  {Teitelboim}(1992)}]{Henneaux:1992ig}%
  \BibitemOpen
  \bibfield  {author} {\bibinfo {author} {\bibfnamefont {M.}~\bibnamefont
  {Henneaux}}\ and\ \bibinfo {author} {\bibfnamefont {C.}~\bibnamefont
  {Teitelboim}},\ }\href@noop {} {\emph {\bibinfo {title} {{Quantization of
  gauge systems}}}}\ (\bibinfo {year} {1992})\BibitemShut {NoStop}%
\bibitem [{\citenamefont {Chatrchyan}\ \emph {et~al.}(2012)\citenamefont
  {Chatrchyan} \emph {et~al.}}]{CMS:2012qbp}%
  \BibitemOpen
  \bibfield  {author} {\bibinfo {author} {\bibfnamefont {S.}~\bibnamefont
  {Chatrchyan}} \emph {et~al.} (\bibinfo {collaboration} {CMS}),\ }\bibfield
  {title} {\bibinfo {title} {{Observation of a New Boson at a Mass of 125 GeV
  with the CMS Experiment at the LHC}},\ }\href
  {https://doi.org/10.1016/j.physletb.2012.08.021} {\bibfield  {journal}
  {\bibinfo  {journal} {Phys. Lett. B}\ }\textbf {\bibinfo {volume} {716}},\
  \bibinfo {pages} {30} (\bibinfo {year} {2012})},\ \Eprint
  {https://arxiv.org/abs/1207.7235} {arXiv:1207.7235 [hep-ex]} \BibitemShut
  {NoStop}%
\bibitem [{\citenamefont {Barker}\ and\ \citenamefont
  {Zell}(2024{\natexlab{b}})}]{Barker:2024dhb}%
  \BibitemOpen
  \bibfield  {author} {\bibinfo {author} {\bibfnamefont {W.}~\bibnamefont
  {Barker}}\ and\ \bibinfo {author} {\bibfnamefont {S.}~\bibnamefont {Zell}},\
  }\bibfield  {title} {\bibinfo {title} {{Consistent particle physics in
  metric-affine gravity from extended projective symmetry}},\ }\href@noop {} {\
   (\bibinfo {year} {2024}{\natexlab{b}})},\ \Eprint
  {https://arxiv.org/abs/2402.14917} {arXiv:2402.14917 [hep-th]} \BibitemShut
  {NoStop}%
\bibitem [{\citenamefont {Karananas}(2025{\natexlab{b}})}]{Karananas:2025xcv}%
  \BibitemOpen
  \bibfield  {author} {\bibinfo {author} {\bibfnamefont {G.~K.}\ \bibnamefont
  {Karananas}},\ }\bibfield  {title} {\bibinfo {title} {{Geometrical origin of
  inflation in Weyl-invariant Einstein-Cartan gravity}},\ }\href
  {https://doi.org/10.1016/j.physletb.2025.139343} {\bibfield  {journal}
  {\bibinfo  {journal} {Phys. Lett. B}\ }\textbf {\bibinfo {volume} {862}},\
  \bibinfo {pages} {139343} (\bibinfo {year} {2025}{\natexlab{b}})},\ \Eprint
  {https://arxiv.org/abs/2501.16416} {arXiv:2501.16416 [gr-qc]} \BibitemShut
  {NoStop}%
\bibitem [{\citenamefont {Ade}\ \emph {et~al.}(2021)\citenamefont {Ade} \emph
  {et~al.}}]{BICEP:2021xfz}%
  \BibitemOpen
  \bibfield  {author} {\bibinfo {author} {\bibfnamefont {P.~A.~R.}\
  \bibnamefont {Ade}} \emph {et~al.} (\bibinfo {collaboration} {BICEP, Keck}),\
  }\bibfield  {title} {\bibinfo {title} {{Improved Constraints on Primordial
  Gravitational Waves using Planck, WMAP, and BICEP/Keck Observations through
  the 2018 Observing Season}},\ }\href
  {https://doi.org/10.1103/PhysRevLett.127.151301} {\bibfield  {journal}
  {\bibinfo  {journal} {Phys. Rev. Lett.}\ }\textbf {\bibinfo {volume} {127}},\
  \bibinfo {pages} {151301} (\bibinfo {year} {2021})},\ \Eprint
  {https://arxiv.org/abs/2110.00483} {arXiv:2110.00483 [astro-ph.CO]}
  \BibitemShut {NoStop}%
\bibitem [{\citenamefont {Laureijs}\ \emph {et~al.}(2011)\citenamefont
  {Laureijs} \emph {et~al.}}]{arXiv:1110.3193}%
  \BibitemOpen
  \bibfield  {author} {\bibinfo {author} {\bibfnamefont {R.}~\bibnamefont
  {Laureijs}} \emph {et~al.} (\bibinfo {collaboration} {EUCLID}),\ }\bibfield
  {title} {\bibinfo {title} {{Euclid Definition Study Report}},\ }\href@noop {}
  {\  (\bibinfo {year} {2011})},\ \Eprint {https://arxiv.org/abs/1110.3193}
  {arXiv:1110.3193 [astro-ph.CO]} \BibitemShut {NoStop}%
\bibitem [{\citenamefont {Weltman}\ \emph {et~al.}(2020)\citenamefont {Weltman}
  \emph {et~al.}}]{Weltman:2018zrl}%
  \BibitemOpen
  \bibfield  {author} {\bibinfo {author} {\bibfnamefont {A.}~\bibnamefont
  {Weltman}} \emph {et~al.},\ }\bibfield  {title} {\bibinfo {title}
  {{Fundamental physics with the Square Kilometre Array}},\ }\href
  {https://doi.org/10.1017/pasa.2019.42} {\bibfield  {journal} {\bibinfo
  {journal} {Publ. Astron. Soc. Austral.}\ }\textbf {\bibinfo {volume} {37}},\
  \bibinfo {pages} {e002} (\bibinfo {year} {2020})},\ \Eprint
  {https://arxiv.org/abs/1810.02680} {arXiv:1810.02680 [astro-ph.CO]}
  \BibitemShut {NoStop}%
\bibitem [{\citenamefont {Abazajian}\ \emph {et~al.}(2019)\citenamefont
  {Abazajian} \emph {et~al.}}]{Abazajian:2019eic}%
  \BibitemOpen
  \bibfield  {author} {\bibinfo {author} {\bibfnamefont {K.}~\bibnamefont
  {Abazajian}} \emph {et~al.},\ }\bibfield  {title} {\bibinfo {title} {{CMB-S4
  Science Case, Reference Design, and Project Plan}},\ }\href@noop {} {\
  (\bibinfo {year} {2019})},\ \Eprint {https://arxiv.org/abs/1907.04473}
  {arXiv:1907.04473 [astro-ph.IM]} \BibitemShut {NoStop}%
\bibitem [{\citenamefont {Allys}\ \emph {et~al.}(2023)\citenamefont {Allys}
  \emph {et~al.}}]{LiteBIRD:2022cnt}%
  \BibitemOpen
  \bibfield  {author} {\bibinfo {author} {\bibfnamefont {E.}~\bibnamefont
  {Allys}} \emph {et~al.} (\bibinfo {collaboration} {LiteBIRD}),\ }\bibfield
  {title} {\bibinfo {title} {{Probing Cosmic Inflation with the LiteBIRD Cosmic
  Microwave Background Polarization Survey}},\ }\href
  {https://doi.org/10.1093/ptep/ptac150} {\bibfield  {journal} {\bibinfo
  {journal} {PTEP}\ }\textbf {\bibinfo {volume} {2023}},\ \bibinfo {pages}
  {042F01} (\bibinfo {year} {2023})},\ \Eprint
  {https://arxiv.org/abs/2202.02773} {arXiv:2202.02773 [astro-ph.IM]}
  \BibitemShut {NoStop}%
\bibitem [{\citenamefont {Handley}\ \emph {et~al.}(2019)\citenamefont
  {Handley}, \citenamefont {Lasenby}, \citenamefont {Peiris},\ and\
  \citenamefont {Hobson}}]{Handley:2019fll}%
  \BibitemOpen
  \bibfield  {author} {\bibinfo {author} {\bibfnamefont {W.~J.}\ \bibnamefont
  {Handley}}, \bibinfo {author} {\bibfnamefont {A.~N.}\ \bibnamefont
  {Lasenby}}, \bibinfo {author} {\bibfnamefont {H.~V.}\ \bibnamefont
  {Peiris}},\ and\ \bibinfo {author} {\bibfnamefont {M.~P.}\ \bibnamefont
  {Hobson}},\ }\bibfield  {title} {\bibinfo {title} {{Bayesian inflationary
  reconstructions from Planck 2018 data}},\ }\href
  {https://doi.org/10.1103/PhysRevD.100.103511} {\bibfield  {journal} {\bibinfo
   {journal} {Phys. Rev. D}\ }\textbf {\bibinfo {volume} {100}},\ \bibinfo
  {pages} {103511} (\bibinfo {year} {2019})},\ \Eprint
  {https://arxiv.org/abs/1908.00906} {arXiv:1908.00906 [astro-ph.CO]}
  \BibitemShut {NoStop}%
\bibitem [{\citenamefont {Weyl}(1918)}]{Weyl:1918ib}%
  \BibitemOpen
  \bibfield  {author} {\bibinfo {author} {\bibfnamefont {H.}~\bibnamefont
  {Weyl}},\ }\bibfield  {title} {\bibinfo {title} {{Gravitation and
  electricity}},\ }\href@noop {} {\bibfield  {journal} {\bibinfo  {journal}
  {Sitzungsber. Preuss. Akad. Wiss. Berlin (Math. Phys. )}\ }\textbf {\bibinfo
  {volume} {1918}},\ \bibinfo {pages} {465} (\bibinfo {year}
  {1918})}\BibitemShut {NoStop}%
\bibitem [{\citenamefont {Weyl}(1931)}]{Weyl:1931}%
  \BibitemOpen
  \bibfield  {author} {\bibinfo {author} {\bibfnamefont {H.}~\bibnamefont
  {Weyl}},\ }\bibfield  {title} {\bibinfo {title} {{Geometrie und Physik}},\
  }\href@noop {} {\bibfield  {journal} {\bibinfo  {journal}
  {Naturwissenschaften}\ }\textbf {\bibinfo {volume} {49}},\ \bibinfo {pages}
  {19} (\bibinfo {year} {1931})}\BibitemShut {NoStop}%
\bibitem [{\citenamefont {Obukhov}(1982)}]{Obukhov:1982zn}%
  \BibitemOpen
  \bibfield  {author} {\bibinfo {author} {\bibfnamefont {Y.~N.}\ \bibnamefont
  {Obukhov}},\ }\bibfield  {title} {\bibinfo {title} {{CONFORMAL INVARIANCE AND
  SPACE-TIME TORSION}},\ }\href {https://doi.org/10.1016/0375-9601(82)90037-8}
  {\bibfield  {journal} {\bibinfo  {journal} {Phys. Lett. A}\ }\textbf
  {\bibinfo {volume} {90}},\ \bibinfo {pages} {13} (\bibinfo {year}
  {1982})}\BibitemShut {NoStop}%
\bibitem [{\citenamefont {Karananas}\ \emph
  {et~al.}(2021{\natexlab{b}})\citenamefont {Karananas}, \citenamefont
  {Shaposhnikov}, \citenamefont {Shkerin},\ and\ \citenamefont
  {Zell}}]{Karananas:2021gco}%
  \BibitemOpen
  \bibfield  {author} {\bibinfo {author} {\bibfnamefont {G.~K.}\ \bibnamefont
  {Karananas}}, \bibinfo {author} {\bibfnamefont {M.}~\bibnamefont
  {Shaposhnikov}}, \bibinfo {author} {\bibfnamefont {A.}~\bibnamefont
  {Shkerin}},\ and\ \bibinfo {author} {\bibfnamefont {S.}~\bibnamefont
  {Zell}},\ }\bibfield  {title} {\bibinfo {title} {{Scale and Weyl invariance
  in Einstein-Cartan gravity}},\ }\href
  {https://doi.org/10.1103/PhysRevD.104.124014} {\bibfield  {journal} {\bibinfo
   {journal} {Phys. Rev. D}\ }\textbf {\bibinfo {volume} {104}},\ \bibinfo
  {pages} {124014} (\bibinfo {year} {2021}{\natexlab{b}})},\ \Eprint
  {https://arxiv.org/abs/2108.05897} {arXiv:2108.05897 [hep-th]} \BibitemShut
  {NoStop}%
\bibitem [{\citenamefont {{Barker}}\ \emph {et~al.}()\citenamefont {{Barker}},
  \citenamefont {{Hobson}}, \citenamefont {{Lasenby}}, \citenamefont {{Lin}},\
  and\ \citenamefont {{Wei}}}]{Barker:2024}%
  \BibitemOpen
  \bibfield  {author} {\bibinfo {author} {\bibfnamefont {W.}~\bibnamefont
  {{Barker}}}, \bibinfo {author} {\bibfnamefont {M.}~\bibnamefont {{Hobson}}},
  \bibinfo {author} {\bibfnamefont {A.}~\bibnamefont {{Lasenby}}}, \bibinfo
  {author} {\bibfnamefont {Y.-C.}\ \bibnamefont {{Lin}}},\ and\ \bibinfo
  {author} {\bibfnamefont {Z.}~\bibnamefont {{Wei}}},\ }\href@noop {} {\emph
  {\bibinfo {title} {{Supplemental materials hosted at
  \href{https://github.com/wevbarker/SupplementalMaterials-2406}{https://github.com/wevbarker/SupplementalMaterials-2406}}}}}\BibitemShut
  {NoStop}%
\bibitem [{\citenamefont {Curtright}(1985)}]{Curtright:1980yk}%
  \BibitemOpen
  \bibfield  {author} {\bibinfo {author} {\bibfnamefont {T.}~\bibnamefont
  {Curtright}},\ }\bibfield  {title} {\bibinfo {title} {{GENERALIZED GAUGE
  FIELDS}},\ }\href {https://doi.org/10.1016/0370-2693(85)91235-3} {\bibfield
  {journal} {\bibinfo  {journal} {Phys. Lett. B}\ }\textbf {\bibinfo {volume}
  {165}},\ \bibinfo {pages} {304} (\bibinfo {year} {1985})}\BibitemShut
  {NoStop}%
\bibitem [{\citenamefont {Capper}\ and\ \citenamefont
  {Duff}(1974)}]{Capper:1974ic}%
  \BibitemOpen
  \bibfield  {author} {\bibinfo {author} {\bibfnamefont {D.~M.}\ \bibnamefont
  {Capper}}\ and\ \bibinfo {author} {\bibfnamefont {M.~J.}\ \bibnamefont
  {Duff}},\ }\bibfield  {title} {\bibinfo {title} {{Trace anomalies in
  dimensional regularization}},\ }\href {https://doi.org/10.1007/BF02748300}
  {\bibfield  {journal} {\bibinfo  {journal} {Nuovo Cim. A}\ }\textbf {\bibinfo
  {volume} {23}},\ \bibinfo {pages} {173} (\bibinfo {year} {1974})}\BibitemShut
  {NoStop}%
\bibitem [{\citenamefont {Barker}\ \emph {et~al.}(2024)\citenamefont {Barker},
  \citenamefont {Marzo},\ and\ \citenamefont {Rigouzzo}}]{Barker:2024juc}%
  \BibitemOpen
  \bibfield  {author} {\bibinfo {author} {\bibfnamefont {W.}~\bibnamefont
  {Barker}}, \bibinfo {author} {\bibfnamefont {C.}~\bibnamefont {Marzo}},\ and\
  \bibinfo {author} {\bibfnamefont {C.}~\bibnamefont {Rigouzzo}},\ }\bibfield
  {title} {\bibinfo {title} {{PSALTer: Particle Spectrum for Any Tensor
  Lagrangian}},\ }\href@noop {} {\  (\bibinfo {year} {2024})},\ \Eprint
  {https://arxiv.org/abs/2406.09500} {arXiv:2406.09500 [hep-th]} \BibitemShut
  {NoStop}%
\bibitem [{\citenamefont
  {Mart\'\i{}n-Garc\'\i{}a}(2008)}]{Martin-Garcia:2008ysv}%
  \BibitemOpen
  \bibfield  {author} {\bibinfo {author} {\bibfnamefont {J.~M.}\ \bibnamefont
  {Mart\'\i{}n-Garc\'\i{}a}},\ }\bibfield  {title} {\bibinfo {title} {{xPerm:
  fast index canonicalization for tensor computer algebra}},\ }\href
  {https://doi.org/10.1016/j.cpc.2008.05.009} {\bibfield  {journal} {\bibinfo
  {journal} {Comput. Phys. Commun.}\ }\textbf {\bibinfo {volume} {179}},\
  \bibinfo {pages} {597} (\bibinfo {year} {2008})},\ \Eprint
  {https://arxiv.org/abs/0803.0862} {arXiv:0803.0862 [cs.SC]} \BibitemShut
  {NoStop}%
\bibitem [{\citenamefont {Barker}(2023{\natexlab{b}})}]{Barker:2023bmr}%
  \BibitemOpen
  \bibfield  {author} {\bibinfo {author} {\bibfnamefont {W.}~\bibnamefont
  {Barker}},\ }\bibfield  {title} {\bibinfo {title} {{Particle spectra of
  gravity based on internal symmetry of quantum fields}},\ }\href@noop {} {\
  (\bibinfo {year} {2023}{\natexlab{b}})},\ \Eprint
  {https://arxiv.org/abs/2311.11790} {arXiv:2311.11790 [hep-th]} \BibitemShut
  {NoStop}%
\bibitem [{\citenamefont {Blixt}\ \emph {et~al.}(2022)\citenamefont {Blixt},
  \citenamefont {Ferraro}, \citenamefont {Golovnev},\ and\ \citenamefont
  {Guzm\'an}}]{Blixt:2022rpl}%
  \BibitemOpen
  \bibfield  {author} {\bibinfo {author} {\bibfnamefont {D.}~\bibnamefont
  {Blixt}}, \bibinfo {author} {\bibfnamefont {R.}~\bibnamefont {Ferraro}},
  \bibinfo {author} {\bibfnamefont {A.}~\bibnamefont {Golovnev}},\ and\
  \bibinfo {author} {\bibfnamefont {M.-J.}\ \bibnamefont {Guzm\'an}},\
  }\bibfield  {title} {\bibinfo {title} {{Lorentz gauge-invariant variables in
  torsion-based theories of gravity}},\ }\href
  {https://doi.org/10.1103/PhysRevD.105.084029} {\bibfield  {journal} {\bibinfo
   {journal} {Phys. Rev. D}\ }\textbf {\bibinfo {volume} {105}},\ \bibinfo
  {pages} {084029} (\bibinfo {year} {2022})},\ \Eprint
  {https://arxiv.org/abs/2201.11102} {arXiv:2201.11102 [gr-qc]} \BibitemShut
  {NoStop}%
\bibitem [{\citenamefont {Blixt}\ \emph {et~al.}(2023)\citenamefont {Blixt},
  \citenamefont {Hohmann}, \citenamefont {Koivisto},\ and\ \citenamefont
  {Marzola}}]{Blixt:2023qbg}%
  \BibitemOpen
  \bibfield  {author} {\bibinfo {author} {\bibfnamefont {D.}~\bibnamefont
  {Blixt}}, \bibinfo {author} {\bibfnamefont {M.}~\bibnamefont {Hohmann}},
  \bibinfo {author} {\bibfnamefont {T.}~\bibnamefont {Koivisto}},\ and\
  \bibinfo {author} {\bibfnamefont {L.}~\bibnamefont {Marzola}},\ }\bibfield
  {title} {\bibinfo {title} {{Teleparallel bigravity}},\ }\href
  {https://doi.org/10.1140/epjc/s10052-023-12247-7} {\bibfield  {journal}
  {\bibinfo  {journal} {Eur. Phys. J. C}\ }\textbf {\bibinfo {volume} {83}},\
  \bibinfo {pages} {1120} (\bibinfo {year} {2023})},\ \Eprint
  {https://arxiv.org/abs/2305.03504} {arXiv:2305.03504 [gr-qc]} \BibitemShut
  {NoStop}%
\end{thebibliography}%

\appendix
\section{Detailed tree-level spectra}\label{ParticleSpectra}

\begin{figure*}[h]
\includegraphics[width=\textwidth,height=\textheight,keepaspectratio]{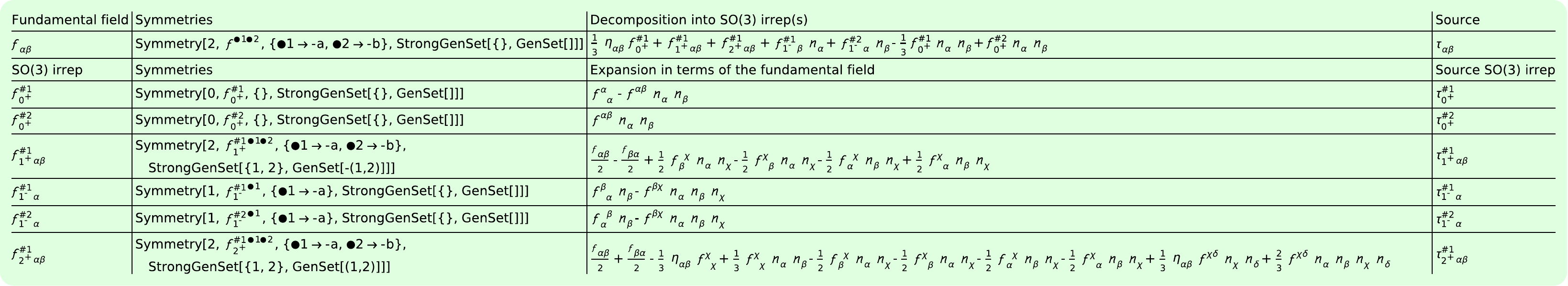}
	\caption{\label{FieldKinematicsF} Kinematic structure of the tetrad perturbation~$\FieldPert{_{\mu\nu}}$, as used in~\cref{PGTAction,eWGTAction}. These definitions are used in~\crefrange{ParticleSpectrographGeneralCase}{ParticleSpectrographGeneralCasePGTNoVector}. See~\cite{Barker:2024juc} for further notational details. This is a vector graphic: all details are visible under magnification.}
\end{figure*}

\begin{figure*}[h]
\includegraphics[width=\textwidth,height=\textheight,keepaspectratio]{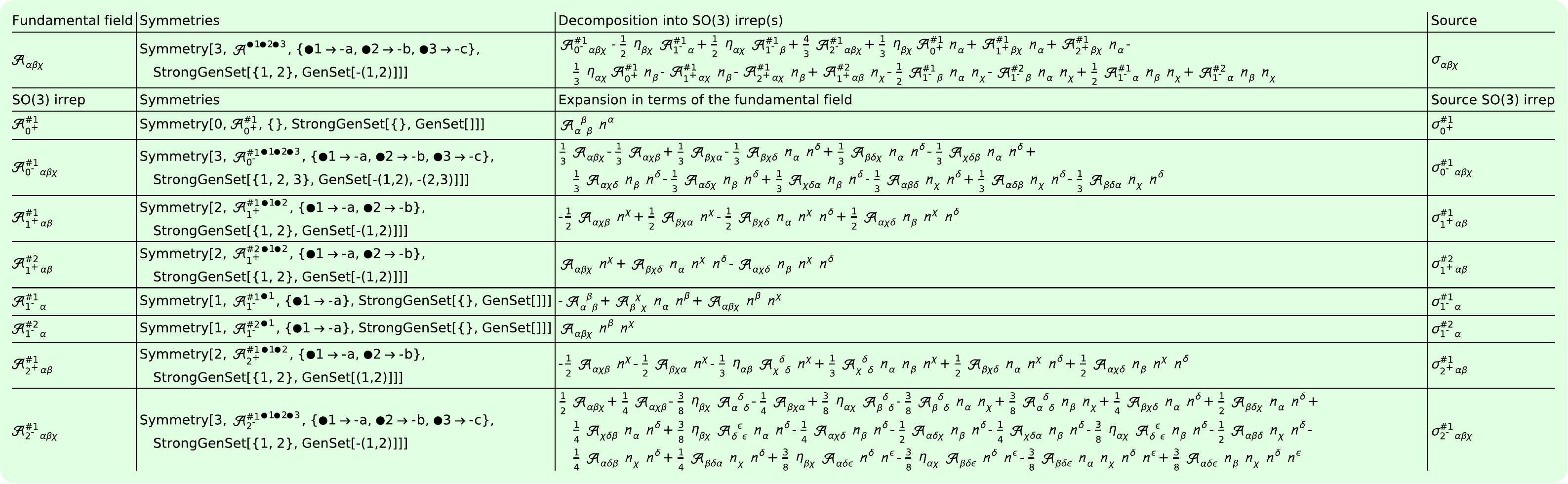}
	\caption{\label{FieldKinematicsA} Kinematic structure of the spin connection perturbation~$\FieldA{_{\mu\nu\sigma}}$, as used in~\cref{PGTAction,eWGTAction}. Note that the~$2^-$ state has a hidden multi-term cyclic symmetry on all its indices, which is not accommodated by the C language implementation of the Butler--Portugal algorithm~\cite{Martin-Garcia:2008ysv}. These definitions are used in~\crefrange{ParticleSpectrographGeneralCase}{ParticleSpectrographGeneralCasePGTNoVector}. See~\cite{Barker:2024juc} for further notational details. This is a vector graphic: all details are visible under magnification.}
\end{figure*}

The particle spectra are obtained using the \emph{Particle Spectrum for Any Tensor Lagrangian} (\PSALTer{}) software~\cite{Barker:2023bmr,Barker:2024ydb,Barker:2024dhb,Barker:2024juc}. Recall that the kinematic variables of PGT were~$\big\{\FieldB{^i_\mu},\FieldA{^{ij}_\mu}\big\}$, and that the inverse~$\FieldH{_i^\mu}$ is determined by~$\FieldB{^i_\mu}$. This means that we are free to use~$\big\{\FieldH{_i^\mu},\FieldA{^{ij}_\mu}\big\}$ instead. Working in the weak-field regime, we take~$\FieldA{^{ij}_\mu}$ to be inherently perturbative. We define the exact perturbation~$\FieldH{_i^\mu}\equiv\FieldDelta{_i^\mu}+\FieldPert{_i^\mu}$, where we make the `Kronecker' choice of Minkowski vacuum (see alternative vacua~\cite{Barker:2023bmr,Blixt:2022rpl,Blixt:2023qbg}). There are a priori 16 d.o.f in~$\FieldPert{_i^\mu}$, and 24 d.o.f in~$\FieldA{^{ij}_\mu}$. Conjugate to~$\FieldPert{_i^\mu}$ and~$\FieldA{^{ij}_\mu}$ are the translational source (i.e. the asymmetric stress-energy tensor)~$\FieldTau{^i_\mu}$, and the matter spin current~$\FieldSpin{_{ij}^\mu}$~\cite{Rigouzzo:2023sbb,Rigouzzo:2022yan,Karananas:2021zkl}. To lowest order in the field perturbations, the Greek and Roman indices are interchangeable. For this reason the final set of fields used in the linearised analysis is~$\big\{\FieldPert{_{\mu\nu}},\FieldA{_{\mu\nu\sigma}}\big\}$, with conjugate sources~$\big\{\FieldTau{^{\mu\nu}},\FieldSpin{^{\mu\nu\sigma}}\big\}$. The various~$\mathrm{SO}(3)$ irreducible parts of these quantities are presented in~\cref{FieldKinematicsF,FieldKinematicsA}, and they have spin-parity ($J^P$) labels to identify them. Duplicate~$J^P$ states can arise, and these are distinguished by extra labels~$\#1$,~$\#2$, etc.

\begin{figure*}[ht]
\includegraphics[width=\textwidth]{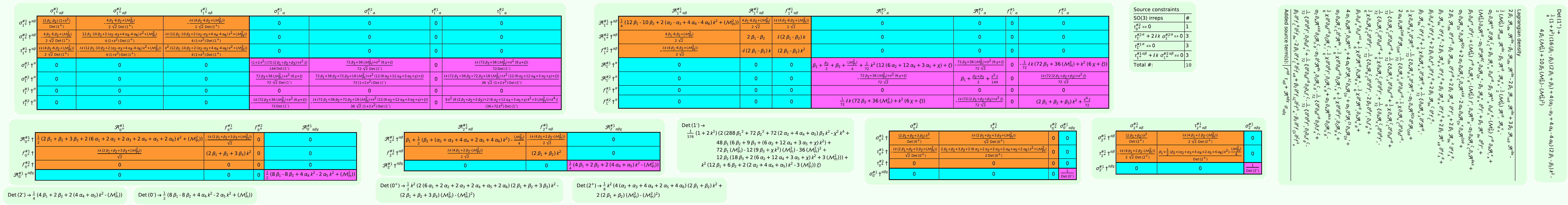}
	\caption{\label{ParticleSpectrographGeneralCase} Partial particle spectrograph of the completely general theory in~\cref{eWGTAction}. From the~$1^-$ sector of the saturated propagator we may read off the quartic pole in~\cref{MasterConstraint}. All quantities are defined in~\cref{FieldKinematicsF,FieldKinematicsA}. See~\cite{Barker:2024juc} for further notational details. This is a vector graphic: all details are visible under magnification.}
\end{figure*}

\begin{figure*}[ht]
\includegraphics[width=\textwidth]{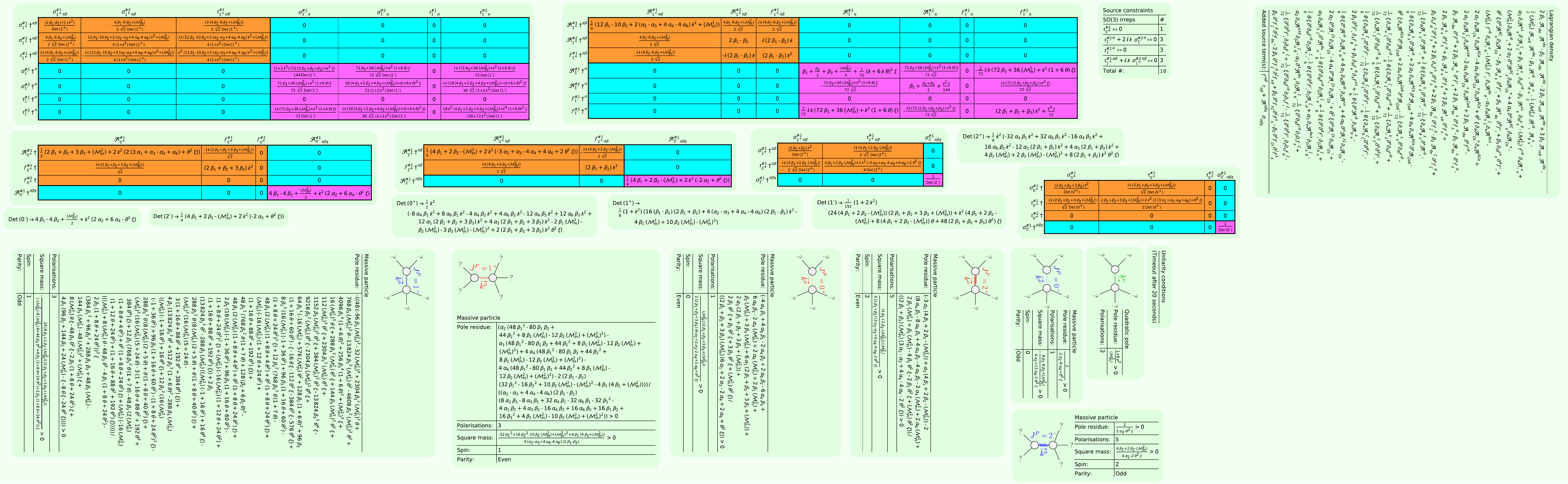}
	\caption{\label{ParticleSpectrographGeneralCaseNoQuartic} Particle spectrograph of the branch~$\Alp{5}=\chi^2/\xi-2\left(\Alp{2}+2\Alp{4}\right)$ of~\cref{eWGTAction}, which gives rise to the mass spectra in~\crefrange{Mass0p}{Mass2m}. All quantities are defined in~\cref{FieldKinematicsF,FieldKinematicsA,Eliminations}. See~\cite{Barker:2024juc} for further notational details. This is a vector graphic: all details are visible under magnification.}
\end{figure*}

\begin{figure*}[ht]
\includegraphics[width=\textwidth]{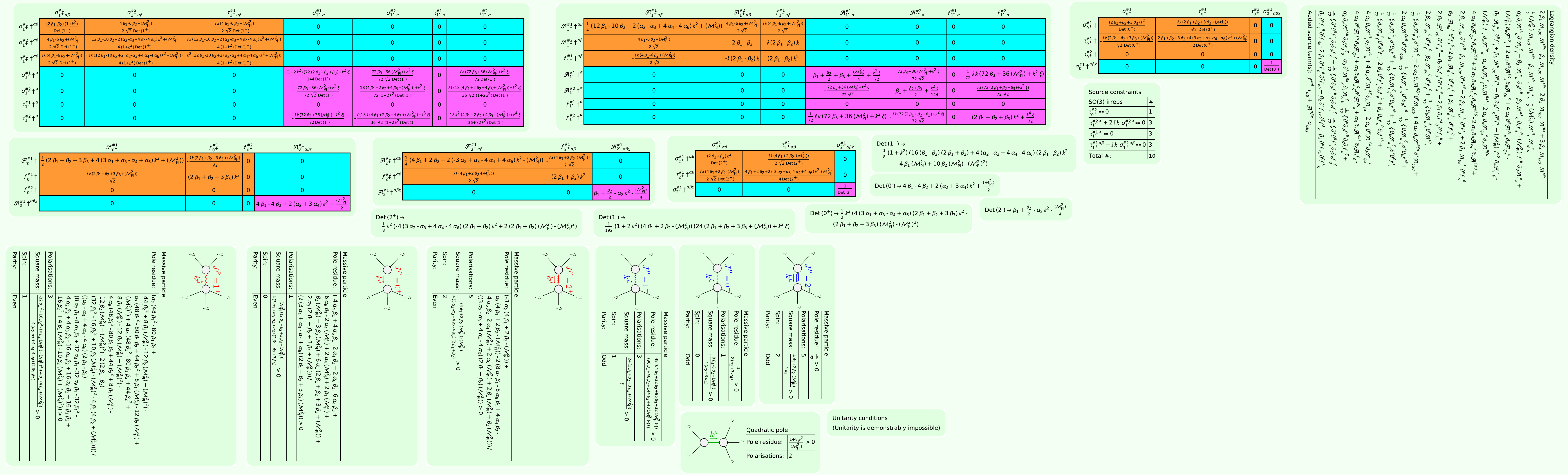}
	\caption{\label{ParticleSpectrographGeneralCaseNoQuarticSplit} Particle spectrograph of the branch~$\Alp{5}=-2\left(\Alp{2}+2\Alp{4}\right)$ with~$\chi=0$ of~\cref{eWGTAction}, which gives rise to the~$\chi\mapsto 0$ limit of the mass spectra in~\crefrange{Mass0p}{Mass2m} (see~\cref{ParticleSpectrographGeneralCaseNoQuartic}). All quantities are defined in~\cref{FieldKinematicsF,FieldKinematicsA,Eliminations}. See~\cite{Barker:2024juc} for further notational details. This is a vector graphic: all details are visible under magnification.}
\end{figure*}

\begin{figure*}[ht]
\includegraphics[width=\textwidth]{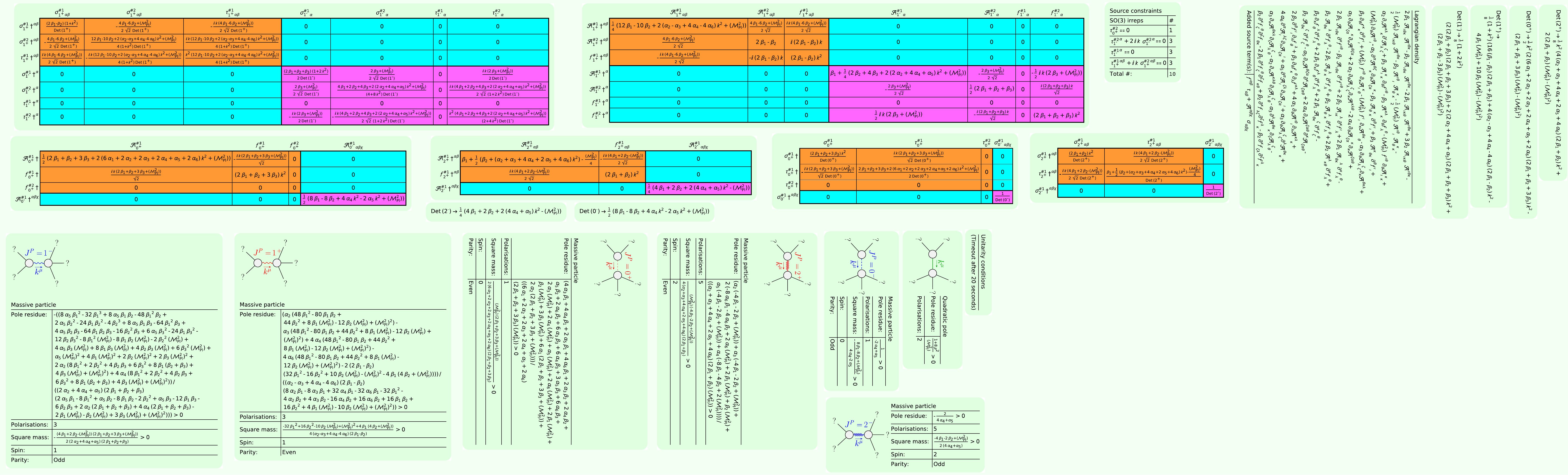}
	\caption{\label{ParticleSpectrographGeneralCasePGT} Particle spectrograph of the branch~$\xi=\chi=0$ of~\cref{eWGTAction}, which gives rise to the mass spectra in~\crefrange{NewMass0p}{NewMass2m}, as first presented in~\cite{Hayashi:1980qp}. All quantities are defined in~\cref{FieldKinematicsF,FieldKinematicsA}. See~\cite{Barker:2024juc} for further notational details. This is a vector graphic: all details are visible under magnification.}
\end{figure*}

\begin{figure*}[ht]
\includegraphics[width=\textwidth]{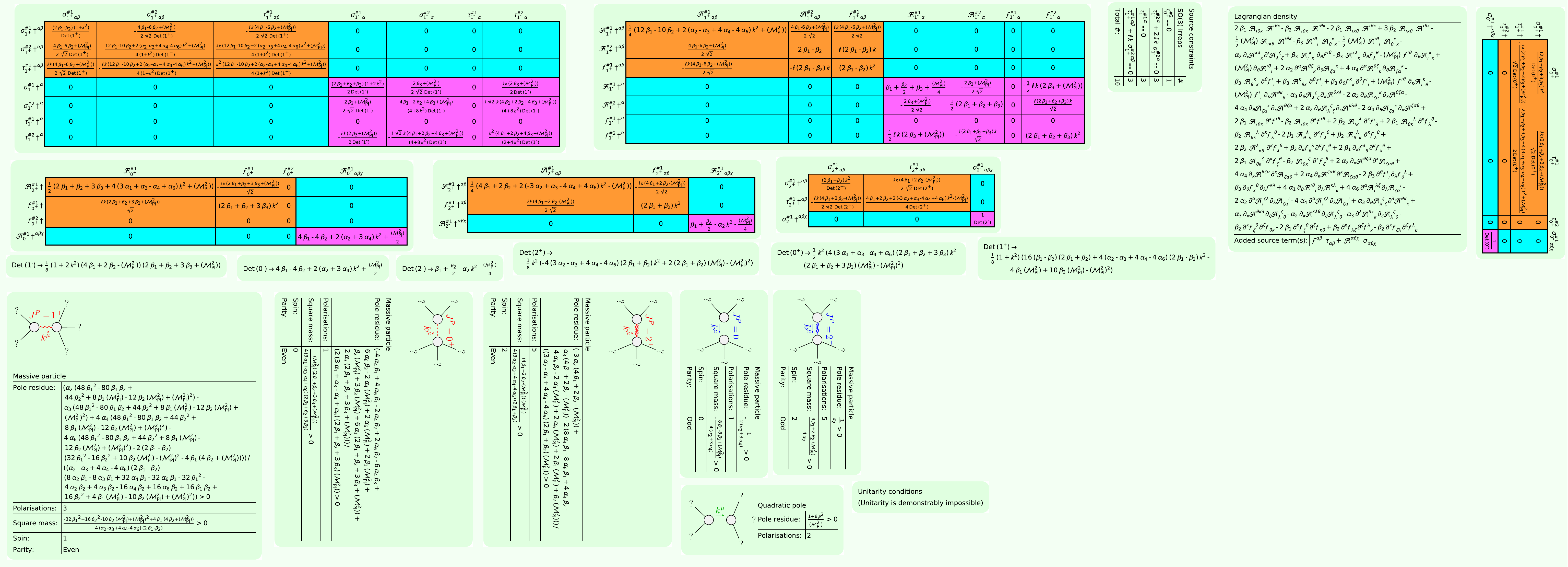}
	\caption{\label{ParticleSpectrographGeneralCasePGTNoVector} Particle spectrograph of the branch~$\Alp{5}=-2\left(\Alp{2}+2\Alp{4}\right)$ of~\cref{PGTAction}, confirming that the~$1^-$ mode is removed. All quantities are defined in~\cref{FieldKinematicsF,FieldKinematicsA}. See~\cite{Barker:2024juc} for further notational details. This is a vector graphic: all details are visible under magnification.}
\end{figure*}

The actual analysis of the theory in~\cref{eWGTAction} was performed across 64 AMD\textsuperscript{\textregistered} \emph{Ryzen Threadripper} CPUs. For the completely unrestricted case in~\cref{eWGTAction} the mass expressions associated with the roots of the quartic pole are expected to be very cumbersome, so we abort the computation before they are computed: the structure of the wave operator and saturated propagator, without the mass spectrum, is shown in~\cref{ParticleSpectrographGeneralCase}. Before the evaluation of the `general' branch for which the~$k^4$ coefficient in~\cref{MasterConstraint} is removed, the condition~$\Alp{5}=\chi^2/\xi-2\left(\Alp{2}+2\Alp{4}\right)$ must be imposed on the linearised action. To avoid Lagrangian couplings appearing on the denominator of the linearised action\footnote{As a mild technical limitation, coupling coefficients in the denominator can cause \PSALTer{} to become slow.} we impose the constraint by introducing a new coupling~$\theta$ and setting in~\cref{eWGTAction}
\begin{equation}\label{Eliminations}
	\Alp{5}\mapsto\theta^2\xi-2\left(\Alp{2}+2\Alp{4}\right), \quad \chi\mapsto\theta\xi.
\end{equation}
The results are shown in~\cref{ParticleSpectrographGeneralCaseNoQuartic}, we see that apart from the two massless polarisations of the Einstein graviton, up to six torsion particles can propagate with the following square masses, where we eliminate~$\theta$ in terms of~$\chi$ and~$\xi$ using~\cref{Eliminations}
\begin{widetext}
\begin{subequations}
\begin{align}
	\MassSquared{0^+}&\equiv\frac{\MPl{}^2\left(2\Bet{1}+\Bet{2}+3\Bet{3}+\MPl{}^2\right)}{2\left(2\Bet{1}+\Bet{2}+3\Bet{3}\right)\left(6\Alp{1}+2\Alp{3}-2\Alp{4}+2\Alp{6}+\chi^2/\xi\right)},\label{Mass0p}\\
	\MassSquared{0^-}&\equiv-\frac{8\Bet{1}-8\Bet{2}+\MPl{}^2}{4\Alp{2}+12\Alp{4}-2\chi^2/\xi},\label{Mass0m}\\
	\MassSquared{1^+}&\equiv\frac{-32\Bet{1}^2+16\Bet{2}^2-10\Bet{2}\MPl{}^2+\MPl{}^4+4\Bet{1}\left(4\Bet{2}+\MPl{}^2\right)}{4\left(\Alp{2}-\Alp{3}+4\Alp{4}-4\Alp{6}\right)\left(2\Bet{1}-\Bet{2}\right)} ,\label{Mass1p}\\
	\MassSquared{1^-}&\equiv-\frac{24\left(4\Bet{1}+2\Bet{2}-\MPl{}^2\right)\left(2\Bet{1}+\Bet{2}+3\Bet{3}+\MPl{}^2\right)}{\xi\left(48\Bet{3}\chi^3/\xi^3+4\Bet{1}\left(1+8\chi/\xi+24\chi^2/\xi^2\right)+2\Bet{2}\left(1+8\chi/\xi+24\chi^2/\xi^2\right)-\MPl{}^2-8\MPl{}^2\chi/\xi\right)},\label{Mass1m}\\
	\MassSquared{2^+}&\equiv\frac{\left(4\Bet{1}+2\Bet{2}-\MPl{}^2\right)\MPl{}^2}{4\left(2\Bet{1}+\Bet{2}\right)\left(3\Alp{2}-\Alp{3}+4\Alp{4}-4\Alp{6}-2\chi^2/\xi\right)},\label{Mass2p}\\
	\MassSquared{2^-}&\equiv\frac{4\Bet{1}+2\Bet{2}-\MPl{}^2}{4\Alp{2}-2\chi^2/\xi}.\label{Mass2m}
\end{align}
\end{subequations}
\end{widetext}
By adding up the~$1+1+3+3+5+5$ spin multiplicities of~\crefrange{Mass0p}{Mass2m} we recover the~$16+24$ kinematic d.o.f in~$\big\{\FieldB{^i_\mu},\FieldA{^{ij}_\mu}\big\}$, less the~$2\times(4+6)$ gauge d.o.f associated with Poincar\'e gauge symmetry --- the only surviving symmetry of the embedded theory. The branch~$2\Alp{2}+4\Alp{4}+\Alp{5}=\chi=0$ is likewise found by setting~$\Alp{5}\mapsto-2\left(\Alp{2}+2\Alp{4}\right)$ and~$\chi\mapsto 0$ in~\cref{eWGTAction}. The analysis is presented in~\cref{ParticleSpectrographGeneralCaseNoQuarticSplit}, and shows that the mass spectrum is fully consistent with imposing~$\chi\mapsto 0$ in~\crefrange{Mass0p}{Mass2m}. Finally, the case~$\xi=\chi=0$ in~\cref{eWGTAction} without any further restrictions leads in~\cref{ParticleSpectrographGeneralCasePGT} to the general case of PGT in~\cref{PGTAction}, which was found previously in~\cite{Hayashi:1980qp}, namely
\begin{widetext}
\begin{subequations}
\begin{align}
	\MassSquared{0^+}&\equiv\frac{\MPl{}^2\left(2\Bet{1}+\Bet{2}+3\Bet{3}+\MPl{}^2\right)}{2\left(6\Alp{1}+2\Alp{2}+2\Alp{3}+2\Alp{4}+\Alp{5}+2\Alp{6}\right)\left(2\Bet{1}+\Bet{2}+3\Bet{3}\right)},\label{NewMass0p}\\
	\MassSquared{0^-}&\equiv-\frac{8\Bet{1}-8\Bet{2}+\MPl{}^2}{4\Alp{4}-2\Alp{5}},\label{NewMass0m}\\
	\MassSquared{1^-}&\equiv-\frac{\left(4\Bet{1}+2\Bet{2}-\MPl{}^2\right)\left(2\Bet{1}+\Bet{2}+3\Bet{3}+\MPl{}^2\right)}{2\left(2\Alp{2}+4\Alp{4}+\Alp{5}\right)\left(2\Bet{1}+\Bet{2}+\Bet{3}\right)},\label{NewMass1m}\\
	\MassSquared{2^+}&\equiv\frac{\MPl{}^2\left(-4\Bet{1}-2\Bet{2}+\MPl{}^2\right)}{4\left(\Alp{2}+\Alp{3}+4\Alp{4}+2\Alp{5}+4\Alp{6}\right)\left(2\Bet{1}+\Bet{2}\right)},\label{NewMass2p}\\
	\MassSquared{2^-}&\equiv\frac{-4\Bet{1}-2\Bet{2}+\MPl{}^2}{2\left(4\Alp{4}+\Alp{5}\right)}.\label{NewMass2m}
\end{align}
\end{subequations}
\end{widetext}
Whilst we confirmed already that the~$\chi\to 0$ limit of~\crefrange{Mass0p}{Mass2m} was continuous, it is evident from~\cref{Mass1m} that once~$\chi\to 0$ has been taken the limit~$\xi\to 0$ eliminates the~$1^-$ vector in which we have been so interested. In fact, this is not surprising, because taking~$\chi\to 0$ at finite~$\xi$ corresponds to setting~$2\Alp{2}+4\Alp{4}+\Alp{5}=0$. From~\cref{NewMass1m} we see how this latter condition would equivalently kill the~$1^-$ mode in~\cref{PGTAction} (we confirm this in~\cref{ParticleSpectrographGeneralCasePGTNoVector}), and the~\cref{PGTAction} theory-space is reached by~$\xi\to 0$.

\begin{figure*}[ht]
\includegraphics[width=\textwidth]{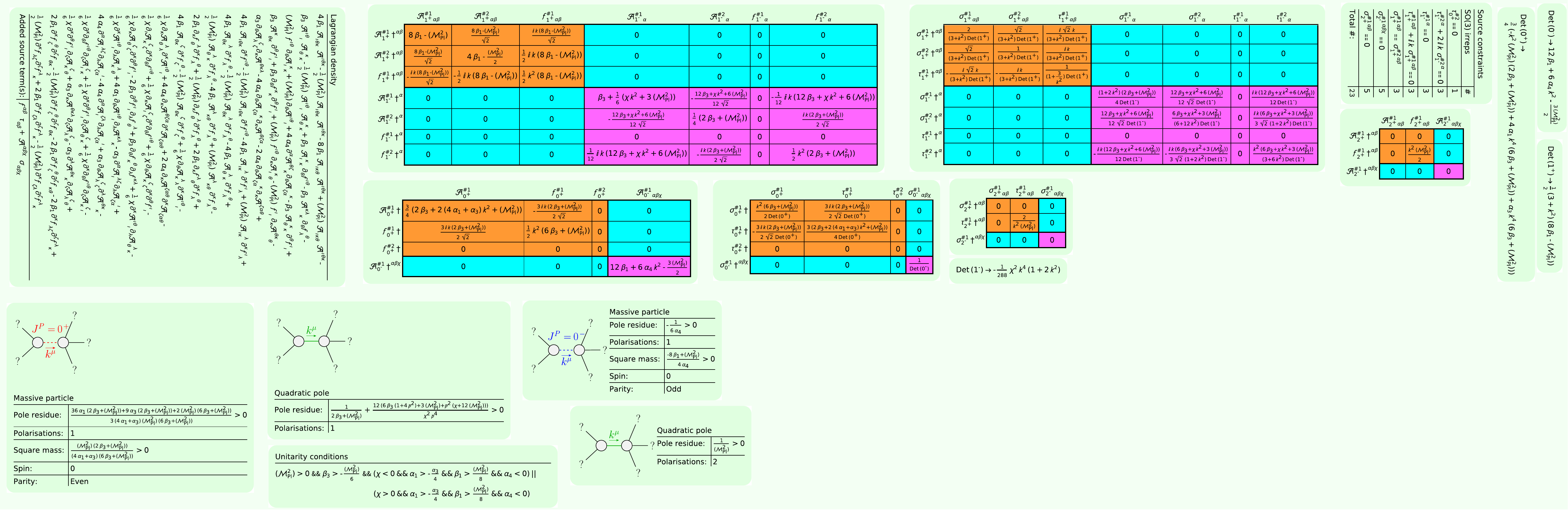}
	\caption{\label{ParticleSpectrographBranch1Conservative} Particle spectrograph of the branch~$\Alp{2}=\Alp{3}-4\Alp{4}+\Alp{6}=4\Alp{4}+\Alp{5}=4\Bet{1}+2\Bet{2}-\MPl{}^2=0$ of~\cref{eWGTAction}. All quantities are defined in~\cref{FieldKinematicsF,FieldKinematicsA}. See~\cite{Barker:2024juc} for further notational details. This is a vector graphic: all details are visible under magnification.}
\end{figure*}
\begin{figure*}[ht]
\includegraphics[width=\textwidth]{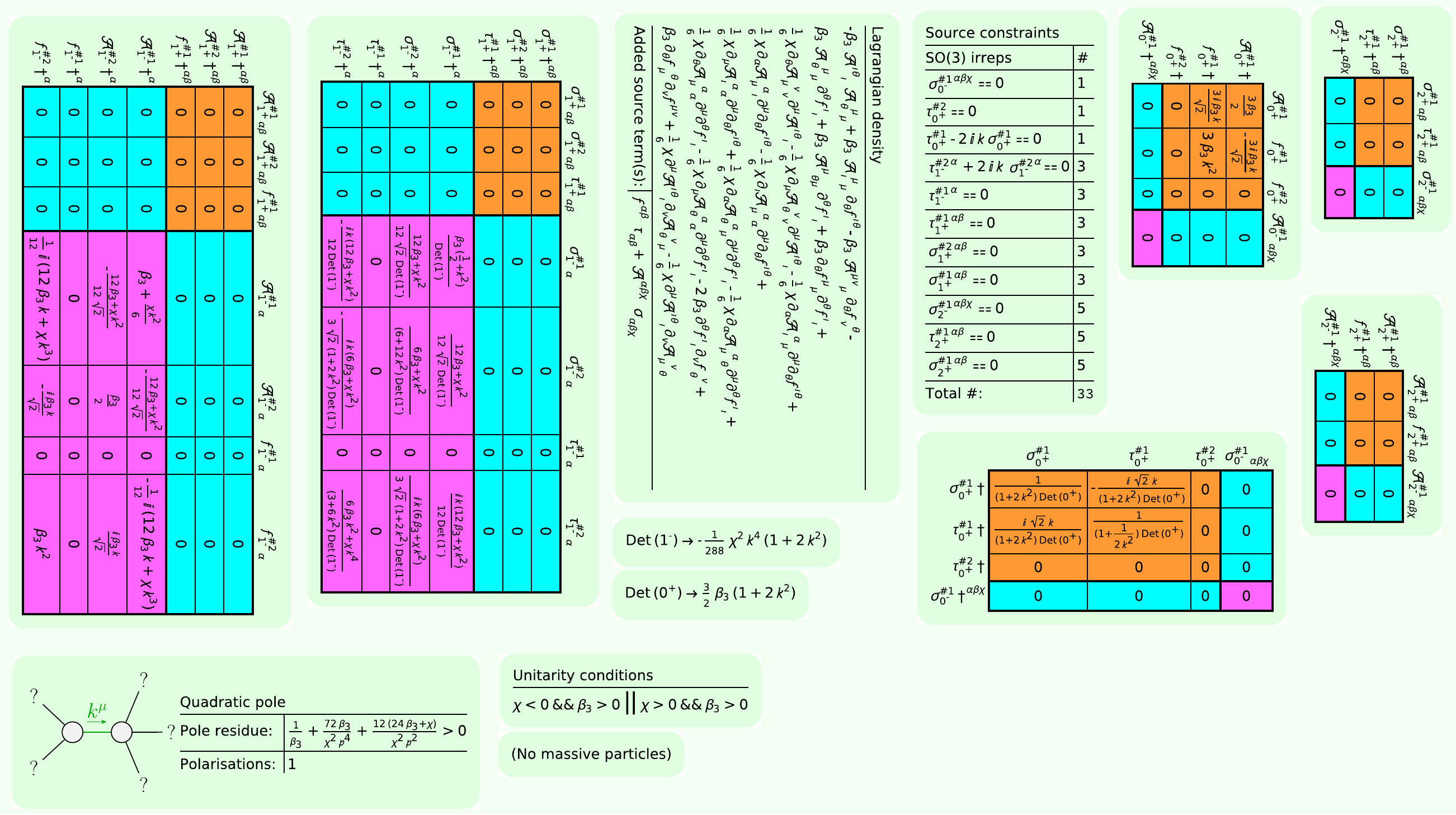}
	\caption{\label{ParticleSpectrographSwitchonPGT} Particle spectrograph of~\cref{SwitchonPGT}, i.e. the minimal PGT realisation of the novel massless scalar particle which appears in~\cref{ParticleSpectrographBranch1Conservative}. All quantities are defined in~\cref{FieldKinematicsF,FieldKinematicsA}. See~\cite{Barker:2024juc} for further notational details. This is a vector graphic: all details are visible under magnification.}
\end{figure*}
\begin{figure*}[ht]
\includegraphics[width=\textwidth]{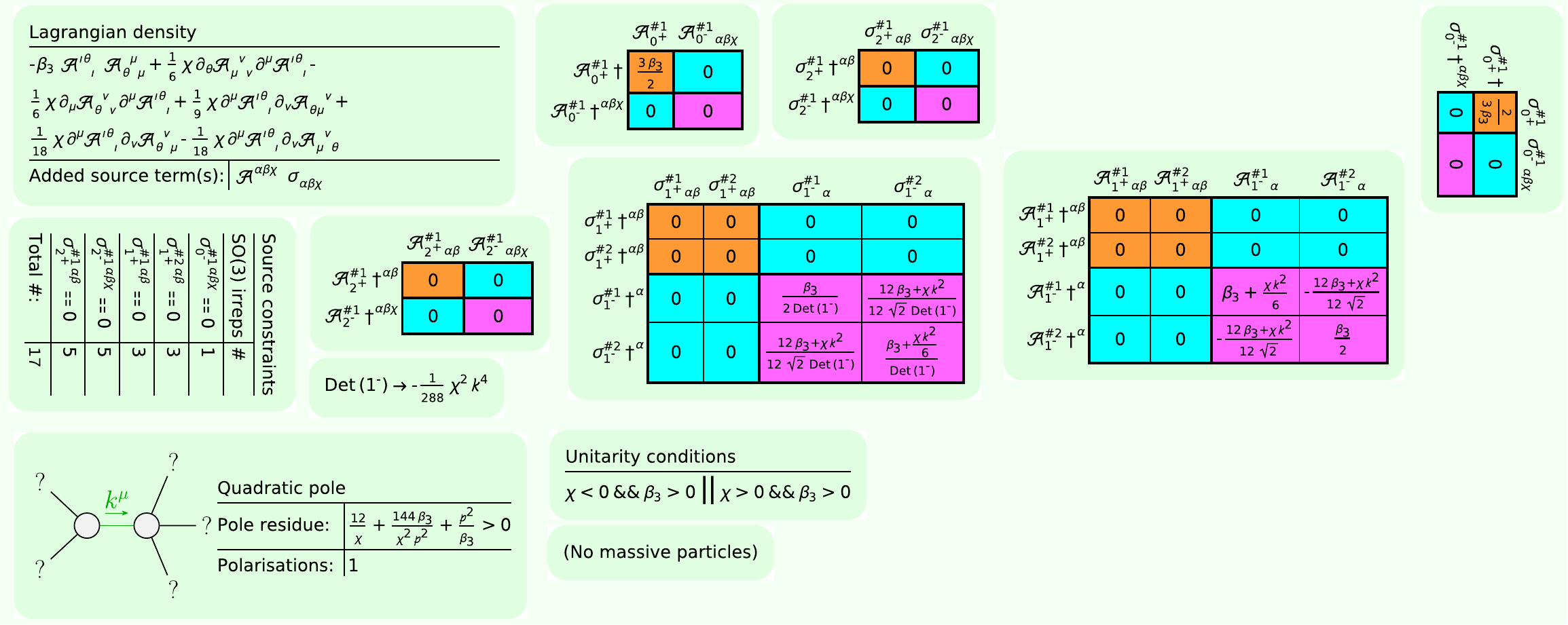}
	\caption{\label{ParticleSpectrographSwitchonNoAxialVector} Particle spectrograph of~\cref{NewCurtright}, i.e. the minimal model of the novel scalar particle which appears in~\cref{ParticleSpectrographBranch1Conservative,ParticleSpectrographSwitchonPGT}. We use the rotational gauge field perturbation~$\FieldA{_{\mu\nu\sigma}}\equiv\FieldA{_{[\mu\nu]\sigma}}$ as a proxy for the Curtright field~$\Curtright{_{\mu\nu\sigma}}\equiv\Curtright{_{[\mu\nu]\sigma}}$ with~$\Curtright{_{[\mu\nu\sigma]}}\equiv 0$, because the latter multi-term symmetry is a difficult attribute to declare in the \PSALTer{} framework. It can be confirmed (by applying transformation rules to the Lagrangian density in this figure, and repeating the analysis) that the axial part~$\FieldA{_{[\mu\nu\sigma]}}$ does \emph{not} contribute to the spectrum. All quantities are defined in~\cref{FieldKinematicsF,FieldKinematicsA}. See~\cite{Barker:2024juc} for further notational details. This is a vector graphic: all details are visible under magnification.}
\end{figure*}

\end{document}